\documentclass[twocolumn,aps,prd,showpacs,nofootinbib,superscriptaddress,preprintnumbers,floatfix,10pt]{revtex4-1}

\usepackage{amsmath}
\usepackage{amsfonts}
\usepackage{amssymb}
\usepackage{epsfig}
\usepackage{graphicx}
\usepackage{color}
\usepackage{slashed}
\usepackage{epstopdf}

\usepackage{hyperref}

\newcommand{\STr}{\text{STr}}

\newcommand{\Eqref}[1]{Eq.~\eqref{#1}}

\newcommand{\pt}{\partial_t}
\newcommand{\yb}{\bar{\psi}}

\newcommand{\scpotential}{single-scale potential}
\newcommand{\scpotentialem}{\textit{single-scale} potential}
\newcommand{\Ueffsc}{U_{\text{eff}}^{\text{S}}}

\newcommand{\Umf}{U^{\text{MF}}}

\newcommand{\mL}{m_\Lambda^2}
\newcommand{\lL}{\lambda_{2,\Lambda}}
\newcommand{\mH}{m_{\text{H}}}
\newcommand{\mtop}{m_{\text{t}}}

\newcommand{\Np}{N_{\text{p}}}

\begin{document}

\preprint{}

\title {Global flow of the Higgs potential in a Yukawa model} 

\author{Julia Borchardt}
\email{julia.borchardt@uni-jena.de}
\affiliation{Theoretisch-Physikalisches Institut, %
Friedrich-Schiller-Universit\"at Jena, Max-Wien-Platz 1, D-07743 Jena, Germany}
\affiliation{Abbe Center of Photonics, Friedrich Schiller University Jena, Max Wien 
Platz 1, 07743 Jena, Germany}
\author{Holger Gies}
\email{holger.gies@uni-jena.de}
\affiliation{Theoretisch-Physikalisches Institut, %
Friedrich-Schiller-Universit\"at Jena, Max-Wien-Platz 1, D-07743 Jena, Germany}
\affiliation{Abbe Center of Photonics, Friedrich Schiller University Jena, Max Wien 
Platz 1, 07743 Jena, Germany}
\affiliation{Helmholtz-Institut Jena, Fr\"obelstieg 3, D-07743 Jena, Germany}
\author{Ren\'{e} Sondenheimer}
\email{rene.sondenheimer@uni-jena.de}
\affiliation{Theoretisch-Physikalisches Institut,  %
Friedrich-Schiller-Universit\"at Jena, Max-Wien-Platz 1, D-07743 Jena, Germany}

\begin{abstract}
We study the renormalization flow of the Higgs potential as a function
of both field amplitude and energy scale. This overcomes limitations
of conventional techniques that rely, e.g., on an identification of
field amplitude and RG scale, or on local field expansions. Using a
Higgs-Yukawa model with discrete chiral symmetry as an example, our
global flows in field space clarify the origin of possible
meta-stabilities, the fate of the pseudo-stable
phase, and provide new information about the renormalization of the
tunnel barrier. Our results confirm the relaxation of the lower bound
for the Higgs mass in the presence of more general microscopic
interactions (higher-dimensional operators) to a high quantitative accuracy.

\end{abstract}


\maketitle

\section{Introduction}
\label{intro}

The discovery of a Higgs boson at the LHC
\cite{Aad:2012tfa,Chatrchyan:2012xdj} completed the search for the
building blocks of the standard model of particle physics. While the
mass of the Higgs boson is in principle a \textit{free} parameter of
the standard model, it has long been known that it is not necessarily
an \textit{arbitrary} parameter. For instance, assuming that the
description of fundamental physics in terms of standard-model degrees
of freedom is valid at a high energy scale $\Lambda$ and that the
theory is sufficiently weakly coupled, the range of possible Higgs
boson masses is restricted by a finite range, the so-called infrared
(IR) window
\cite{Maiani:1977cg,Krasnikov:1978pu,Lindner:1985uk,Wetterich:1987az,Altarelli:1994rb,Schrempp:1996fb,Hambye:1996wb}. The
fact that this IR window shrinks for increasing $\Lambda$ can be
traced back to the fixed-point structure of the RG flow
\cite{Wetterich:1987az,Wetterich:1981ir} which connects the mass of
the Higgs boson to that of the heaviest quark.\footnote{This implies
  that the results of the present paper could equally well be
  rephrased in terms of bounds on the top mass. This might even be the
  more relevant viewpoint \cite{Bezrukov:2014ina}, as the top mass is
  presently known less precisely. However, we stick to the ``Higgs
  boson mass'' perspective in order to conform with a larger part of
  the literature.}

The edges of the IR window
\cite{Krive:1976sg,Hung:1979dn,Linde:1979ny,Cabibbo:1979ay,Politzer:1978ic,Kuti:1987nr,Sher:1988mj,Hasenfratz:1987eh,Luscher:1988uq,Lindner:1988ww,Ford:1992mv,Heller:1993yv,Arnold:1989cb,Sher:1993mf,Altarelli:1994rb,Casas:1994qy,Espinosa:1995se,Bergerhoff:1999jj,Isidori:2001bm,Ellis:2009tp,Holthausen:2011aa,EliasMiro:2011aa,Degrassi:2012ry,Alekhin:2012py,Masina:2012tz,Buttazzo:2013uya,Gabrielli:2013hma,Bednyakov:2015sca},
i.e., the upper and lower admissible values, for the Higgs mass are
actually not sharply defined, but depend on a number of additional
assumptions. This is most obvious for the upper ``triviality bound'',
as the Higgs sector becomes strongly coupled at high scales for large
values of the Higgs mass. Perturbative estimates of this bound, e.g.,
depend on an ad hoc choice of coupling value up to which perturbation
theory is trusted. Nonperturbative methods have shown that this upper
bound relaxes considerably if one allows the system to start
microscopically with a strong Higgs self-coupling
\cite{Gies:2013fua,Gies:2014xha}.  

As is less well appreciated, a similar fuzziness also holds for the
lower edge of the IR window on which we concentrate in the present
work. This indeterminacy arises from the (implicit) assumptions
imposed on the precise form of the microscopic theory at the high
scale $\Lambda$. For instance, by considering only renormalizable
operators in conventional perturbative estimates of the lower bound,
the couplings of all higher-order operators are implicitly chosen to
vanish at the high scale $\Lambda$. Strictly speaking, this
corresponds to fixing infinitely many further parameters, in addition
to the parameters of the standard model.\footnote{This statement holds
  for any finite value of $\Lambda$. If the limit $\Lambda\to \infty$
  could be taken, e.g., at an asymptotically free or asymptotically
  safe fixed point, all these parameters would be predictions of the
  theory. Whether the standard model could have this property is not
  fully understood, hence $\Lambda$ presumably remains a finite though
  unknown parameter of the model.}

By contrast, if one defines the standard model more agnostically in
terms of its symmetries, field content and measured IR parameters, the
microscopic interactions quantified by the bare action at the high
scale $\Lambda$ remains largely unspecified. It can contain infinitely
many operators and corresponding coupling constants which would be
computable if the underlying theory was known. It is reasonable to
assume that these couplings if measured in units of the scale
$\Lambda$ are of order $\mathcal{O}(1)$. Still, an even more strongly
coupled UV regime is not excluded by observation. The fact that the
perturbative description of electroweak collider data works so well
indicates that colliders so far probe a regime where Nature is close
to the Gau\ss{}ian fixed point of the renormalization group
(RG). Here, power-counting arguments hold and higher-dimensional
operators are irrelevant, exerting a negligible influence on IR
observables. In turn, the IR observables dominantly constrain only the
marginal and relevant (renormalizable) operators of the bare action
and put hardly any bound on the irrelevant couplings.

In a series of works, it has recently been shown that these
unconstrained higher-dimensional operators in fact can relax the lower
edge of the IR window, i.e., can lower the lower (stability) bound on
the Higgs mass without introducing metastability
\cite{Gies:2013fua,Gies:2014xha,Eichhorn:2015kea}. Comparatively
simple modifications of the bare action, e.g., in terms of a
dimension-six operator at the Planck scale can lower the lower mass
bound by $\sim 1$GeV, while preserving absolute stability on all
scales \cite{Eichhorn:2015kea}. The mechanism behind this relaxation
of the lower edge has two aspects: first, negative couplings of
renormalizable operators which seem to introduce an instability from a
perturbative viewpoint can still be associated with a fully stable
potential in the presence of the higher-dimensional operators. Second,
the higher-dimensional operators take influence on the running of the
renormalizable part over a range of scales thus modifying the
\textit{approach} to the perturbative region while leaving the IR region
itself intact as addressed by perturbation theory.

These results can already be observed in a simple mean-field study
(large-$N$ limit) as well as in extended mean-field approximation
(1/$N$ corrections), but unfold more comprehensively in a controlled
nonperturbative study using the functional RG. They have been
confirmed by lattice simulations in the range of scales accessible to
current lattice sizes
\cite{Hegde:2013mks,Chu:2015nha,Chu:2015ula,Akerlund:2015fya}. Recent functional RG
studies also including higher-order fermionic operators have shown the
same features \cite{Eichhorn:2014qka,Jakovac:2015kka,Jakovac:2015iqa}; for further
studies of higher-dimensional operators in this context, see, e.g.,
 \cite{Datta:1996ni,Barbieri:1999tm,Grzadkowski:2001vb,Burgess:2001tj,Barger:2003rs,Blum:2015rpa}.

While controlled quantitative results have so far been obtained for
a small class of operators represented by simple
low-order polynomials of the field, a possible metastable regime with
competing vacua has not been explored so far.

The present work is devoted to a first step in this direction, namely
to study the full functional renormalization of the Higgs effective
potential as a function of both field amplitude and RG scale. This is
facilitated by the development of pseudo-spectral methods for
functional flows \cite{Borchardt:2015rxa,Borchardt:2016}. If applied
to the fully stable regimes studied before, our results confirm the
lowering of the Higgs mass bounds to a high accuracy. In addition, the
full potential solver allows to address the fate of the RG flow in
potential metastable regimes and the renormalization of the tunnel
barrier.

\begin{figure}
 \includegraphics[width=0.45\textwidth]{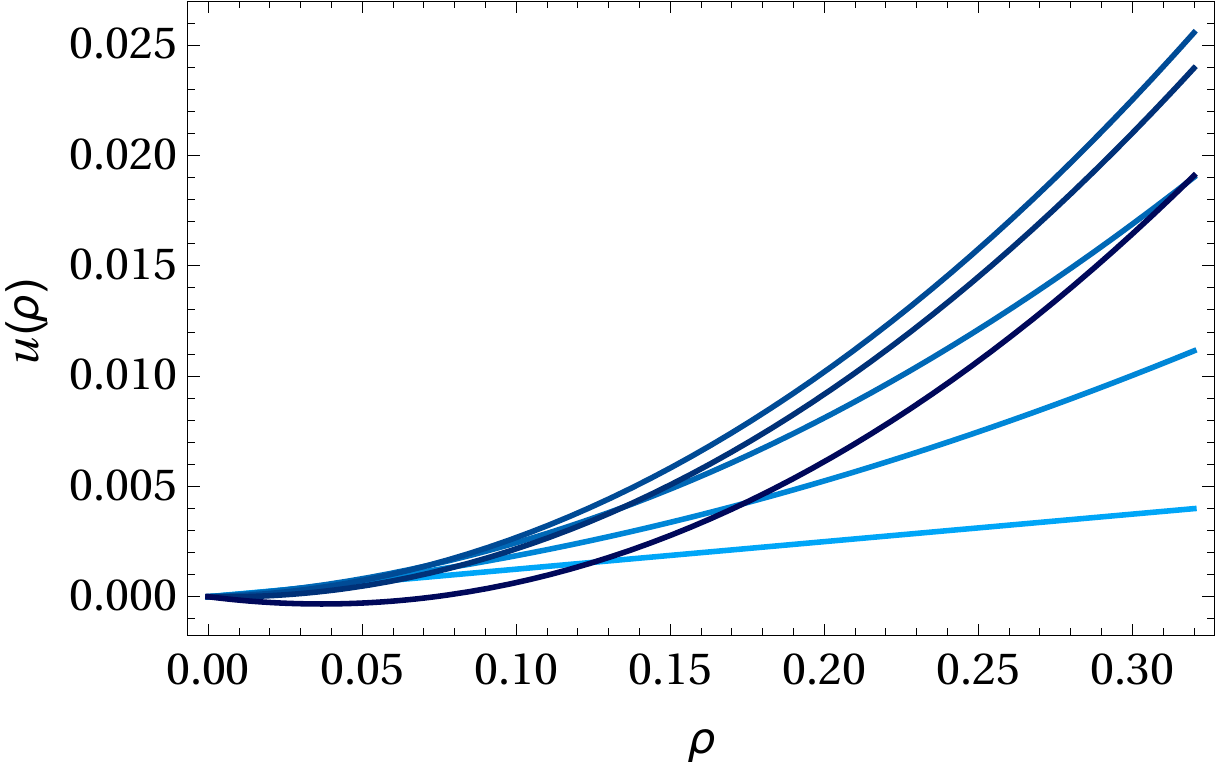}
 \caption{Example of the RG flow (from blue to black) of the
   dimensionless effective potential $u$ in dependence of the
   dimensionless field amplitude $\rho\sim \phi^2$. Here, the UV
   potential has been chosen as linear in $\rho$; see
   \Eqref{eq:rhoconv} below for conventions. The flow results from a
   $\beta$ functional of the potential $u$, defined for each value of
   the field amplitude $\beta[u](\rho)$, see \Eqref{eq:FlowPotential}.}
 \label{fig:potflow}
\end{figure}

Our present study is performed within a simple Higgs-Yukawa model with
a discrete chiral symmetry, which has proved useful for addressing the
qualitative properties of the IR window. Our main results are read off
from the properties of the Higgs effective potential as a function of
field amplitude $\phi$ and RG scale $k$. A first illustration for such
a fully stable flow is shown in Fig.~\ref{fig:potflow},
where the dimensionless potential $u$ is depicted as a function of the
dimensionless field amplitude $\rho\sim\phi^2$ for various values of
$k$ ranging from the UV towards the IR (blue to black) where a vacuum
expectation value has developed.

The paper is organized as follows: in Sect.~\ref{sec:model}, we
introduce our toy model and briefly summarize some recent results
which are of relevance for our present
study. Section~\ref{sec:CW-vs-MF} is devoted to an extensive
mean-field study, for the first time also including the meta-stable
regime. Here, we also clarify the fate of a recently discovered
pseudo-stable phase \cite{Eichhorn:2015kea} and discuss the connection
with the effective \scpotential\ obtained from conventional
perturbative constructions. In Sect.~\ref{sec:PotentialFlow} we
investigate the full RG flow of the entire scalar potential by means
of the functional RG. We compare the local behavior around the
electroweak minimum as well as the global behavior with the results
obtained by polynomial expansion of the potential and the mean-field
estimates. While we still span the bare potential in terms of a few
polynomial operators in this work, the techniques used here will
facilitate future studies of general classes of bare actions.

\section{Higgs-Yukawa model with discrete symmetry}
 \label{sec:model}

\subsection{The model}
All mechanisms relevant for the present work can already be studied in
a simple Higgs-Yukawa model corresponding to the reduction of the
standard model to the top quark $\psi$ with the largest Yukawa
coupling, and a real scalar Higgs degree of freedom $\phi$. The
classical euclidean action of this model is given by:
\begin{align}
 S = \int_x \left[ \frac{1}{2} (\partial_\mu \phi)^2 + U_{\Lambda}(\phi) + \bar{\psi}i\slashed{\partial}\psi + i\bar{h} \, \phi\bar{\psi}\psi \right].
 \label{eq:action}
\end{align}
The model features a discrete chiral $\mathbb{Z}_2$ symmetry, 
$
 \psi \to e^{i\frac{\pi}{2}\gamma_5} \psi
$, 
$\bar{\psi} \to \bar{\psi} e^{i\frac{\pi}{2}\gamma_5}
$, 
$
 \phi \to -\phi,
$
%
mimicking the global part of the electroweak symmetry group, and
protecting the fermion against acquiring a mass term. No massless
Goldstone bosons appear after spontaneous symmetry breaking as
the symmetry is discrete. Hence, the particle spectrum is gapped in
the broken phase as in the standard model.  This toy model was
intensively discussed in the context of stability of the effective
potential in the literature, e.g.,
\cite{Holland:2003jr,Holland:2004sd,Branchina:2005tu,Gies:2013fua,Krajewski:2014vea}.

In order to make semi-quantitative contact with the standard model, we
impose Coleman-Weinberg renormalization conditions
\cite{Coleman:1973jx} on the effective potential obtained after
integrating out all fluctuations down to the IR,
\begin{eqnarray}
 U_{\text{eff}}'(\phi_0)=0, \quad \mH^2 &=& \phi_0^2 {U_\text{eff}}''(\phi_0), \quad \mtop^2=\phi_0^2
 h^2,
\label{eq:rencond}
\end{eqnarray}
where $\phi_0$ denotes the (renormalized) field value at the minimum
of the potential\footnote{For the plots below, the physically irrelevant
  zero point is chosen such that either $U(0)=0$ or $U(\phi_0)=0$
  depending on numerical convenience.}, and all couplings are also considered to be
renormalized at a suitable renormalization point $\mu$, e.g.,
$\mu=\phi_0$. In the present work, we choose
for the observable parameters: $\mtop=173$GeV for the top mass, and
$\phi_0\equiv v = 246$GeV. The Higgs mass $\mH$ then is treated as a
function of the cutoff and a functional of the bare action, $\mH=\mH[S_\Lambda;\Lambda]$.

Despite this apparent physical fixing, the simplified model, of
course, deviates quantitatively from the standard model in essential
aspects: for instance, whereas the center of the IR window for the
Higgs mass is near $\sim 150$GeV for a Planck scale cutoff in the
standard model \cite{Buttazzo:2013uya}, it is near $\sim 215$GeV for
the present simple model at high energy scales \cite{Eichhorn:2014qka} 
mainly due to the absence of the gauge sectors.

\subsection{Perturbative effective \scpotential}

In order to make contact with the conventional perturbative treatment,
we briefly sketch the standard line of argument to obtain an estimate
of the effective potential. For simplicity, we consider only the
one-loop level. Perturbatively, only the renormalizable operators of
the bare potential are considered,
\begin{align}
 U_{\Lambda} = \frac{m_{\Lambda}^2}{2}\phi^2 + \frac{\lambda_{2,\Lambda}}{8}\phi^4 
, \label{eq:Ubare}
\end{align}
featuring the bare mass parameter $m_{\Lambda}^2$ and bare $\phi^4$
coupling $\lambda_{2,\Lambda}$. The estimate for the effective
potential is based on the $\beta$ function for the renormalized
running coupling $\lambda_2$,
\begin{align}
 k \frac{d\lambda_2}{dk} \equiv 
 \pt \lambda_2 = \frac{1}{16\pi^2}(9\lambda_2^2 + 8h^2\lambda_2 - 16h^4),
 \label{eq:1loopbetafunc}
\end{align}
depending as well on the renormalized running Yukawa coupling $h$. For
the present line of argument, it suffices to ignore the running of $h$
(it will be fully included in our detailed studies later). The
discussion can even be simplified further by noting that the
$\lambda_2$-terms in \Eqref{eq:1loopbetafunc} are small compared to
the $h^4$ term for small Higgs masses and large top masses. In this
limit which corresponds to ignoring scalar fluctuations, the
integration of the $\beta$ function yields
\begin{align}
 \pt \lambda_2 &= -\frac{h^4}{\pi^2} \quad \Rightarrow \quad
 \lambda_2(k) = \lambda_{2,\mu} - \frac{h^4}{2 \pi^2} \ln{\frac{k^2}{\mu^2}},
\label{eq:1looph4}
\end{align}
with $\mu$ denoting the renormalization point for $\lambda_2$.

The conventional perturbative estimate of the effective potential is
then inspired by the Coleman-Weinberg form of the effective potential
\cite{Coleman:1973jx}. One assumes that the effective potential is
well approximated by identifying the dependence of the integrated
scalar self-coupling on the RG scale $k$ with the scalar field itself,
$\lambda_2(k=\phi)$. We emphasize, that the identification $k=\phi$
mixes momentum scale information $k$ with the field amplitude. In
general, the full effective action in field theory would provide
separate information about the two scales which need not be the
same. By this identification, we obtain a \scpotentialem\ which in our
simple approximation reads
\begin{align}
\Ueffsc(\phi) &= \frac{1}{2}m_{\mu}^2\phi^2 + \frac{\lambda_2(k=\phi)}{8}\phi^4 \notag \\
 &=\frac{1}{2} m_{\mu}^2\phi^2 + \frac{\lambda_{2,\mu}}{8}\phi^4 - \frac{h^4\phi^4}{16\pi^2}\ln{\frac{\phi^2}{\mu^2}}.
\end{align}
Imposing the renormalization conditions
\eqref{eq:rencond} together with the choice $\mu=\phi_0=v$, we can
write the \scpotential\ as
\begin{align}
\Ueffsc(\phi) =& - \frac{1}{4}\left[ \mH^2 + \frac{\mtop^4}{2\pi^2 v^2} \right] \phi^2
 + \frac{1}{8}\left[ \frac{\mH^2}{v^2} + \frac{3\mtop^4}{4\pi^2 v^4} \right] \phi^4 \notag \\
 & - \frac{\mtop^4 \, \phi^4}{16\pi^2 v^4} \ln{\frac{\phi^2}{v^2}}.
 \label{eq:r1loopPotential}
\end{align}
Note also, that the bare potential
\eqref{eq:Ubare} remains completely unspecified in this
derivation. The implicit use of only renormalizable operators together
with the limit $\Lambda\to\infty$ permitted by perturbative
renormalizability seems to suggest that the details of the bare
potential are irrelevant.

\begin{figure}[t]
\includegraphics[width=0.45\textwidth]{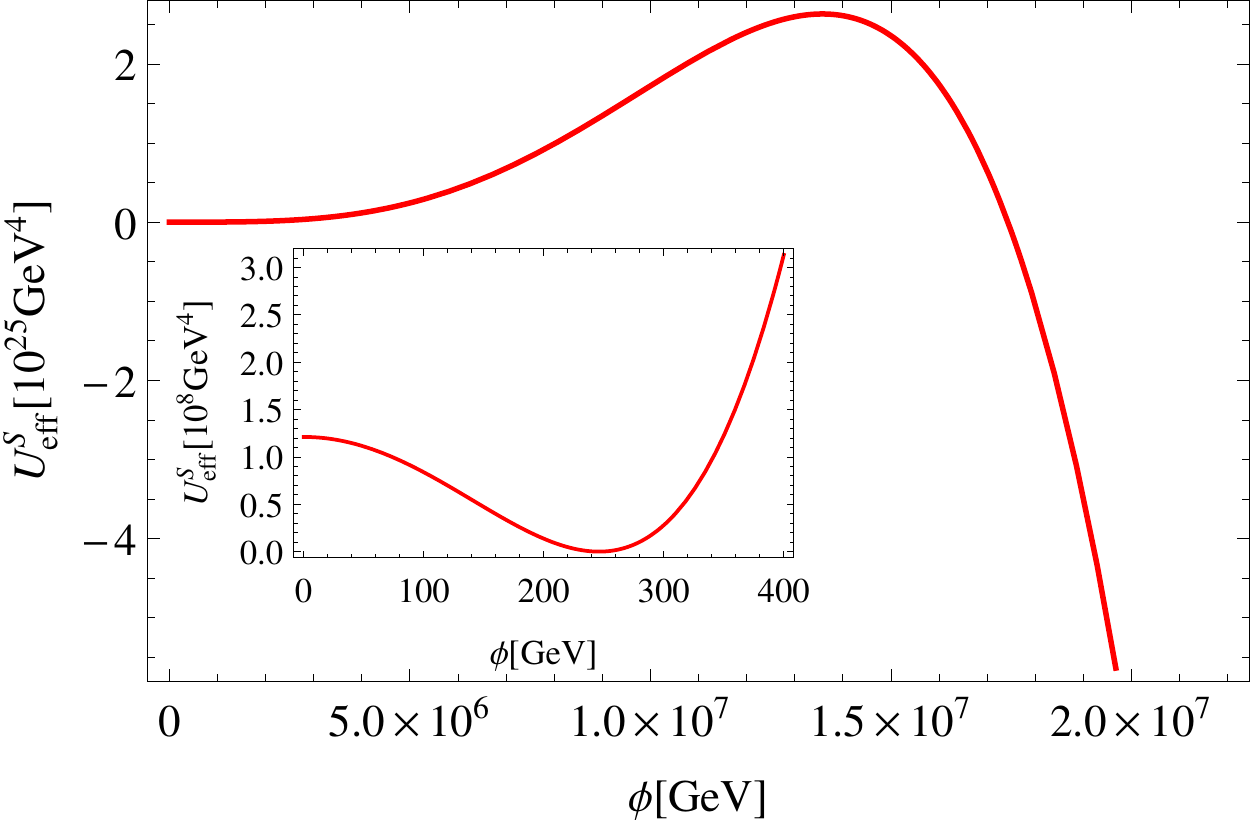}
\caption{Conventional effective \scpotential\ $\Ueffsc$ as a function
  of the field amplitude $\phi$. While the potential looks stable
  around the electroweak minimum, it develops an instability at large
  field values within our toy model. This instability seems to be
  driven by top fluctuations which turn the scalar self-coupling
  negative at large scales, cf. \Eqref{eq:1looph4}.}
\label{fig:CWt-Potential}
\end{figure}

 Clearly, this \scpotential\ develops an
instability for large Yukawa couplings, i.e., large $\mtop$. For the
present choice of parameters, the instability occurs at a scale of
$\sim 10^7$GeV in our toy model, see
Fig.~\ref{fig:CWt-Potential}. This instability is related to the
running of $\lambda_2(k)$, which turns negative at sufficiently large
$k$, cf. \Eqref{eq:1looph4}.

In the full standard model, the corresponding instability scale is of order $\sim
10^{10}$GeV. Though current state-of-the-art calculations
\cite{Buttazzo:2013uya,Degrassi:2012ry,Bednyakov:2015sca} determine the \scpotential\ to
NNLO precision, including two-loop threshold corrections,
and self-consistent resummations
\cite{Andreassen:2014gha,Andreassen:2014eha}, the present rather
cartoon-like presentation in a toy model still captures the essence of
the origin of the instability occurring in the perturbative estimate of
the \scpotential.

A qualitative difference arises in the standard model from the
electroweak gauge fluctuations, which render the $\phi^4$ coupling
positive again at even higher scales such that the
\scpotential\ becomes bounded from below and a second minimum arises
beyond the Planck scale which turns out to be the global
one. Therefore, the absolute instability of the \scpotential\ is a
particularity of our model. Below, this will actually be useful to
make one of our main points more transparent.

\section{Mean-field effective potential and stability}
\label{sec:CW-vs-MF}

In the following, we use mean-field methods to study the effective
potential. We stick to the same simplifications as before, ignoring
bosonic fluctuations and the running of the Yukawa coupling, but keep
track of all scales involved, the momentum scale of fluctuations $k$,
the field amplitude $\phi$ and the UV cutoff scale $\Lambda$. Parts of
this discussion follows \cite{Gies:2013fua,Gies:2014xha}, where also
more technical details can be found. Here, we focus on the new aspects
arising for un-/metastable scenarios.

\subsection{Mean-field potential}

With these prerequisites, the mean-field potential is directly related
to the fermion determinant. More precisely, working with an explicit
UV cutoff $\Lambda$ and an IR regulator scale $k$, the mean-field potential reads,
\begin{equation}
 U_k^{\text{MF}} = U_{\Lambda} - \frac{1}{\Omega} \ln \det{}_{\Lambda,k} (i\slashed{\partial} + ih\phi),
\label{eq:lndet}
\end{equation}
where $\Omega$ denotes the spacetime volume, and irrelevant field
independent constants are ignored. If we introduced $N$ fermion
flavors, the mean-field potential would become exact in the limit
$N\to\infty$. The notation $\det_{\Lambda,k}$ indicates that the
determinant is regularized and includes momentum modes $p$ in the
range $k^2\leq p^2\leq \Lambda^2$. The result is regularization
dependent. As long as we do not send $\Lambda\to\infty$, this
dependence is physical and can be viewed as a model for the details of
the embedding into a more fundamental underlying UV complete
theory. For a close contact with later sections, we use a piece-wise
linear regulator familiar from functional RG studies
\cite{Litim:2000ci,Litim:2001up}; we emphasize that all conclusions
remain the same also for a sharp momentum cutoff, propertime or
zeta-function regularization, see \cite{Gies:2014xha}. We obtain,
\begin{equation}
 U_k^{\text{MF}} = U_{\Lambda} - \frac{h^2 (\Lambda^2-k^2)\phi^2}{16\pi^2} 
 + \frac{h^4 \phi^4}{16 \pi^2} \ln{\frac{\Lambda^2+h^2\phi^2}{k^2+h^2\phi^2}},
\label{eq:MFeffPot-k}
\end{equation}
which makes all scale dependencies explicit. By varying the RG scale
$k$, we can observe how the mean-field effective potential as a full
function of the field amplitude $\phi$,
\begin{equation}
\begin{split}
 U_{\text{eff}}^{\text{MF}}(\phi) = U_{k=0}^{\text{MF}}(\phi)
 =& \frac{1}{2}\left( \mL - \frac{h^2 \Lambda^2}{8\pi^2} \right) \phi^2 + \frac{\lL}{8}\phi^4 
 \\
 &\quad + \frac{h^4 \phi^4}{16 \pi^2} \ln{\left(1+\frac{\Lambda^2}{h^2\phi^2}\right)},
\label{eq:meanfield}
\end{split}
\end{equation}
is built up from fermionic fluctuations renormalizing the bare potential $U_{\Lambda}$ while running from $k=\Lambda$ to $k\to0$.

\begin{figure}[t]
\includegraphics[width=0.45\textwidth]{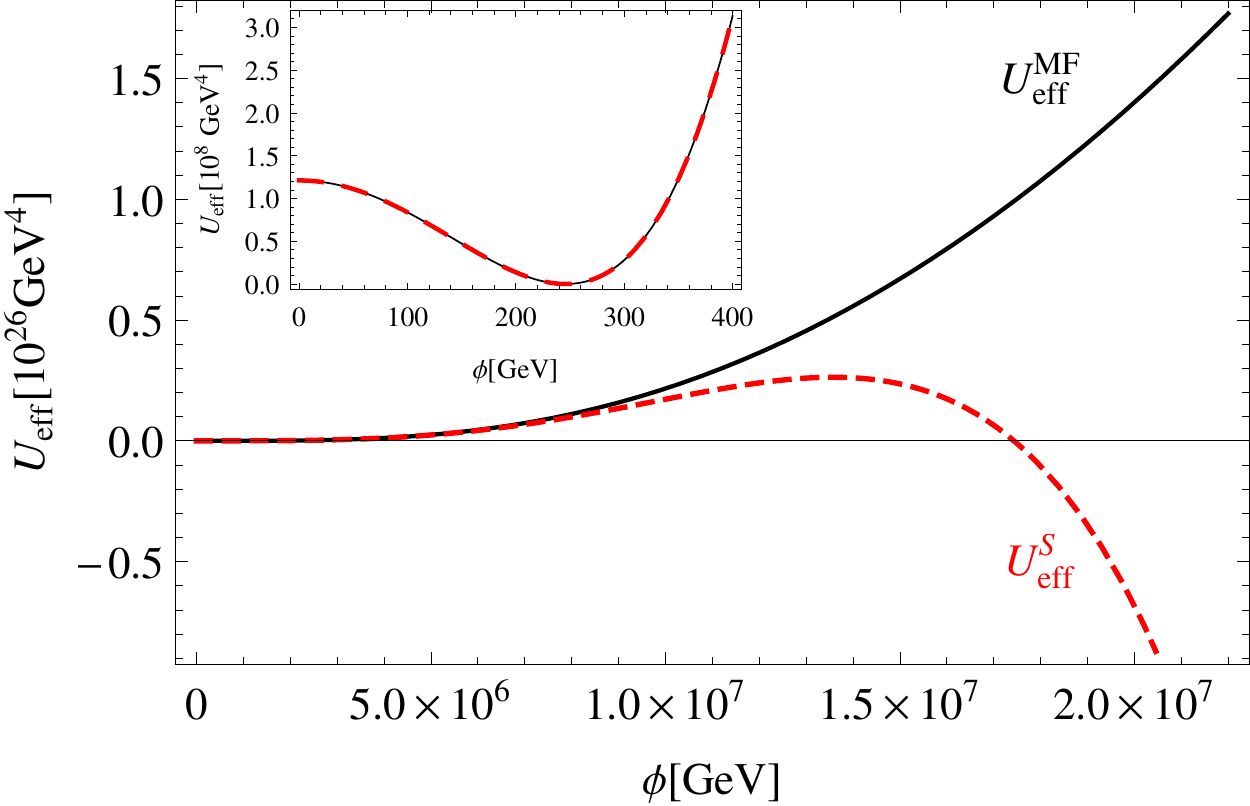}
\caption{Comparison
  between the effective potential where the cutoff is kept finite
  (black solid line, $\Lambda=\Lambda_{\text{cr}}=1.22 \cdot 10^7$GeV) and 
  the \scpotential\ (red dashed line). Both approaches describe the same low energy
  physics around the Fermi scale as they should, while at high
  energies a seeming instability appears in $\Ueffsc$.}
\label{fig:CW-vs-MF}
\end{figure}

Apart from the induced mass term $\sim h^2\Lambda^2\phi^2$, the whole
interaction part of the determinant (2nd line of \Eqref{eq:meanfield})
is positive. The bare mass term $m_{\Lambda}^2$ can now be fixed by
the renormalization condition $U_{\text{eff}}^{\text{MF}}{}'(\phi_0=v)=0$, fixing the
Fermi scale,
\begin{align}
 m_{\Lambda}^2 =& \frac{h^2 \Lambda^2}{8\pi^2} - \frac{h^4v^2}{8\pi^2}\left[ 2\ln{\left(1+\frac{\Lambda^2}{h^2v^2}\right)} - \frac{\Lambda^2}{\Lambda^2 + h^2v^2} \right]
 \notag \\
 &- \frac{1}{2}\lambda_{2,\Lambda}v^2.
 \label{eq:mass-bare}
\end{align}
Inserting \Eqref{eq:mass-bare} into \Eqref{eq:meanfield} yields a
globally stable effective potential for any value of the UV cutoff
$\Lambda$ and any admissible non-negative value of the bare $\phi^4$
coupling $\lambda_{2,\Lambda}\geq 0$, cf. solid black line in
Fig.~\ref{fig:CW-vs-MF}. It is important to stress that a bare
potential of quartic type, Eq.~\eqref{eq:Ubare}, with negative
$\lambda_{2,\Lambda}$ would be inconsistent right from the beginning,
as the functional integral over the scalar field would be ill-defined.

For completeness of the presentation, we recall that the mass of the
scalar particle now becomes a function of the cutoff and
$\lambda_{2,\Lambda}$, cf.~\cite{Gies:2013fua,Gies:2014xha},
\begin{align}
 \mH^2 =& v^2 U_{\text{eff}}^{\text{MF}}{}''(v) \notag \\
 =& \frac{\mtop^4}{4\pi^2 v^2} \left[2 \ln \left( 1+ \frac{\Lambda^2}{\mtop^2} \right) 
 - \frac{3\Lambda^4+2\mtop^2\Lambda^2}{(\Lambda^2+\mtop^2)^2}\right] + v^2 \lambda_{2,\Lambda}. 
 \label{eq:MFmH}
\end{align}
This demonstrates that a lower bound for the Higgs mass is obtained by
the physical restriction that the bare potential of $\phi^4$-type at a
given UV cutoff $\Lambda$ must be bounded from below, i.e.,
$\lambda_{2,\Lambda}\geq 0$. Thus, the lower bound (lower edge of the
IR window) is given by $\lambda_{2,\Lambda} = 0$ for this class of
bare potentials. This way of determining the lower bound has been
suggested in \cite{Holland:2003jr,Holland:2004sd}, and has been used
in full non-perturbative lattice simulations
\cite{Fodor:2007fn,Gerhold:2007yb,Gerhold:2007gx,Gerhold:2009ub}. Generically,
one observes a strong quantitative agreement with mean-field theory
for this lower bound. In this fashion, strong constraints on the
existence of a heavy fourth generation arise
\cite{Gerhold:2010wv,Bulava:2012rb,Bulava:2013ep,Djouadi:2012ae}.

For the purpose of the present work, we reverse the line of argument:
for a given Higgs mass of, say $\mH = 125$GeV, this implies that a
maximal scale of UV extension $\Lambda$ is obtained. Choosing the
minimal admissible value $\lambda_{2,\Lambda} = 0$ a cutoff of
$\Lambda_{\text{cr}} = 1.22 \cdot 10^7$GeV is obtained by writing
$\Lambda=\Lambda(\mH^2,\lambda_{2,\Lambda})$. For larger values of the
UV cutoff, $\Lambda>\Lambda_{\text{cr}}$ no physical (mean-field) RG
trajectory can be found that connects an admissible bounded bare
potential to an IR Higgs mass of $125$GeV. As long as
$\Lambda\leq\Lambda_{\text{cr}}$, the bare potential as well as the
effective potential do not exhibit an instability.
Figure~\ref{fig:CW-vs-MF} shows a comparison between the mean-field
potential (solid/black line) where the cutoff is kept finite and the
\scpotential\ (red/dashed line) where the cutoff has implicitly been
sent to infinity. The \scpotential\ approximation starts to break down
for field amplitudes, where $h\phi/\Lambda\gtrsim\mathcal{O}(1)$,
i.e., where terms which are dropped in the implicit $\Lambda\to\infty$
limit are actually sizable.

\begin{figure}[t]
\includegraphics[width=0.45\textwidth]{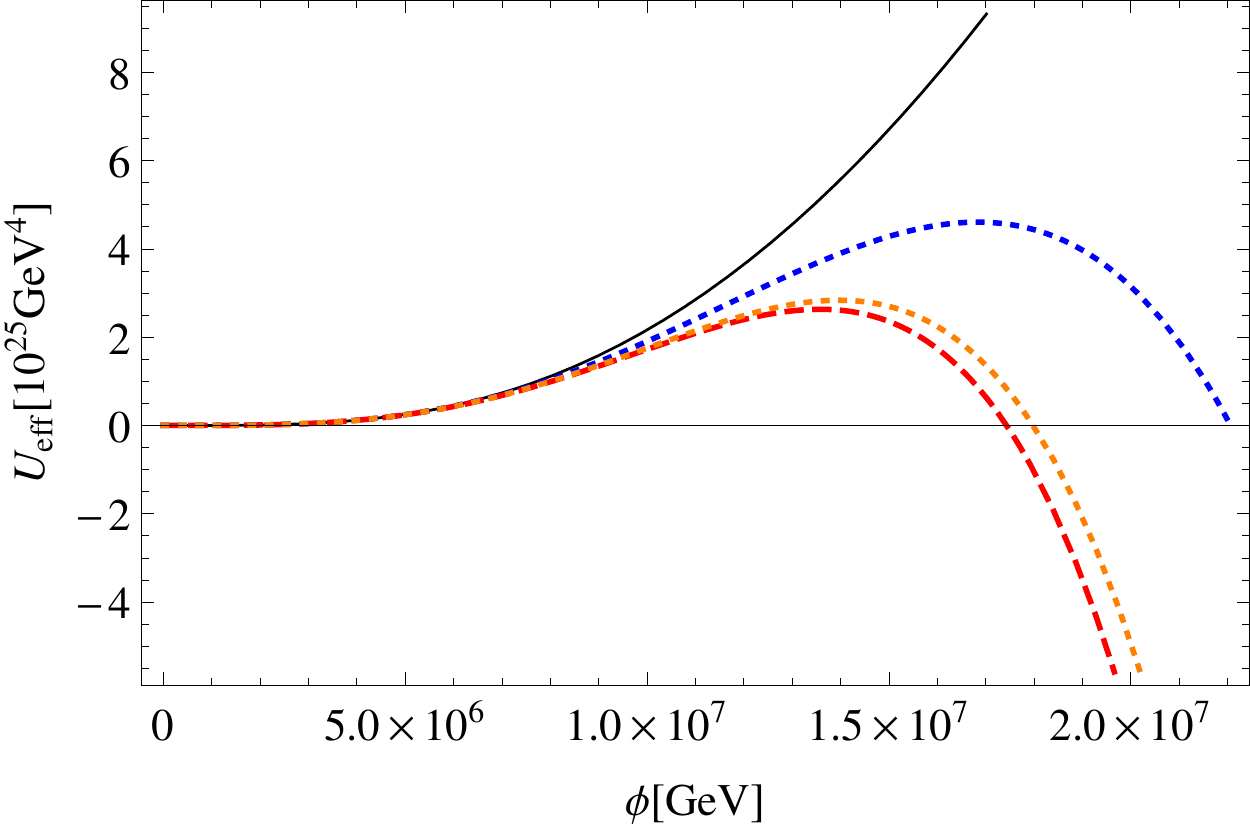}
\caption{Approach of
  the mean-field potential to the \scpotential\ if we blindly allow
  for cutoffs $\Lambda$ larger than the critical value of
  $\Lambda_{\text{cr}} = 1.22 \cdot 10^7$GeV (black solid) with IR
  physics kept fixed; $\Lambda = 2 \cdot 10^7$GeV (blue dotted line),
  and $\Lambda = 5 \cdot 10^7$GeV (orange dotted line). For $\Lambda >
  10^8$GeV, there is no visible difference between the mean-field
  potential and the \scpotential\ (red dashed line) in this plot.}
\label{fig:CW-vs-MF2}
\end{figure}

It is, of course, possible to reduce the multi-scale mean-field
potential to the \scpotential. First, we blindly enforce all
renormalization conditions. In particular the first condition in
\Eqref{eq:rencond} for large cutoffs implies
\begin{align}
 \lambda_{2,\Lambda} = \frac{\mH^2}{2\kappa} - \frac{h^4}{2\pi^2} \left[ \ln \frac{\Lambda^2}{\mtop^2} -\frac{3}{2} \right] + \mathcal{O}\left( \! \frac{1}{\Lambda^2} \! \right).
 \label{eq:lambda-bare}
\end{align}
Inserting Eq.~\eqref{eq:lambda-bare} and Eq.~\eqref{eq:mass-bare} into
the mean-field effective potential finally leads to a potential with
the requested minimum at $v$ and Higgs mass of $\mH$ by
construction. The cutoff remains still a free parameter. In the naive
large cutoff limit, we obtain
\begin{align*}
 \Umf_{\text{``}\Lambda\to\infty\text{''}}  =& - \frac{1}{4}\left[ \mH^2 + \frac{\mtop^4}{2\pi^2v^2} \right] \phi^2
 + \left[ \frac{\mH^2}{v^2} + \frac{3\mtop^4}{4\pi^2 v^4} \right]\frac{1}{8} \phi^4 \notag \\
 & - \frac{\mtop^4 \, \phi^4}{16\pi^2 v^4} \ln{ \left( \frac{\Lambda^2}{\Lambda^2+h^2\phi^2} \frac{\phi^2}{v^2} \right) } 
 + \mathcal{O}\left( \! \frac{1}{\Lambda^2} \! \right).
\end{align*}
For a cutoff larger than the critical value $\Lambda_{\text{cr}}$, the
potential develops an instability and rapidly approaches the
\scpotential, see Fig.~\ref{fig:CW-vs-MF2}. For $\Lambda>10^8$GeV, the
difference between the mean-field effective potential with a finite
cutoff and the \scpotential\ with implicit limit $\Lambda \to \infty$
becomes very small. Taking the naive limit $\Lambda \to \infty$, the
\scpotential\ \eqref{eq:r1loopPotential} is obtained as expected.

We emphasize that the consistency condition that the bare potential
should be bounded from below for a well-defined generating functional
is no longer fulfilled for all $\Lambda>\Lambda_{\text{cr}}$. This can
be directly read off from expression \eqref{eq:lambda-bare}:
$\lambda_{2,\Lambda}$ has to be chosen negative for
$\Lambda>\Lambda_{\text{cr}}$, and thus already the bare potential is
unstable. At this point, we conclude that the apparent instability of
the \scpotential\ appears due to an inconsistent UV boundary condition
for the theory. As long as the consistency condition $\lL\geq 0$ is
fulfilled, no instability can be found within the class of quartic
bare potentials.

\subsection{Generalized bare potentials}

As already emphasized in
\cite{Gies:2013fua,Gies:2014xha,Eichhorn:2015kea}, these observations
do not imply that in- or metastabilities are completely
excluded. Whether or not an in-/metastability occurs is not a matter
of the fermionic fluctuations but has to be seeded by the microscopic
underlying theory. A specific example from string phenomenology is
given in \cite{Hebecker:2013lha}. 

From the perspective of the standard model as an effective field
theory, the embedding into a UV complete theory is parametrized by the
bare action at the cutoff $\Lambda$. Of course, the bare action is
expected to host all operators compatible with the symmetry with
couplings of order $\mathcal{O}(1)$ in units of the cutoff $\Lambda$.

In the following, we consider the simplest extension of the bare
potential by including a higher-dimensional $\phi^6$ operator as an
example,
\begin{align}
 U_{\Lambda} = \frac{\mL}{2} \phi^2 + \frac{\lL}{8} \phi^4 + \frac{\lambda_{3,\Lambda}}{48\Lambda^2} \phi^6.
 \label{eq:gen-bare-Potential}
\end{align}
Within the same mean-field approximation as used before, we can
straightforwardly compute the mass of the Higgs boson in our model as
a function of $\Lambda$ and the parameters $\lL$ and
$\lambda_{3,\Lambda}$, cf.~\Eqref{eq:MFmH},
\begin{equation}
\begin{split} 
 \mH^2
 =& \frac{\mtop^4}{4\pi^2 v^2} \left[2 \ln \left( 1+ \frac{\Lambda^2}{\mtop^2} \right) 
 - \frac{3\Lambda^4+2\mtop^2\Lambda^2}{(\Lambda^2+\mtop^2)^2}\right] 
 \\
 &+ v^2 \lambda_{2,\Lambda} 
 + \frac{v^4}{2\Lambda^2}\lambda_{3,\Lambda}.
 \label{eq:MFmHg}
\end{split}
\end{equation}
It is obvious that the previous lower bound of ~\Eqref{eq:MFmH} can be
relaxed by a negative value for $\lL$, while a positive
$\lambda_{3,\Lambda}$ can stabilize the bare potential. For small
negative $\lL$ and sufficiently large $\lambda_{3,\Lambda}$ the
effective potential as well as the potential at intermediate scales
$k$ are globally stable and have a unique minimum. In this regime, it
is easily possible to obtain Higgs masses below the perturbative lower
bound, i.e., decrease the edge of the IR window.

For even smaller $\lL$, i.e., larger absolute values of a negative
$\lL$, the effective potential $U^\text{MF}_k$ starts to develop a
second minimum towards lower RG scales $k$ and becomes metastable,
while the bare potential $U_\Lambda$ is still stable. For even smaller values of $\lL$, also the bare potential can become
metastable.

\begin{figure}[t]
\includegraphics[width=0.45\textwidth]{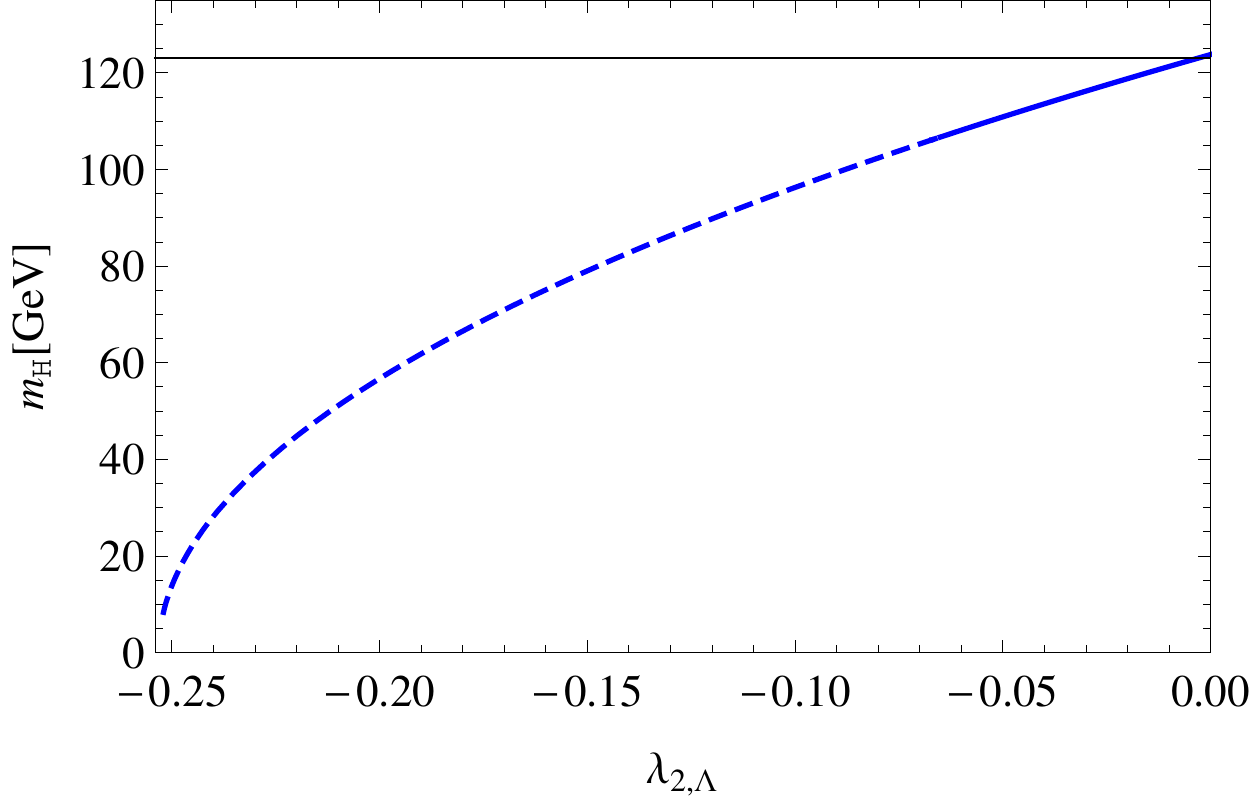}
\caption{Higgs masses for the class of generalized bare potentials for
  $\Lambda=10^7$GeV. The bare potential is stabilized by
  $\lambda_{3,\Lambda}=3$. The horizontal black solid line marks the
  lower Higgs mass consistency bound within quartic bare
  potentials. The blue solid line indicates values for $\lL$ where the
  IR potential is stable while for the blue dashed line a
  metastability occurs.}
\label{fig:HiggsMass-below}
\end{figure}

For an illustration, let us assume a fixed cutoff $\Lambda=10^7$GeV.
Within the class of quartic bare potentials \eqref{eq:Ubare}, the
lowest Higgs mass according to \Eqref{eq:MFmH} is given by
$\mH=123.8$GeV for $\lL=0$. Stabilizing the more general class of bare
potentials \eqref{eq:gen-bare-Potential} with a fixed value of
$\lambda_{3,\Lambda}=3$, we can choose negative values of $\lL$,
yielding also smaller values of the Higgs mass, see
Fig.~\ref{fig:HiggsMass-below}. The resulting mean-field potentials
are stable with a unique (electroweak) minimum on all scales (blue
solid line) until we reach a value for the bare quartic coupling of
$\lL=-0.065$.  For even smaller values of $\lL$, a second minimum
arises in the course of the mean-field flow, while the bare potential
still has a unique minimum. This second minimum is a local minimum
only for a small range of $\lL$ values, $ -0.0671< \lL < -0.065$. For
$\lL<-0.0671$, the second minimum becomes the global one (blue dashed
line), which renders the electroweak minimum in the effective
potential metastable. Within this regime of metastability, the Higgs
mass can be made arbitrarily small by a suitable choice of parameters
even without any metastability in the bare potential.

It is important to emphasize that the metastability observed here in
this model is a consequence of the shape of the bare potential encoded
in both renormalizable and non-renormalizable operators. In the
present model, this metastability remains invisible in the
perturbatively estimated \scpotential\ which would predict complete
instability. We conclude that metastability properties of the model
can only be reliably calculated if the bare potential at a UV scale is
known. The \scpotential\ is not sufficient as a matter of principle. 

In the present model, this conclusion becomes obvious as the
\scpotential\ does not even exhibit a metastable region. This is
different from the standard model, where the \scpotential\ itself
predicts metastability for light Higgs masses, as the \scpotential\ is
stabilized by electroweak fluctuations again at high field
amplitudes. Still, the same conclusion about the reliability of the
metastability estimate of the \scpotential\ holds as for the simple
model.

\begin{figure*}[t]
\includegraphics[width=0.4\textwidth]{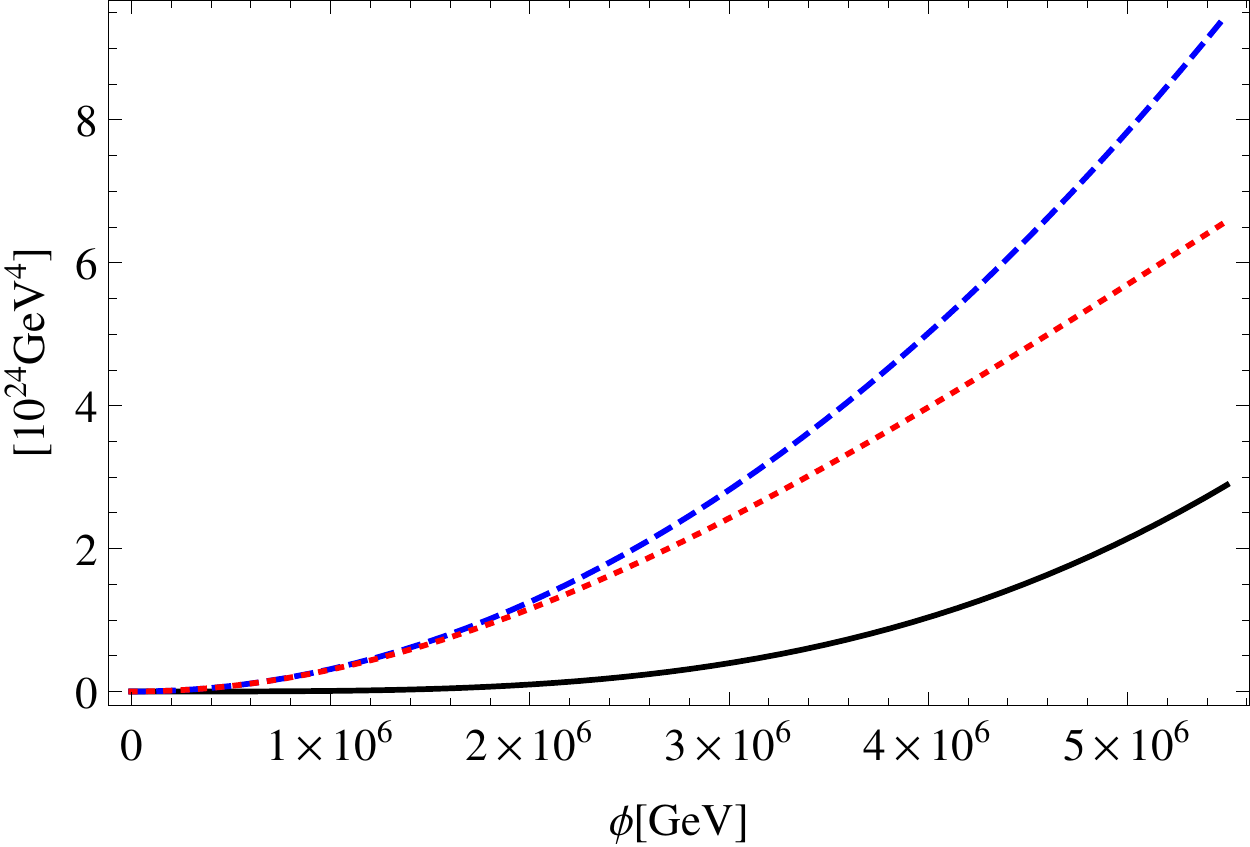}
\hspace*{0.075\textwidth}
\includegraphics[width=0.4\textwidth]{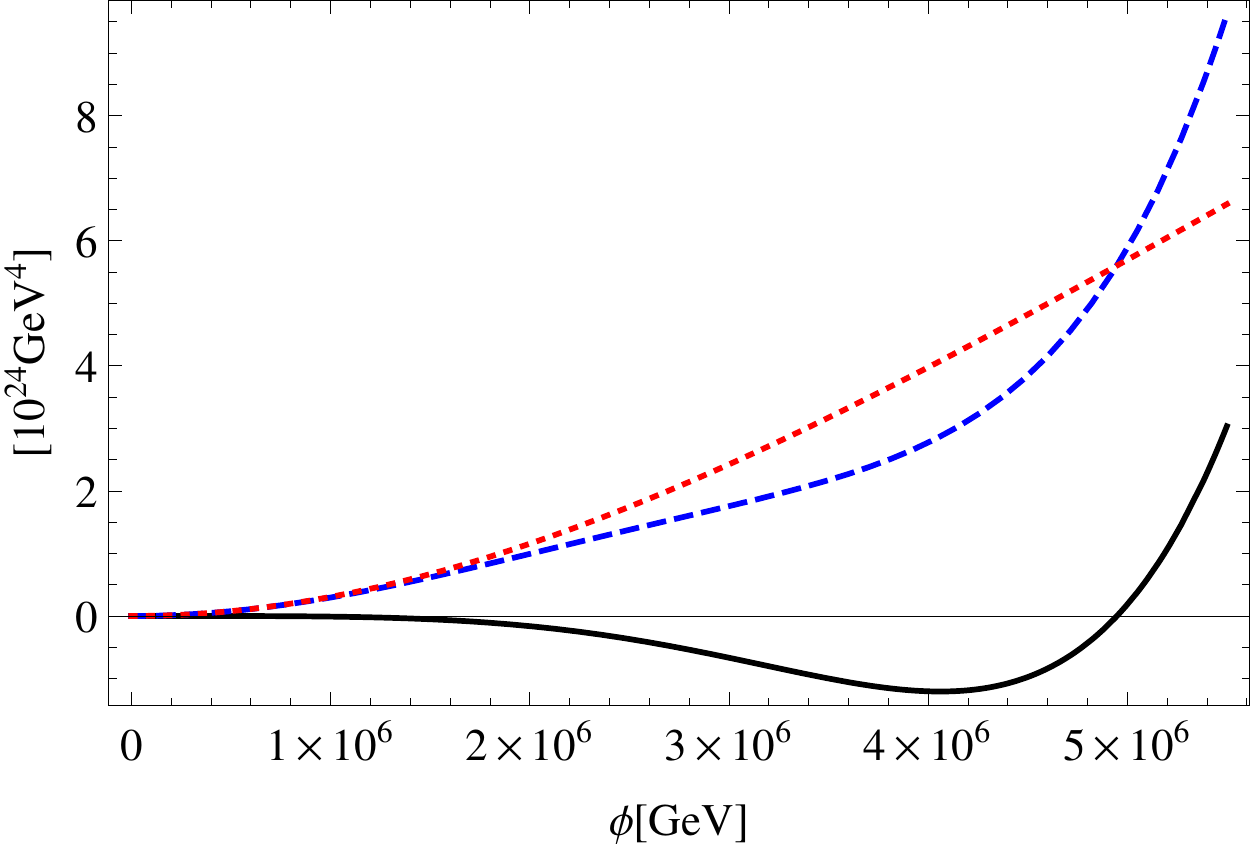}
\caption{Mean-field potential (black solid) as the difference between
  the bare potential (blue dashed) and the absolute value of the
  fermion determinant (red dotted). The Fermi minimum at $\phi=246$GeV
  is hardly visible on the scale of the plot. Left panel: the quartic
  bare potential always exceeds the contributions from the fermion
  loop for field values above the Fermi scale ($\Lambda=10^7$GeV, $\lL=0$,
  $\lambda_{3,\Lambda}=0$ ). Right panel: a
  metastability seeded by the bare potential develops in the course of
  the RG flow ($\Lambda=10^7$GeV, $\lL=-0.15$, $\lambda_{3,\Lambda}=3$).}
\label{fig:MFPot-metastable}
\end{figure*}

The fact that the metastability in the effective potential is seeded
in the bare potential is illustrated in
Fig.~\ref{fig:MFPot-metastable}. Here, the effective mean-field
potential (black solid line) is shown as the difference between the
bare potential (blue dashed line) and the absolute value of the
fermion determinant (red dotted line). The left panel depicts the case
with stable bare as well as effective potential (initial parameters:
$\Lambda=10^7$GeV, $\lL=0$, $\lambda_{3,\Lambda}=0$). By contrast, the right panel shows the case where a
second minimum arises in the effective potential (initial parameters:
$\Lambda=10^7$GeV, $\lL=-0.15$, $\lambda_{3,\Lambda}=3$). One clearly sees how the modified structure of the
generalized bare potential with a negative $\lL$ is responsible for
the second minimum at large scales besides the electroweak one (the
latter at $\phi=246$GeV is hardly visible on the scales of the
plot). We emphasize again that there is no possibility for the
mean-field potential to develop a second minimum for the case of
quartic bare potentials because the bare potential always exceeds the
fermion determinant.

\begin{figure*}[t]
\includegraphics[width=0.3\textwidth]{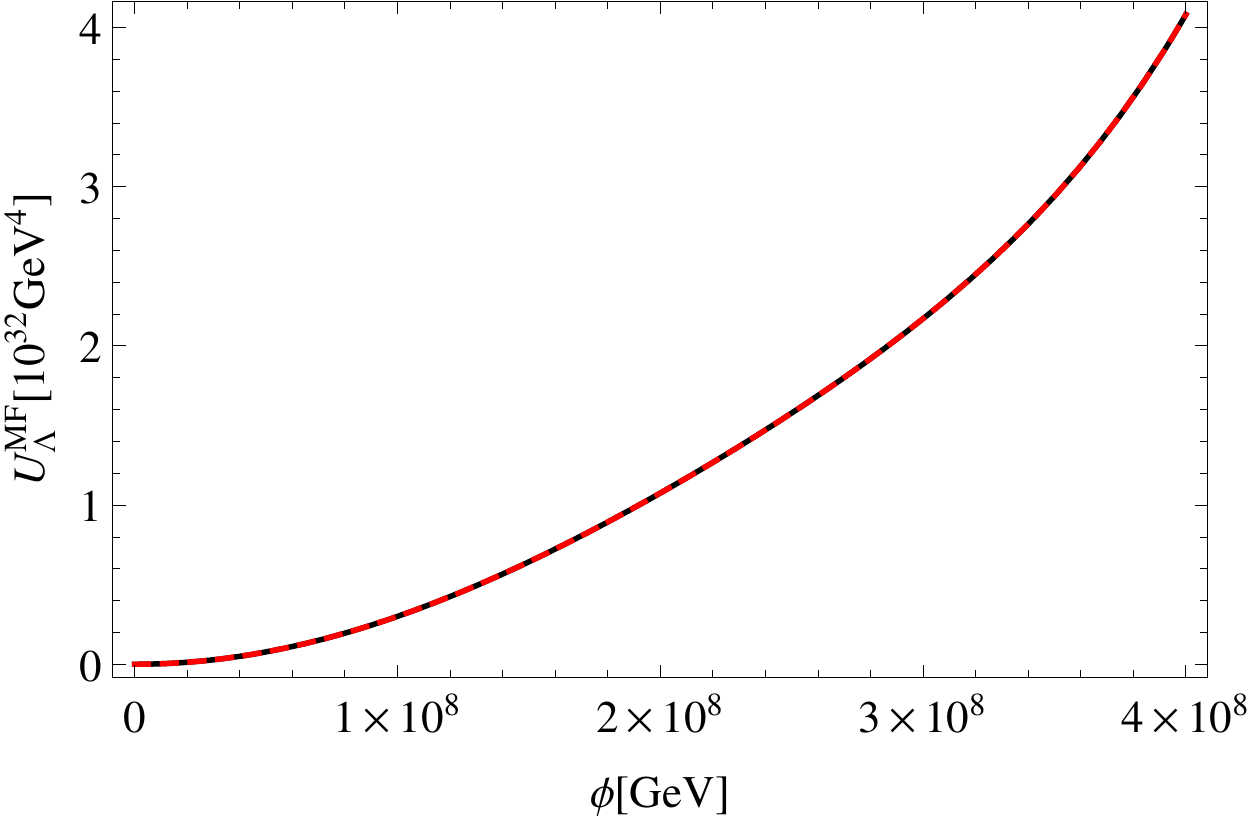}
\hspace*{0.02\textwidth}
\includegraphics[width=0.3\textwidth]{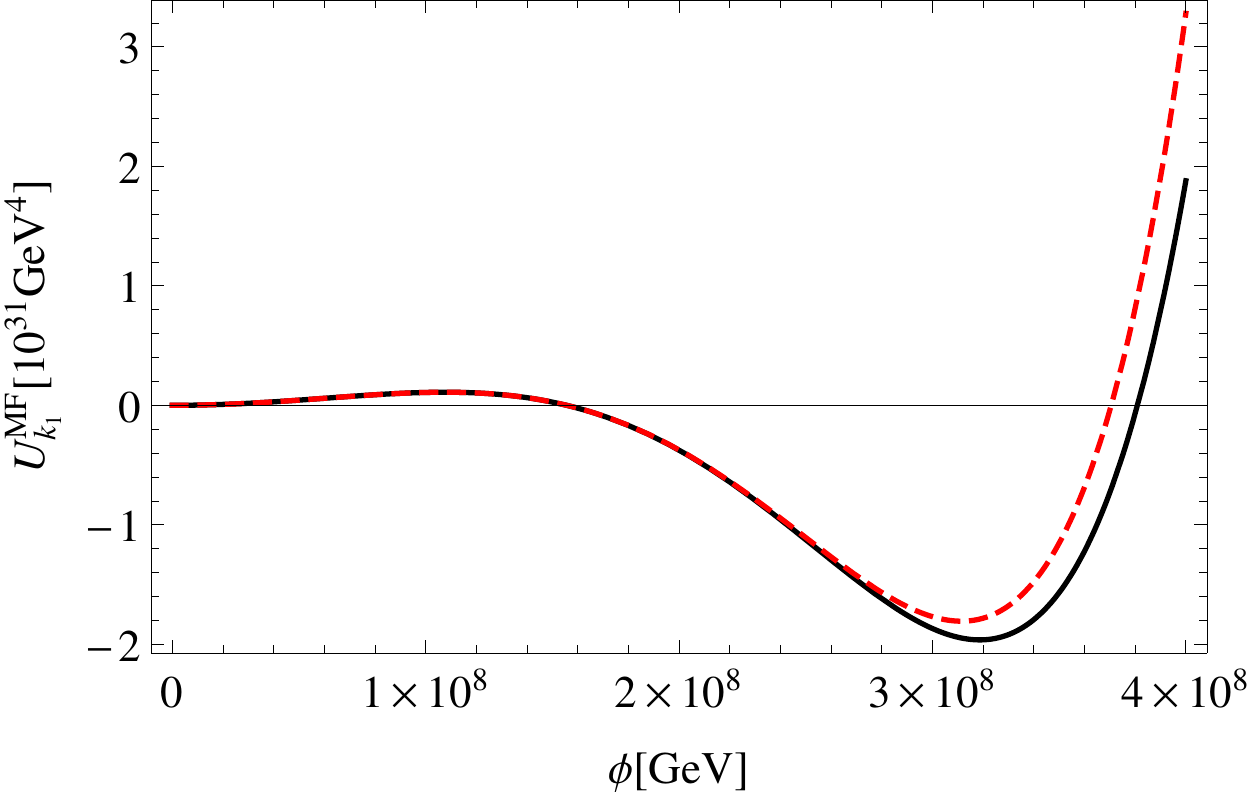}
\hspace*{0.02\textwidth}
\includegraphics[width=0.3\textwidth]{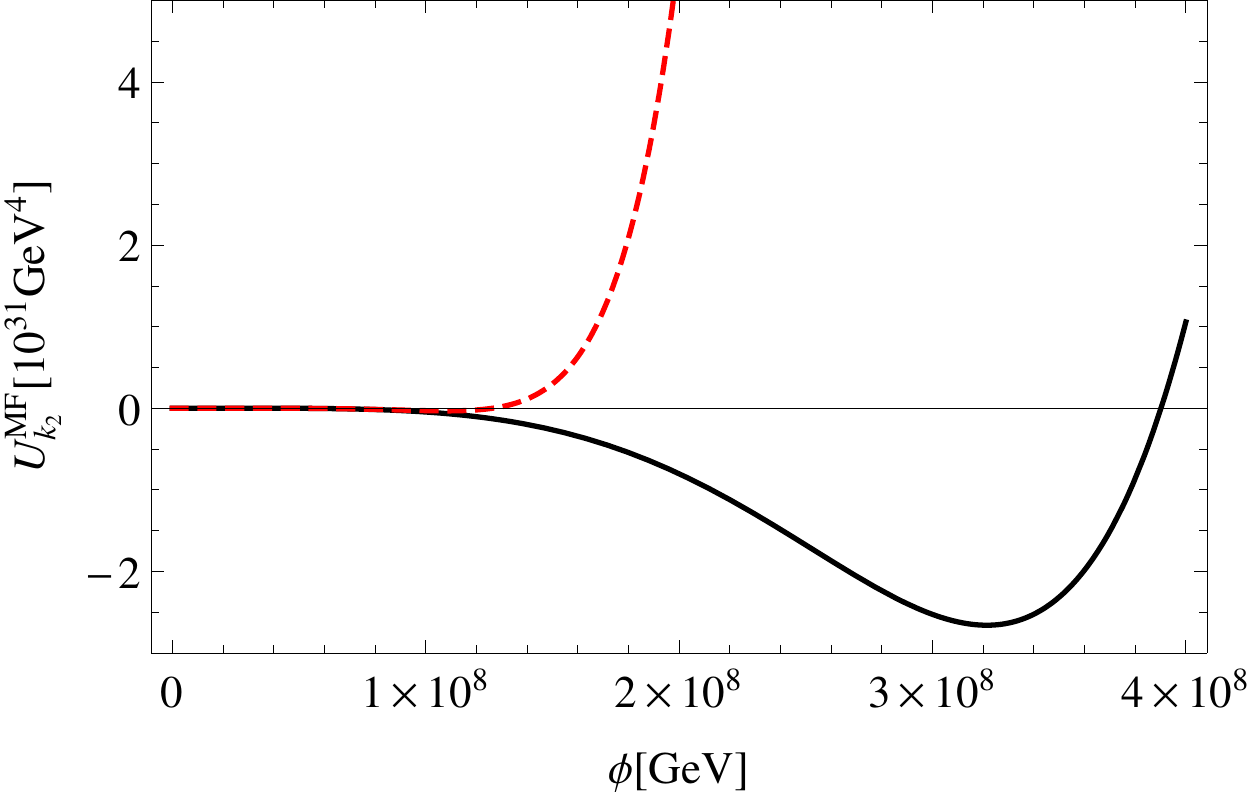}
\caption{Full mean-field potential (black solid line) and the
  potential approximated by a Taylor expansion (red dashed line) around the origin up to
  $\phi^8$ for different values of the RG scale $k$. The left panel
  shows the bare potential for $\Lambda=10^9$GeV ($\lambda_{3,\Lambda}=3$, and $\lL$ is chosen such that $\mH=125$GeV), where the Taylor approximation fits the
  full potential as it should. The middle plot shows the scalar
  potential slightly below the UV cutoff, $k_1=2.5\cdot 10^8$GeV $<\Lambda$, where the
  second minimum is built up. Toward the IR, $k_2=5\cdot 10^7$GeV, the
  second minimum settles while it disappears within the Taylor
  expansion (right plot).}
\label{fig:Taylor-vs-Full}
\end{figure*}

With the full mean-field potential at hand, we can also clarify the
nature of the \textit{pseudo-stable phase} observed in a polynomial
expansion of the effective potential in \cite{Eichhorn:2015kea}. In
this approximation, RG flows were observed that start at $k=\Lambda$
with a globally stable bare potential, then run trough a metastable
regime with two minima and finally end up in the IR $k=0$ with one
stable minimum at the Fermi scale.  In the same spirit, we now expand
the mean-field effective average potential \eqref{eq:MFeffPot-k}
around the minimum at the origin and follow its flow in comparison
with the flow of the full mean-field effective potential. This is
depicted in Fig.~\ref{fig:Taylor-vs-Full}. Indeed, the potential
approximated by a polynomial expansion shows the same pseudo-stable
behavior as observed in \cite{Eichhorn:2015kea}. A second minimum
appears but disappears again after a short RG time. The polynomial
expansion thus looks stable again in the IR. This is in contrast to
the full mean-field potential where the second minimum survives the RG
flow towards the IR. We conclude that the pseudo-stable phase is an
artifact of the finite convergence radius of the polynomial
expansion. The global effective mean-field potential exhibits a
metastability also in this phase.

\section{Nonperturbative flow of the scalar potential}
\label{sec:PotentialFlow}

\subsection{Functional renormalization group}

While the mean-field approximation is highly convenient for first
analytically controllable estimates, we have to go beyond for
quantitative accuracy. The functional RG is an ideal tool for this
task, as it resums large classes of higher-order diagrams,
automatically accounts for threshold corrections and provides
information about the global RG flow of the effective potential, i.e.,
all relevant scales can be studied independently. We use the Wetterich
equation \cite{Wetterich:1992yh} in order to compute the RG flow of
the effective action $\Gamma_k$,
\begin{align}
 \pt \Gamma_k = \frac{1}{2} \STr \left[\pt R_k\Big(\Gamma_k^{(2)}+R_k\Big)^{-1} \right], \quad t = \ln\frac{k}{\Lambda}. 
 \label{eq:Wetterich}
\end{align}
The solution to this equation interpolates between the bare action
$S_\Lambda=\Gamma_{k=\Lambda}$ and the full effective action
$\Gamma=\Gamma_{k\to 0}$ that accounts for all quantum
fluctuations. The regulator function $R_k$ implements the
shell-by-shell integration at the momentum scale $k$.
$\Gamma_k^{(2)}$ denotes the Hessian of the effective average action with
respect to the fields in the theory, $(\hat\phi,\psi,\bar{\psi})$. For
detailed reviews see, \cite{Berges:2000ew,Pawlowski:2005xe,Gies:2006wv,Delamotte:2007pf,Braun:2011pp}.

We solve the Wetterich equation within a systematic derivative
expansion of the action at next-to-leading order,
\begin{align}
\Gamma_{k}  =  \int_x \left[
 \frac{Z_{\phi,k}}{2} \, ( \partial_{\mu} \hat\phi )^2 +  U_{k}(\rho)
 + Z_{\psi,k} \, \yb i \slashed{\partial}
  \psi + i \bar{h}_{k} \, \hat\phi  \yb \psi \right].
\label{eq:trunc}
\end{align}
The Wetterich equation $\eqref{eq:Wetterich}$ provides flow equations
for the potential, the Yukawa coupling, and the wave function
renormalizations for the fields $Z_{\phi,k}$ and $Z_{\psi,k}$. The
latter can be summarized by the anomalous dimensions of the fields,
\begin{align}
 \eta_{\phi} = -\frac{\pt Z_{\phi,k}}{Z_{\phi,k}}, \quad \eta_{\psi} = -\frac{\pt Z_{\psi,k}}{Z_{\psi,k}}.
\end{align}
In \Eqref{eq:trunc}, we have introduced the unrenormalized scalar
field $\hat\phi$, which is related to the renormalized field by
$\phi=Z_{\phi,k}^{1/2} \hat\phi$. At mean-field level, the
distinction is irrelevant as $Z_{\phi,k}|_{\text{MF}}=1$. In terms of
dimensionless renormalized quantities,
\begin{eqnarray}
 \rho &=& Z_{\phi,k} k^{2-d}\, \frac{1}{2}\hat{\phi}^2 = \frac{k^{2-d}}{2} \phi^2, \quad
 u = k^{-d}U_k, \label{eq:rhoconv}\\ 
 h^2 &=& Z_{\phi,k}^{-1}Z_{\psi,k}^{-1} k^{d-4} \bar{h}_k^2,\label{eq:hconv}
\end{eqnarray}
the nonperturbative flow equations in agreement with \cite{Gies:2009hq,Gies:2013fua,Gies:2014xha} read
in $d$ spacetime dimensions
\begin{widetext}
\begin{align}
 \pt \, u \equiv& \beta[u]= -d\, u + (d-2+\eta_{\phi}) \rho u' + 2v_{d} \, \Big[ l_{0}^{d} \left(u' + 2\rho u''; \eta_{\phi} \right) 
 - d_{\gamma} \, l_{0}^{(F)d}\left(2\rho h^{2}; \eta_{\psi} \right) \Big],   \label{eq:FlowPotential} 
\\
 \pt\, h^2 =& \left[ \eta_{\phi} + 2\eta_{\psi} + d - 4 \right] h^{2} 
  + 8v_{d} h^{4} \Big[ l_{1,1}^{(FB)d}\left(2 h^2 \kappa,u' + 2\kappa u'';\eta_{\psi},\eta_{\phi}\right) \notag  \\
  &\qquad 
  - \left( 6 \kappa u'' + 4 \kappa^{2} u''' \right) \, l_{1,2}^{(FB)d}\left(2 h^2 \kappa,u' + 2\kappa u'';\eta_{\psi},\eta_{\phi} \right)
  - 4 h^{2} \kappa \, l_{2,1}^{(FB)d}\left(2 h^2 \kappa,u' + 2\kappa u'';\eta_{\psi},\eta_{\phi}\right) \Big]_{\rho=\kappa},
  \label{eq:FlowYukawa}
\\
 \eta_{\phi} =& \frac{8v_{d}}{d} \bigg[ \kappa \left[3 \, u'' + 2 \kappa u''' \right]^{2} m_{4,0}^{d}\left(u' + 2 \kappa u'';\eta_{\phi}\right) 
 + d_{\gamma} \, h^{2} \Big[m_{4}^{(F)d}\left(2h^2 \kappa;\eta_{\psi}\right) -2 h^2 \kappa \, m_{2}^{(F)d}\left(2 h^2 \kappa;\eta_{\psi}\right) \Big] \bigg]_{\rho=\kappa},
 \label{eq:etaphi}
\\
 \eta_{\psi} =& \frac{8 v_{d}}{d} h^2  m_{1,2}^{(FB)d}\left(2 h^2 \kappa,u' + 2\kappa u'';\eta_{\psi},\eta_{\phi}\right)\Big|_{\rho=\kappa}, \label{eq:etapsi}
\end{align}
\end{widetext}
where primes denotes derivatives with respect to $\rho$. Here, we
extract the flow of the Yukawa coupling and the anomalous dimensions
at the (running) minimum of the potential $\kappa= \rho_{\text{min}}
\sim \phi_0^2$; see \cite{Vacca:2015nta} for an extended flow in the 
Yukawa sector. The threshold functions $l$ and $m$ encode the
decoupling of massive modes. Evaluated for a piece-wise linear
regulator function \cite{Litim:2000ci,Litim:2001up}, these are listed,
for instance, in \cite{Gies:2013fua}.  Of course, the perturbative
$\beta$ functions can be recovered from these nonperturbative flow
equations by neglecting resummation and threshold effects. The flow
equation also contains mean-field theory as a simple limit: ignoring
the anomalous dimensions as well as the flow of the Yukawa coupling,
and dropping the bosonic fluctuations $\sim l_0^d$ in
\Eqref{eq:FlowPotential}, the integration of the potential $\pt u$
precisely yields the mean-field potential
\eqref{eq:MFeffPot-k}. Hence, all known limits are unified in the
functional RG framework which we can now use to go beyond the
perturbative/mean-field studies.

Previous work on Higgs boson mass bounds has solved the potential flow
by means of a polynomial expansion
\cite{Gies:2013fua,Gies:2014xha,Jakovac:2015kka,Jakovac:2015iqa} about the flowing minimum. More concretely,
\begin{equation}
\begin{split}
 u &= \sum_{n=1}^{\Np} \frac{\lambda_n}{n!} \rho^n \,\, \text{(SYM)} \quad \text{or} \\
 u &= \sum_{n=2}^{\Np} \frac{\lambda_n}{n!} (\rho-\kappa)^n \,\, \text{(SSB)},
 \label{eq:polyexp}
\end{split}
\end{equation}
approximate the potential by a polynomial of degree $\Np$ in the
symmetric (SYM) or symmetry-broken (SSB) regime. For the present
problem, the quality of this expansion has been confirmed for
potentials \cite{Gies:2013fua} with a single minimum at any given
scale. As the example of the seeming pseudo-stable phase above has
shown, a proper description of metastability doubtlessly requires a
PDE solver for the RG flow of the full potential
\eqref{eq:FlowPotential} as a function of both $k$ and $\rho$.

Solvers for such types of Yukawa systems have been successfully
developed and applied in various functional RG studies from low-energy
QCD models \cite{Bohr:2000gp,Herbst:2013ail}, critical phenomena
\cite{Adams:1995cv,Hofling:2002hj,Bonanno:2000yp} to ultracold-gas systems
\cite{Boettcher:2014tfa}. The particular difficulty in the present case arises
from the necessity to run the RG over many orders of magnitude in the
presence of a relevant operator $\phi^2$ of canonical dimension
2. This requires substantial precision.

Another challenge is the approach to convexity which is expected to
hold for the full effective potential
\cite{ORaifeartaigh:1986hi,Litim:2006nn}, but is spoiled by both the
perturbative \scpotential\ as well as the mean-field
approximation. Whereas the approach to convexity is less interesting
for the case of a single minimum, it may become essential for
metastable scenarios as the tunneling lifetime depends on the shape of
the manifestly non-convex tunnel barrier.

\subsection{Pseudo-spectral flows}

In general, the derivative expansion of the functional RG results in a
system of coupled partial differential equations (PDEs). In the
present case, we have to deal with one such PDE
\eqref{eq:FlowPotential} coupled to an ODE \eqref{eq:FlowYukawa} and
two algebraic equations \eqref{eq:etaphi} and \eqref{eq:etapsi}. 

In order to obtain global information with high accuracy, we advocate
pseudo-spectral methods \cite{Boyd:ChebyFourier} which span the
solutions in terms of global basis functions of high polynomial (or
rational) degree. Under mild analyticity assumptions, the convergence
to the exact result is exponential \cite{Boyd:ChebyFourier}. In particular
in comparison with finite difference methods, pseudo-spectral methods
are memory minimizing, as a certain accuracy requires only a
comparatively small number of grid points.

In the following, we briefly sketch our methods; for more details, see
\cite{Borchardt:2016}. It is worth mentioning that pseudo-spectral
methods have already successfully been applied to various problems in
physics in general \cite{Robson:1993}; for first applications in the
context of the functional RG, see
\cite{Fischer:2004uk,Gneiting:2005}. The present code development is
based on a highly accurate pseudo-spectral solver for computing global
fixed-point solutions within the functional RG
\cite{Borchardt:2015rxa}.

In principle, pseudo-spectral methods include any kind of basis
function system. In the present work, we use Chebyshev polynomials of
the first kind. Boundary effects can be controlled since the
polynomials are defined on a finite interval. The behavior of
higher-order coefficients provides an error estimate of the absolute
remainder of the solution. As a main advantage also in practice,
function values, derivatives, and integrals of the objective function
are easily accessible from the Chebyshev coefficients by means of
recursive algorithms yielding a high precision calculation.

We apply pseudo-spectral methods in both the field and the RG time
direction.  For this, we subdivide the time interval into patches,
apply a Newton-Raphson iteration to each patch, and solve for the
coefficients.  For minimizing the number of coefficients and
increasing the resolution in physically interesting regions, we use
multiple domains in field direction.  All these patches are connected
by additionally demanding matching conditions for the objective
function and a sufficient number of its derivatives. The
Newton-Raphson iteration provides an error estimate for the solution
of the equations.

As a result, this method is stable over many orders of magnitude in
$k$.  This enables us to choose high UV cutoffs.  This choice is
solely restricted by the number of digits needed for tuning the IR
quantities. As the flow of the present problem includes a scalar mass
parameter with canonical scaling of dimension 2, we need to tune
approximately twice as many digits as the number of orders of
magnitude between the UV scale and the Fermi scale. All full potential
computations have been done with \texttt{long double}. Thus, we
restricted ourselves to a maximal UV cutoff of $10^{10}$GeV for the full
potential calculation.  In principle, higher values for $\Lambda$ are
straightforwardly accessible by using a higher accuracy for the
floating-point arithmetics.

\subsection{Higgs mass bounds}

\begin{figure}[t]
\includegraphics[width=0.45\textwidth]{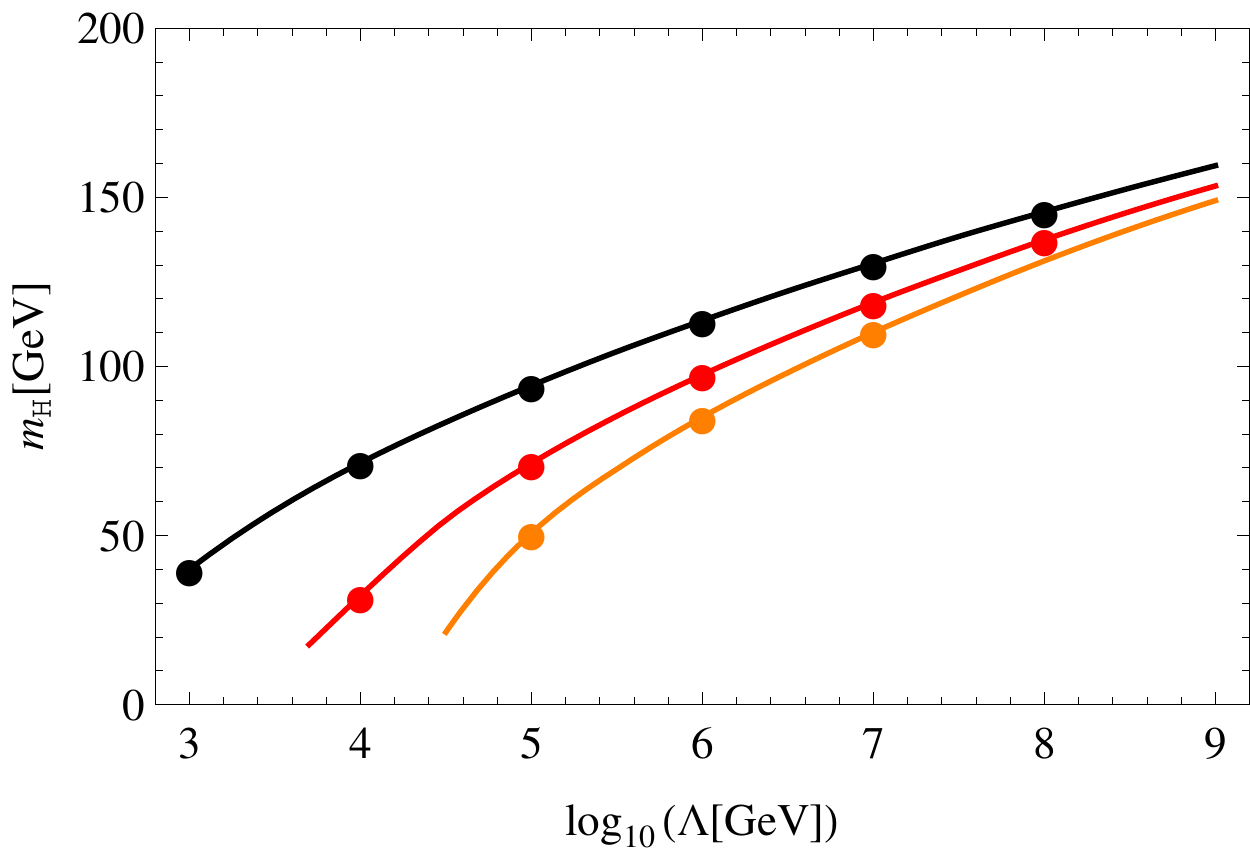}
\caption{Higgs boson mass as a function of the UV cutoff for various
  bare potentials. The filled circles are obtained by solving the full
  PDE system. These match perfectly with the Higgs masses computed
  within a polynomial expansion \eqref{eq:polyexp} of the scalar
  potential for the class of $\phi^4$-type bare potentials (black,
  conventional ``lower bound'' $\lL=0$) as well as generalized bare
  potentials for the case where the effective potential is stable
  for $\Lambda\gtrsim 10^{4}$GeV (red, $\lL=-0.1$, $\lambda_{3,\Lambda}=3$) or
  develops a metastability (orange, $\lL=-0.15$, $\lambda_{3,\Lambda}=3$).}
\label{fig:HiggsMass-Full-vs-Taylor}
\end{figure}

As a first benchmark, we perform a comparison to local polynomial
solutions of the flow equation. For this, we compute Higgs masses for
different initial values for the flow equations over a large range of
cutoff values. In Fig.~\ref{fig:HiggsMass-Full-vs-Taylor}, we depict
the resulting IR Higgs mass as a function of the UV cutoff $\Lambda$:
for the restricted class of $\phi^4$ bare potentials, the black data
shows the resulting lowest possible Higgs mass, i.e., the conventional
lower bound for $\lL=0$. Examples within the class of generalized bare
potentials that lead to a relaxation of the lower bound are shown in
red ($\lL=-0.1$, $\lambda_{3,\Lambda}=3$) and orange
($\lL=-0.15$, $\lambda_{3,\Lambda}=3$). The solid
lines mark the Higgs masses computed within the polynomial truncation,
and filled circles indicate the Higgs masses resulting from the
pseudo-spectral full potential computation of this work. The black and
red line agree with the results of \cite{Gies:2013fua}, and illustrate the relaxation of the conventional lower bound
by higher-dimensional operators. The orange data corresponds to a
potential that develops a metastability (i.e., seems pseudo-stable in
the polynomial expansion). For all cases, the pseudo-spectral data
lies perfectly on top of the polynomial results. The full numerical
PDE solution thus provides a strong confirmation that the polynomial
expansion is suitable for extracting local information such as the
Higgs mass ($\sim$ curvature of the potential at $\phi=v$).

\subsection{Global flows}

\begin{figure}[t]
\includegraphics[width=0.45\textwidth]{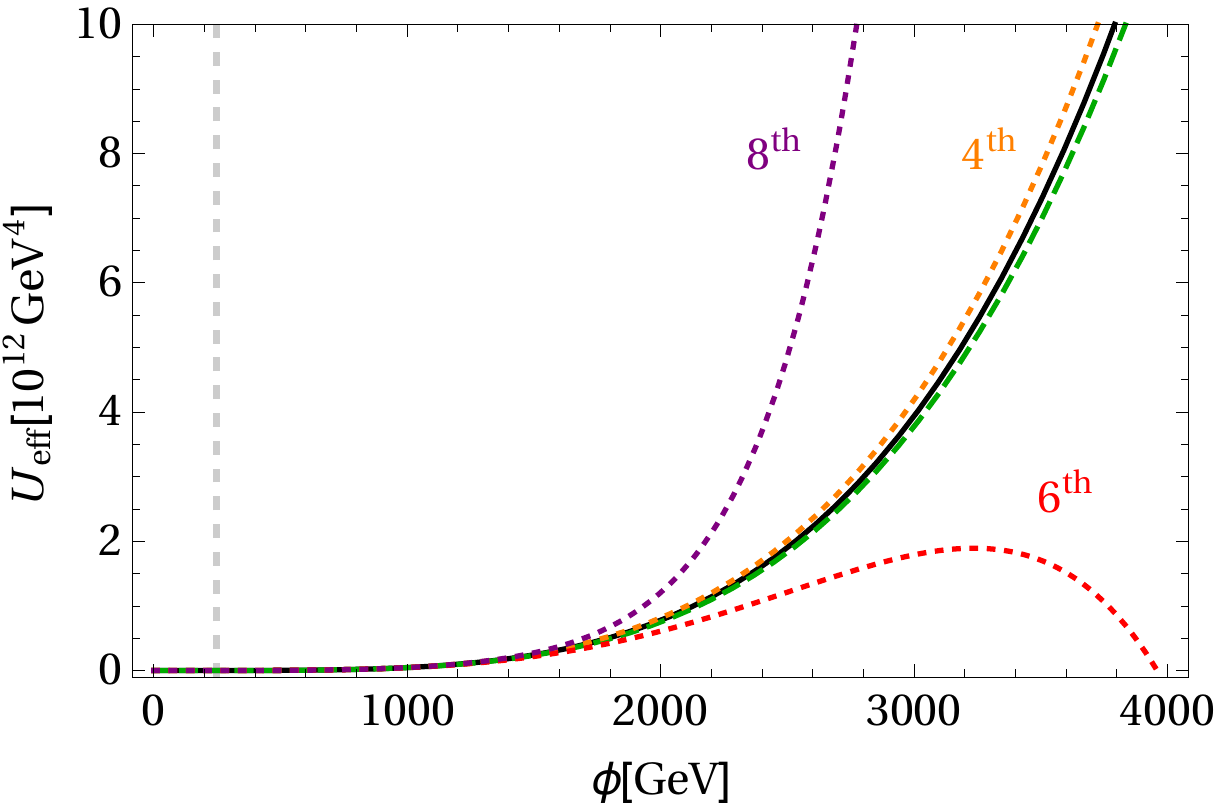}
\caption{Comparison between
  the effective potential of the full calculation (black, $k_\text{IR}=113.9$GeV), the
  mean-field result (green dashed) and polynomial truncated potentials
  (dotted) up to fourth order (orange), sixth order (red) and eighth
  order (purple) for the stable case. The cutoff is chosen to be
  $\Lambda = 10^9$GeV and bare couplings are tuned such that the Higgs
  mass is $159.4$GeV in all cases for reasons of comparison. The
  vertical gray-dashed line indicates the position of the Fermi
  minimum at $\phi=246$GeV.}
\label{fig:IRpotential-Full-vs-Taylor-MF}
\end{figure}

Let us start with a closer look at the global behavior of the flow for
the class of the $\phi^4$ bare potentials.  Obviously, the polynomial
truncation lacks in describing the asymptotic behavior of the
potential which can be seen in
Fig.~\ref{fig:IRpotential-Full-vs-Taylor-MF}.  This is not surprising
since the flow equations suggest the asymptotic behavior to be that of
the UV potential $\sim\phi^4$ because fluctuations for large field
amplitudes are suppressed. By contrast, the asymptotic behavior of the
polynomial expansion by construction is fixed to the highest power of
the field which is taken into account in the truncation, $\sim
\phi^{2\Np}$. These higher order couplings are generated during the RG
flow, even if the bare potential is of $\phi^4$ type.  Therefore,
considering only terms up to $\phi^4$, (accidentally) displays the
asymptotic behavior best.

Naively, the polynomial truncation up to sixth order in $\phi$ seems
to suggest an instability; however, the inflection point is beyond the
radius of convergence of the polynomial expansion around the Fermi
scale. This radius of convergence is approximately of the order of the
curvature around the electroweak minimum \cite{Gies:2013fua}. For
large field values the polynomial expansion behaves like an asymptotic
series with alternating signs between the coefficients. Incidentally,
an alternating series is also obtained from the polynomial expansion
of the mean-field effective potential.  As long as the $\phi^4$ class of
 bare potentials is considered, no hint for an
in-/metastability can be found within the radius of convergence of the
polynomial expansion. This is confirmed by the fully stable potential
obtained from the global pseudo-spectral flow.

We observe that the mean-field potential agrees quite well with the
results for the full potential, for small as well as for larger field
values, see green dashed curve in
Fig.~\ref{fig:IRpotential-Full-vs-Taylor-MF}.  The fluctuations of the
bosons appear to play a minor role in this parameter regime near the
lower edge of the IR window, where the fermionic contributions
dominate.  Moreover, neglecting the anomalous dimensions and
the flow of the Yukawa-coupling does not have a significant
effect. Thus, the simple mean-field effective potential is an
effective tool to get first insights into the global behavior of the
scalar potential, justifying the seemingly severe approximations of
Sect.~\ref{sec:CW-vs-MF}.

The qualitative picture remains the same for full flows also in the
class of generalized bare potentials. For those cases where no second
minimum emerges during the polynomial truncated flow, we observe that
also the full flow does not develop an outer minimum for field values
larger than the Fermi scale. Our new results hence confirm the
existence of a class of stable bare potentials giving rise to Higgs
masses below the conventional lower bound as, for instance, depicted
by the red curve in Fig.~\ref{fig:HiggsMass-Full-vs-Taylor} for $\Lambda\gtrsim 10^{4}$GeV.
Therefore, the mechanism of lowering the lower bound for completely
stable potentials remains active beyond the polynomial expansion and
the mean-field analysis. Thus higher-order operators can diminish the
lower Higgs mass bound of the standard model.

Let us take a closer look at the inner workings of the equations: for
large field amplitudes in the asymptotic regime of the potential, all
fluctuations are suppressed as the threshold functions approaching zero.
In this regime, the flow is dominated by dimensional scaling, i.e.,
the first two terms in Eq.~\eqref{eq:FlowPotential}.  This holds for
the $\phi^4$ as well as for the generalized class of bare potentials
irrespectively of the stability properties.

\begin{figure}[t]
\includegraphics[width=0.45\textwidth]{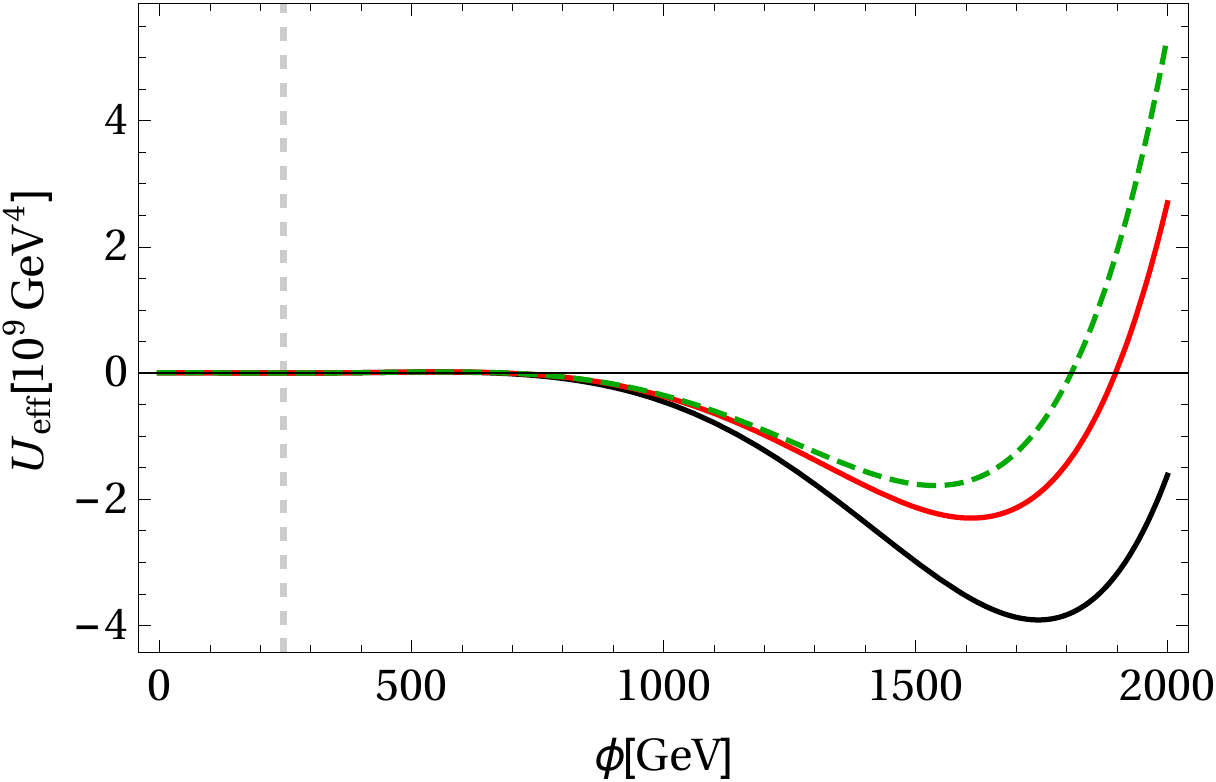}
\caption{Comparison between the effective potential for a metastable
  case obtained by the full pseudo-spectral calculation where the anomalous
  dimensions are computed at the electroweak vev (black/lower solid line, $k_\text{IR}=105.5$GeV), at the outer
  minimum (red/upper solid line, $k_\text{IR}=93.3$GeV) and the mean-field result (green dashed line). Setting the
  anomalous dimensions by hand to zero would yield a rather good agreement with the
  mean field potential.  The Higgs mass is tuned to $\mH=25$GeV where
  the UV-cutoff is $5$TeV and $\lambda_{3,\Lambda} = 1$.}
\label{fig:IRpotential-metastable-Full-vs-Taylor-MF}
\end{figure}

Deviations from the mean-field limit require relevant bosonic
fluctuations. In the small coupling (i.e., small Higgs mass) regime,
this can indeed occur in the full flow due to threshold effects of the
following type: if a second minimum emerges seeded by a suitable bare
potential, the curvature near the top of the barrier between the minima
is negative, such that the bosonic threshold function $l_0^d(u'+2\rho
u''; \eta_\phi)$ is enhanced. This type of bosonic enhancement is
only visible in a full potential flow. For a meaningful comparison
between mean-field and full flow, we tune the bare potential such that
the IR physics including the Higgs mass is kept fixed. For
illustrative purposes, we choose a UV cutoff of only $5$TeV, a
Higgs mass of 25GeV and $\lambda_{3,\Lambda} = 1$. The resulting potentials end up in the metastable
regime as depicted in
Fig.~\ref{fig:IRpotential-metastable-Full-vs-Taylor-MF}. The
mean-field potential (dashed line) differs from the full solutions
(solid lines) integrated down to an IR value of $k_{\text{IR}}\sim 100$GeV. The full solutions correspond to defining the
anomalous dimensions at the local Fermi minimum (black) or at the
second global minimum (red). These curves differ as fields and
couplings are renormalized differently within the two schemes, with
the black curve corresponding to a renormalization choice better
resolving the Fermi minimum and the red curve better suited for the
second minimum. In other words, the axes for the different solid lines
have a different meaning due to the different field
rescaling during the flow. If the
anomalous dimensions were artificially set to zero, mean-field and full
potential results would still match rather well.

\subsection{Convexity of the effective potential}

By definition of the effective action as a Legendre transform of the
Schwinger functional, we expect the full effective action and
particularly the effective potential to be convex functions of the
field. This convexity property cannot be seen neither in the
perturbative construction of the \scpotential\ nor in the mean-field
approximation. The Wetterich equation does have this convexity
property in the limit $k\to 0$ for the bosonic potential, see
e.g. \cite{Berges:2000ew,Litim:2006nn}. However, at finite $k$, the
regulator term $\sim R_k$ sources a non-convex contribution which
vanishes in the limit $k\to0$. 

For potentials with a single minimum, it is known that convexity of
the running potential sets in rather late in RG time, i.e., convexity
is driven by the very deep IR modes which are often no longer relevant
for the IR observables. For instance in the examples above, we have
stopped the flow at scales $k_{\text{IR}}\sim 10\ldots100$GeV, where the IR
Fermi scale observables have already settled to their physical
values. Still, for these values of $k$, the approach to convexity has
not fully set in yet.  Whereas this demonstrates that convexity is not
important for the static observables, it is an interesting question as
to whether the approach to convexity can be important for estimates of
the tunneling rate between two different minima. The relevance of this
question becomes obvious from the fact that any tunneling barrier in a
convex potential is (naively) exactly zero by construction.

In order to understand the onset of convexity in the present model,
let us start with the simpler case with only one minimum at the Fermi
scale. Here, convexity only affects the inner region of the potential
with $\phi<v$. Technically, convexity of the effective potential is
generated by singularities in the bosonic propagators entering the
threshold functions. In the present case, the bosonic threshold
function $l_0^d$ corresponding to the regularized propagator
is proportional to
\begin{equation}
l_0^d \sim \frac{1}{1+u'+2\rho u''},
\end{equation}
exhibiting a
singularity at $u'+2\rho u'' \to -1$, or $u'\to -1$ for small $\rho$
or small $u''$. The flow avoids this singularity by renormalizing the
negative curvature of the potential in the inner region $0\geq
U_k''(\phi)\sim k^2( u'+ 2\rho u'') \gtrsim - k^2\to 0$ with $k\to
0$. This establishes convexity for $k\to 0$.

\begin{figure}
 \includegraphics[width=0.45\textwidth]{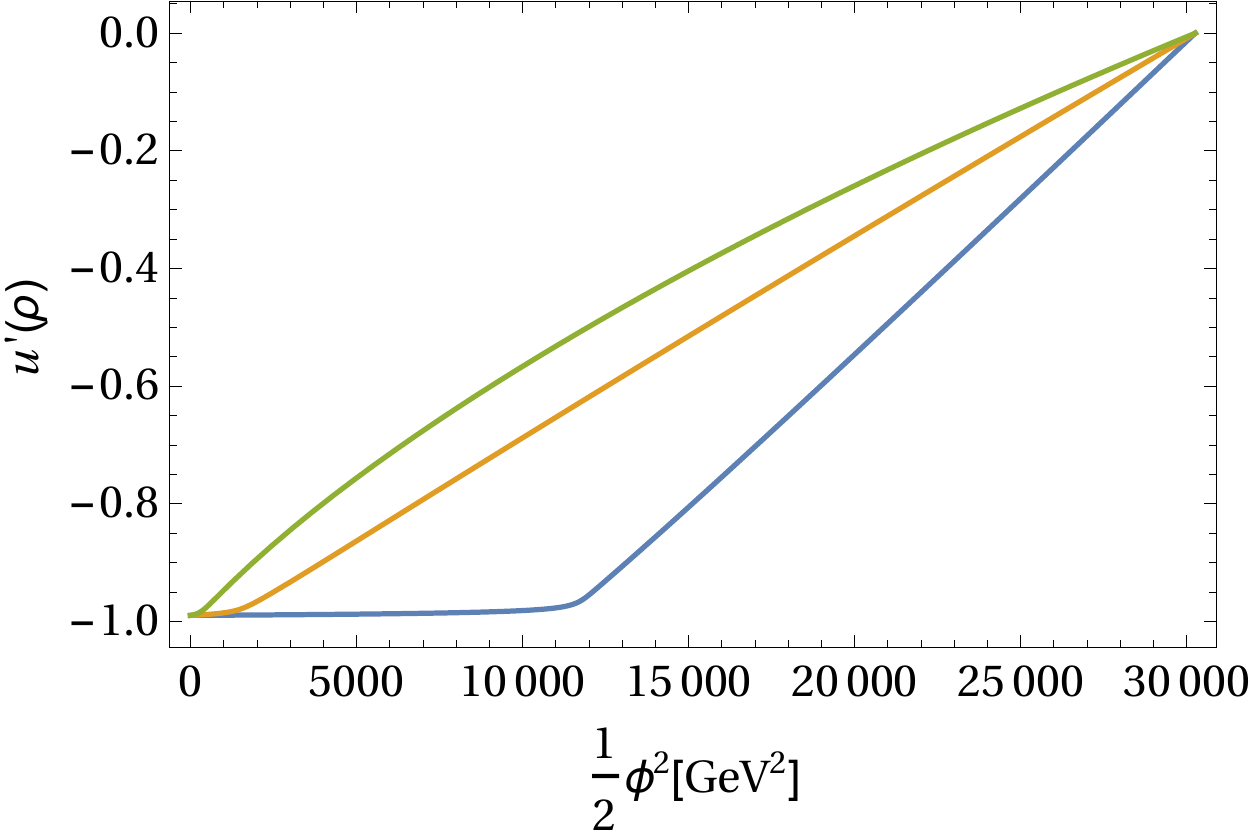}
 \caption{The approach to convexity of the potential is faster if the relation
   between bosonic and fermionic coupling $\mH/\mtop$ increases (from
   green/top to blue/bottom: $\mH/\mtop = 0.23, 0.66, 1.12$.  Here,
   the first derivative of the potential as a function of the
   dimensionful field invariant is depicted. All potentials exhibit a
   minimum (i.e. $u'=0$) at $\phi=246$GeV
   ($\phi^2/2{=}30258\text{GeV}^2$). The approach to convexity becomes
   manifest by a characteristic flattening of the inner region and
   $u'$ approaching $u'\to -1$, see text.  We have chosen $\Lambda =
   10^3$GeV and $\mH=39.7$GeV for the green/top curve stopping the
   flow at $k_\text{IR}=33.4$GeV, and $\Lambda = 10^6$GeV and
   $\mH=113.6$GeV for the orange/middle curve stopping at
   $k_\text{IR}=80$GeV. The blue curve is added for illustration; here
   $\Lambda = 6.5\cdot10^4$GeV, $\text{vev}=246$GeV, $\mtop=426.3$GeV,
   $\mH=476.6$GeV, and $k_\text{IR}=264.5$GeV, such that the curve is
   not tuned to the physical top mass in terms of the renormalization
   conditions \eqref{eq:rencond}.}
 \label{fig:convexity1}
\end{figure}

In comparison to purely bosonic models, fermionic fluctuations delay
convexity since they enter the flow equation with an opposite sign, 
cf.~the last term in \Eqref{eq:FlowPotential}. 
Thus, bosonic fluctuations have to exceed the fermionic fluctuations
first.  As convexity also introduces a non-analyticity at $\phi=v$,
its onset becomes numerically first visible in higher
derivatives. Therefore, we consider the first derivative of the
potential $u'$ in the following. The balancing between bosons and
fermions also implies that the onset of convexity becomes more
pronounced if the boson coupling $\lambda_2$ is enhanced relative to
the fermion coupling $h$. In terms of physical parameters, this
implies that convexity should become more prominent for larger
Higgs-to-top mass ratios. In Fig.~\ref{fig:convexity1}, we plot $u'$
for three different ratios $\mH/\mtop$. The flow has been stopped at a
scale $k_{\text{IR}}$ such that the potentials have the same distance
from the singularity of the bosonic propagator $1-|u'(0)|=0.01$. The
faster approach to convexity then is directly visible in terms of the
position of non-analyticity $\rho_{\text{kink}}$ which we observe to
move towards larger field amplitudes if $\mH/\mtop\sim \lambda_2/h$
increases. By contrast, if $\mH/\mtop$ is small, the characteristic
flat region of $u'$ is hardly visible at this particular scale $k_{\text{IR}}$ and
would increase only towards even smaller scales.

It should be emphasized that convexity is a notoriously difficult
problem for pseudo-spectral methods, since it introduces a
non-analyticity which violates the assumptions on the function space
for which exponential convergence can be proved. Hence, the numerics
will unavoidably face a singularity problem in the very deep IR. For
further adapted methods, see \cite{Bonanno:2004pq,Pelaez:2015nsa}.

\begin{figure*}
 \includegraphics[width=0.4\textwidth]{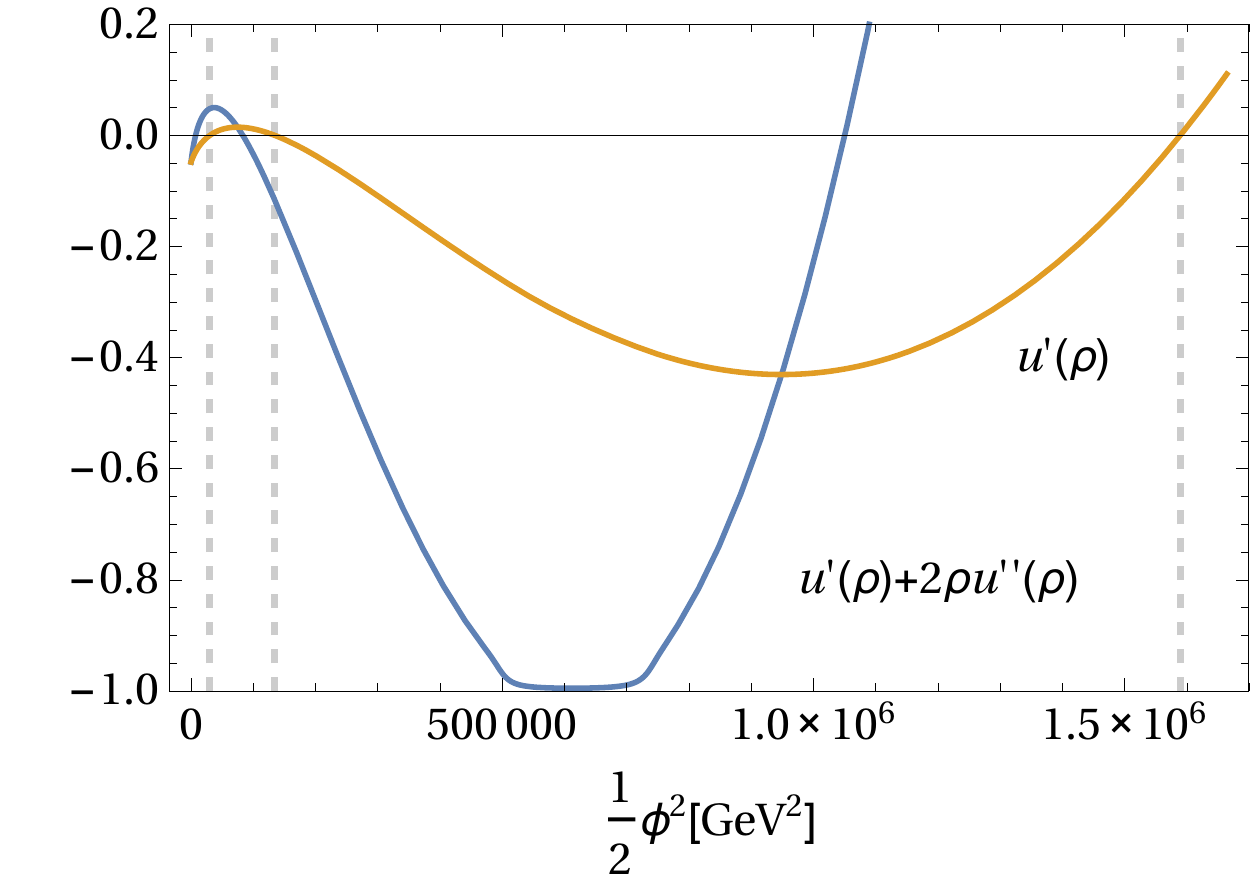}
 \hspace*{0.075\textwidth}
  \includegraphics[width=0.4\textwidth]{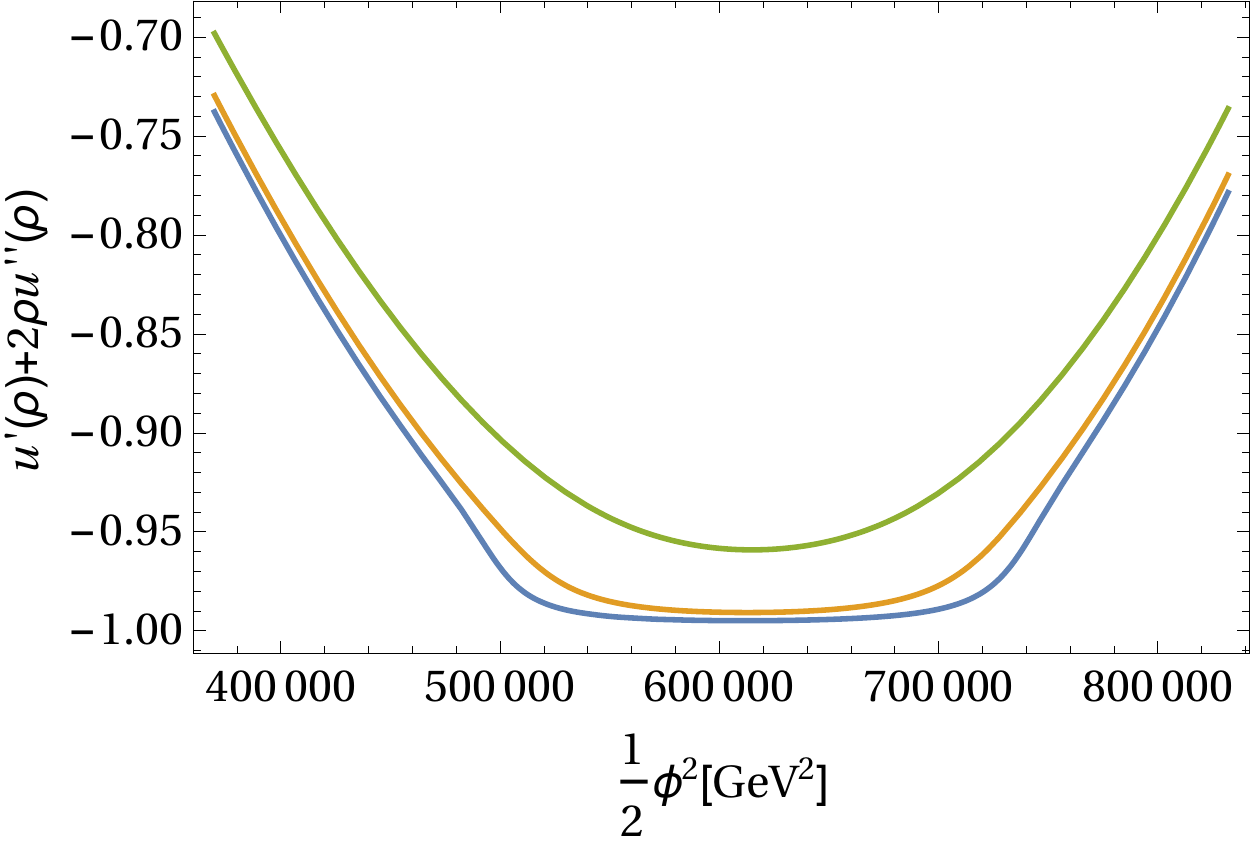}
 \caption{Onset of convexity at a scale $k\sim 100$GeV for the case of two minima.
  Whereas the small-field region is still dominated by fermionic fluctuations,
   the bosons control the flow for larger field amplitudes, especially in the
   region of the minimum of the field-dependent mass term $u'+2\rho u''$.
   Left panel: mass term and the first derivative of the potential $u'$
   over a wide field range. The vertical dashed lines mark the location
   of both minima and the maximum of the tunnel barrier of the potential
   in between. The minimum of the mass term
 approaches the singularity $u'+2\rho u''\to -1$, triggering the approach to convexity. This is shown in the right panel in detail
 for decreasing scale $k\in\{112.5,110,109.5\}$GeV from top (green) to bottom (blue).
 For this example flow, we have used $\mH=24.1$GeV, $\lambda_{3,\Lambda} = 1$ and $\Lambda=5$TeV.
}
 \label{fig:convexity2}
\end{figure*}

Let us now turn to the more interesting case of two minima which is
numerically more challenging since the field-dependent ``mass term'' becomes negative, 
$u'+2\rho u''<0$, not only for small fields but also for a second
region at larger fields.  As an illustrative example, we choose a
similar flow as above.  We plot this mass term in the region of both
minima of the tunnel barrier,
cf. Fig.~\ref{fig:convexity2} (left panel).  For comparison we
show $u'(\rho)$ as well. The dashed vertical lines indicate the position of both minima 
and the maximum in between.  Convexity becomes first visible for 
larger fields at the minimum of the mass term which tends to
$u'+2\rho u'' \to -1$ after $k$ has dropped below the scale of the top
quark, cf. Fig.~\ref{fig:convexity2} (right panel). For the current example, 
this minimum of the mass term is located in between the maximum and the 
outer minimum of the potential, but this relative position may vary 
depending on the scale and the precise choice of parameters.
As the maximum of the potential is situated within the region where $u'+2\rho u''<0$ for larger fields,
the flat region eventually extends beyond the maximum, significantly affecting
the tunnel barrier for small scales $k\lesssim 100\text{GeV}$.
For small fields, the fermions still control the flow at $k\sim
100$GeV.  However, for decreasing scale $k$ the bosonic fluctuations
win out over the fermionic ones and convexity sets in as well, 
similar to Fig.~\ref{fig:convexity1}.  We 
emphasize that the approach to convexity appears to set in at different scales for large field 
amplitudes than for small ones.

In the present
example, convexity affects the tunnel barrier at scales $k$ which are 
more than an order of magnitude smaller than the field amplitude
of the barrier and the outer minimum. A calculation of the tunnel rate
which is dominated by the latter scales hence is expected to be only
weakly influenced by the approach to convexity.
As a general rule, we conclude that the standard recipes for
calculations of the tunnel rate \cite{Coleman:1977py,Callan:1977pt}
remain unaffected as long as the fermion fluctuations dominate the
renormalization of the potential. Whether or not this is the case at
the relevant scales of interest will in general depend on the details
of the scale-dependent potential and thus also on the details of the
bare potential. As soon as the boson fluctuations become important,
the approach to convexity has also be accounted for in estimates of
the tunneling rate.

In the functional RG context, a proposal for this has been worked out
for scalar models in \cite{Strumia:1998qq}. A formalism that can also
systematically deal with further radiative effects in the resulting
inhomogeneous instanton background on top of a radiatively generated
potential has recently been developed with the help of a
self-consistent functional scheme based on the 2PI effective action
\cite{Garbrecht:2015oea,Garbrecht:2015yza}.

We would like to emphasize the necessity of a simultaneous consistent
treatment of the renormalization flow of the potential together with
the fluctuation contributions in a tunnel-rate calculation -- even if
the bare potential was known exactly. Of course, unknown higher
dimensional operators then further add to the indeterminacy of the
vacuum decay rate
\cite{Branchina:2013jra,Branchina:2014usa,Branchina:2014rva,Lalak:2014qua,Eichhorn:2015kea}. For
instance, the influence of gravity-induced higher dimensional
operators has been studied in \cite{Loebbert:2015eea,Bhattacharjee:2012my,Haba:2014qca,Abe:2016irv}.

\subsection{Quantum phase diagram of the Higgs-Yukawa model}

Can the outer global minimum be used to define the electroweak vacuum?
If the occurrence of metastability is rather generic in presence of
higher-dimensional operators, could it be possible to fix physical
parameters with respect to the global minimum as the Fermi scale? In
order to address these questions, we now reconsider the model from a
more general viewpoint.

So far, we have fixed the model with the help of the renormalization
conditions \eqref{eq:rencond} applied to the first or innermost
minimum. Instead, let us now start from a fixed UV cutoff $\Lambda$
with some bare potential bounded from below and read off the IR phases
from the effective potential at some IR scale $k$ where all modes have
decoupled (apart from the approach to convexity). We are most
interested in this quantum phase diagram as a function of the
(super-)renormalizable operators $\sim m_\Lambda^2 \phi^2$ and
$\sim\lL\phi^4$, as the electroweak precision data tells us that the
standard model is sufficiently close to the Gau\ss{}ian fixed point,
where perturbation theory based on these operators works very well. In
other words, higher-dimensional operators do not take a momentarily
measurable influence on collider data.

In the language of critical phenomena, the standard model appears to
be close to a second-order quantum phase transition that effectively
allows to push the UV cutoff to large values (compared to collider
scales). The natural candidate in the standard model is the
electroweak (quantum) phase transition represented by the
order-disorder phase transition of discrete chiral symmetry in our
simple model. It is, in fact, straightforward to verify by means of
perturbation theory, mean-field theory or the functional RG that this
phase transition is of second order for $\phi^4$ type bare potentials
in the stable regime. The ``control parameter'' for the quantum phase
transition is the bare mass term $\mL$.

In the following, we perform this investigation for the class of
generalized bare potentials. For this, we fix $\lambda_{3,\Lambda}=1$
as a representative of a higher-dimensional operator that induces
absolute stability. We expect the following results to hold also for
other polynomial operators that ensure absolute stability for large
field amplitudes. For technical simplicity, we keep the Yukawa
coupling $h^2\sim \mathcal{O}(1)$ fixed and also neglect the anomalous dimensions, as both do
not induce qualitative differences. Still, we keep the full bosonic
fluctuation contribution to the flow of the potential.

\begin{figure}
\vspace*{0.5cm}
 \includegraphics[width=0.49\textwidth]{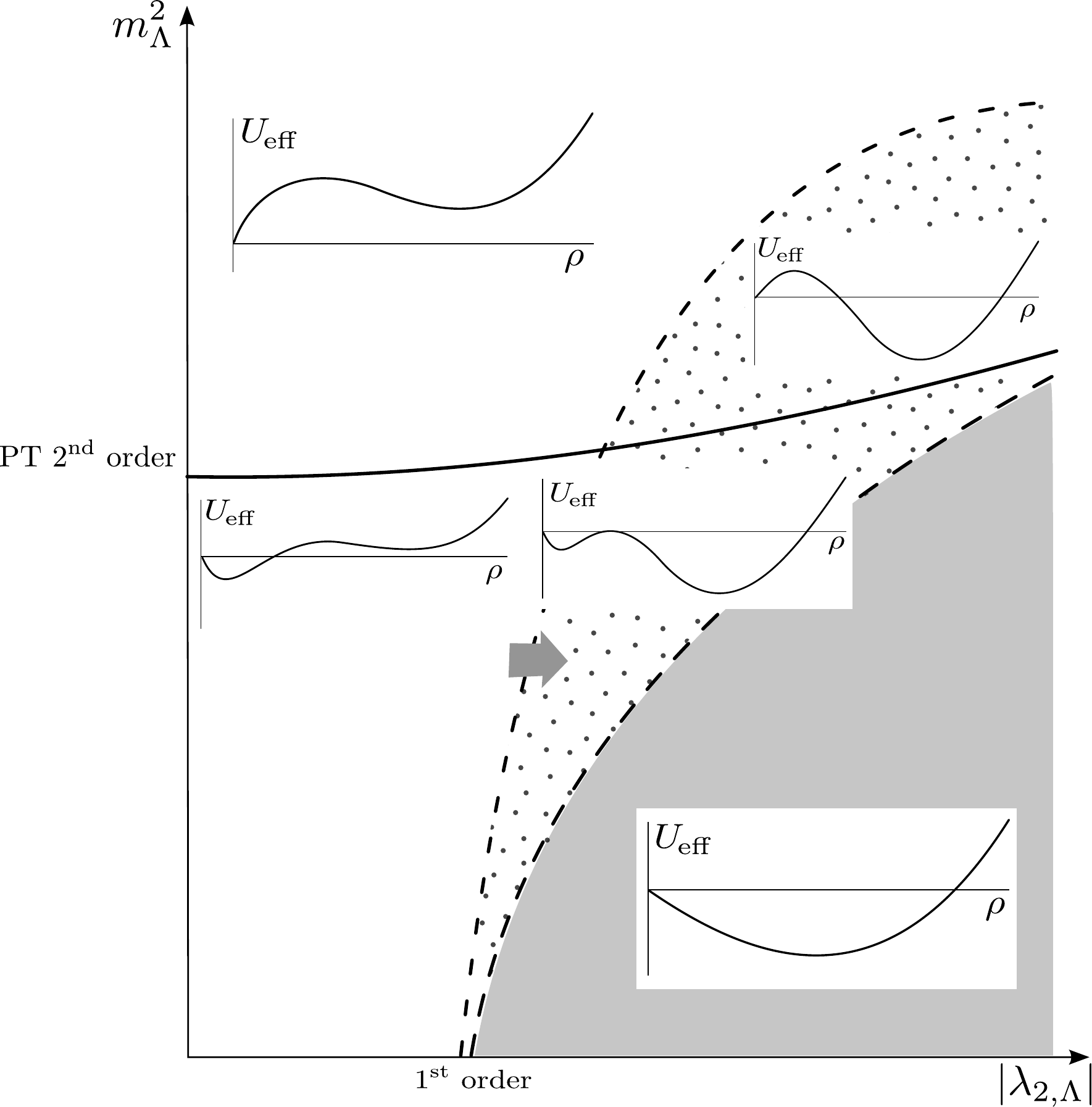}
 \caption{Quantum phase portraits of the IR effective potentials with
   possible metastabilities seeded from the bare action. As an
   example, the large amplitude $\rho\sim\phi^2$ region is stabilized
   by a $\phi^6$ operator in the UV. The phase portraits are sketched
   for various initial values for a negative bare $\lL$ and suitable
   mass parameters $\mL$ near the critical regions.
   In the dotted region the effective potential is metastable,
   while it is stable in the white and gray-shaded region.
   In the gray-shaded region only the outer minimum exists.
 }
 \label{fig:phasediagram}
\end{figure}

Choosing $\lL$ negative but with a small absolute value, the potential
will still show only one minimum and the phase transition as a
function of $m_{\Lambda}^2$ still is of second order as for the
$\phi^4$-class, cf. left-hand side of Fig.~\ref{fig:phasediagram}.
Increasing the absolute value of a negative $\lL$ a bit, and starting
with a large value of $\mL$, the sufficiently negative $\lL$ may seed
a local higher minimum at large field amplitudes. Nevertheless, the
system is in the symmetric phase with the global minimum at $\phi=0$
(upper left part of Fig.~\ref{fig:phasediagram}).  (On the left-hand
side of this figure, we do not further distinguish between the
existence or non-existence of a further local outer minima; potentials
with a local outer minimum shown here only represent possible
examples.)

Decreasing $\mL$, we indeed observe a second-order
phase transition to a broken phase driven by fermion fluctuations
where the order parameter $\phi_0=v$ is switched on continuously, cf.
white region in Fig.~\ref{fig:phasediagram}.  A local higher minimum
at larger field amplitudes may arise by decreasing $\mL$ or persists
if it already existed.  It is this second-order phase transition which
can serve to define a ``continuum limit'' essentially establishing
cutoff independence.

Decreasing $\mL$ further first leads to a lowering
of the outer minimum such that the inner minimum becomes
metastable (dotted region). The phase transition between the two cases is of first
order (dashed lines). For even smaller mass parameters $\mL$, the
inner minimum vanishes discontinuously while the outer remains
(gray-shaded region in Fig.~\ref{fig:phasediagram}). We also
classify this discontinuous change of the system as a first-order
transition, even though it would not correspond to a thermal phase
transition: on both sides of the lower dashed line the system is
dominated by the global minimum in the thermodynamic limit.

This analysis demonstrates that only the transition between the
symmetric and the symmetry-broken phase, which is driven by fermion
fluctuations, is of second order. Therefore, only this transition can
be used to separate the cutoff scale from the IR physics in this
model. This forces us to associate the Fermi scale with the innermost
minimum driven by fermion-fluctuations. The outer one seeded by the
bare action cannot be used for a definition of the Fermi scale as it
is highly unlikely to match with the perturbative description of
electroweak precision data. In our flows, this mismatch becomes
visible in the practical difficulty to push the cutoff beyond
$\sim1$TeV while satisfying all renormalization conditions with $v$
corresponding to the outer minimum.

For negative $\lL$ with an even larger absolute value (right-hand side of
Fig.~\ref{fig:phasediagram}), the phase portraits are similar in the
sense that only the transition driven by the fermionic fluctuations in
the inner region of the potential is a second-order
transition. The difference is that this transition occurs only
after the outer minimum seeded by the bare action has become the
global one. As a consequence, both the symmetric phase for larger
$\mL$ as well as the fermion driven broken phase (inner minimum) are
metastable (dotted region) on both sides of the transition. In this regime of bare
potentials, the separation of IR physics from the UV scale hence goes
along with a metastability.

We can also compare the phase portraits for fixed $\mL$ and use $\lL$
as control parameter. For instance, the transition marked by the gray thick arrow
is a first-order broken-to-broken
transition. 
This is likely to correspond to an
equivalent transition first observed in lattice simulations of a
similar chiral model \cite{Chu:2015nha}.

We emphasize that the phase portraits determined here correspond to
quantum phase transitions with control parameters corresponding to
parameters of the bare action. This is a priori unrelated to the
nature of finite temperature phase transitions in the same model, even
though a relation might be established dynamically because of a
thermal decoupling of the fermions. For recent lattice studies, see
\cite{Akerlund:2015fya,Akerlund:2015gfy}.

\section{Conclusions}
\label{sec:conc}

We have investigated the renormalization flow of the effective scalar
potential keeping the full dependence on all relevant scales: the
field amplitude $\phi$, an RG scale $k$ and the UV cutoff scale
$\Lambda$. This is necessary to overcome the limitations of
conventional approximation schemes, relying on identifications such as
$k=\phi$, or implicit perturbative limits $\Lambda\to\infty$. The
advantage of keeping the full scale dependence becomes obvious for the
analysis of metastabilities.

Using a simple Higgs-Top-Yukawa model as an example, our analysis
demonstrates that metastabilities are not primarily induced by fermion
fluctuations, but have to be seeded by suitable structures in the bare
potential. In particular, metastabilities cannot occur if a
well-defined bare action is restricted to contain only renormalizable
operators. Upon the inclusion of suitable higher-dimensional
operators, metastabilities generically occur for small Higgs masses
and large cutoffs -- at least within the class of simple polynomial bare
potentials studied in this work.

Whereas these latter conclusions are in part reminiscent to results
obtained within perturbative estimates based on a single-scale
effective potential ($k=\phi$), it is worthwhile to note some decisive
differences: our approach allows for addressing arbitrary bare
potentials, defining the model purely in terms of its symmetry, field
content and a minimal set of IR parameters. No assumption about the
manifest absence of an infinity of irrelevant operators is made. While
the occurrence of higher-dimensional operators is conventionally
interpreted as particle physics beyond the standard model (e.g.,
induced by integrating out heavy particle multiplets), our approach
allows to also associate such operators to any origin that can be
parametrized in terms of any effective action, e.g., a coarse-grained
continuum action in a spacetime arising from discrete building blocks.

In this work, we have studied global RG flows of the potential using
pseudo-spectral methods. This facilitates to study the full RG
evolution of competing minima, to analyze the quantum phase diagram of
the model, and to quantify the approach to convexity. The latter is
not accessible in perturbation theory or mean-field theory. The global
flows also serve to confirm earlier results from local flows evaluated
at the running Fermi minimum to a high accuracy, such as, for
instance, the relaxation of the perturbative lower bound on the Higgs
mass. On the other hand, the global flow also reveals the limitations
of the local flow in metastable regimes as competing minima turn out
to be beyond the radius of convergence of local flows. Further, the
global flows also demonstrate the usefulness of the mean-field
approximation in the small-Higgs-mass regime. 

Finally, we emphasize that a full determination of \textit{consistency
  bounds} for the IR observables of the standard model as a function
of the cutoff $\Lambda$ as the scale of maximum UV extension has not
yet been completed. For this, the mapping of a wide range of bare
actions to the IR observables would have to be computed with the RG,
technically corresponding to an extremization problem in an
infinite-dimensional space. The capability of handling global flows
and extending the current studies to nonpolynomial interactions will
be a necessary prerequisite for this.

\section*{Acknowledgments}

We thank Bj\"orn Garbrecht, Tobias Hellwig, Tina K. Herbst, Benjamin
Knorr, Stefan Lippoldt, Peter Millington, Michael M. Scherer, Matthias
Warschinke, and Luca Zambelli for valuable discussions. We acknowledge
support by the DFG under Grants No. GRK1523/2 and No. Gi328/7-1.

\bibliography{bibliography}

\begin{thebibliography}{110}%
\makeatletter
\providecommand \@ifxundefined [1]{%
 \@ifx{#1\undefined}
}%
\providecommand \@ifnum [1]{%
 \ifnum #1\expandafter \@firstoftwo
 \else \expandafter \@secondoftwo
 \fi
}%
\providecommand \@ifx [1]{%
 \ifx #1\expandafter \@firstoftwo
 \else \expandafter \@secondoftwo
 \fi
}%
\providecommand \natexlab [1]{#1}%
\providecommand \enquote  [1]{``#1''}%
\providecommand \bibnamefont  [1]{#1}%
\providecommand \bibfnamefont [1]{#1}%
\providecommand \citenamefont [1]{#1}%
\providecommand \href@noop [0]{\@secondoftwo}%
\providecommand \href [0]{\begingroup \@sanitize@url \@href}%
\providecommand \@href[1]{\@@startlink{#1}\@@href}%
\providecommand \@@href[1]{\endgroup#1\@@endlink}%
\providecommand \@sanitize@url [0]{\catcode `\\12\catcode `\$12\catcode
  `\&12\catcode `\#12\catcode `\^12\catcode `\_12\catcode `\%12\relax}%
\providecommand \@@startlink[1]{}%
\providecommand \@@endlink[0]{}%
\providecommand \url  [0]{\begingroup\@sanitize@url \@url }%
\providecommand \@url [1]{\endgroup\@href {#1}{\urlprefix }}%
\providecommand \urlprefix  [0]{URL }%
\providecommand \Eprint [0]{\href }%
\providecommand \doibase [0]{http://dx.doi.org/}%
\providecommand \selectlanguage [0]{\@gobble}%
\providecommand \bibinfo  [0]{\@secondoftwo}%
\providecommand \bibfield  [0]{\@secondoftwo}%
\providecommand \translation [1]{[#1]}%
\providecommand \BibitemOpen [0]{}%
\providecommand \bibitemStop [0]{}%
\providecommand \bibitemNoStop [0]{.\EOS\space}%
\providecommand \EOS [0]{\spacefactor3000\relax}%
\providecommand \BibitemShut  [1]{\csname bibitem#1\endcsname}%
\let\auto@bib@innerbib\@empty
\bibitem [{\citenamefont {Aad}\ \emph {et~al.}(2012)\citenamefont {Aad} \emph
  {et~al.}}]{Aad:2012tfa}%
  \BibitemOpen
  \bibfield  {author} {\bibinfo {author} {\bibfnamefont {G.}~\bibnamefont
  {Aad}} \emph {et~al.} (\bibinfo {collaboration} {ATLAS}),\ }\href {\doibase
  10.1016/j.physletb.2012.08.020} {\bibfield  {journal} {\bibinfo  {journal}
  {Phys. Lett.}\ }\textbf {\bibinfo {volume} {B716}},\ \bibinfo {pages} {1}
  (\bibinfo {year} {2012})},\ \Eprint {http://arxiv.org/abs/1207.7214}
  {arXiv:1207.7214 [hep-ex]} \BibitemShut {NoStop}%
\bibitem [{\citenamefont {Chatrchyan}\ \emph {et~al.}(2012)\citenamefont
  {Chatrchyan} \emph {et~al.}}]{Chatrchyan:2012xdj}%
  \BibitemOpen
  \bibfield  {author} {\bibinfo {author} {\bibfnamefont {S.}~\bibnamefont
  {Chatrchyan}} \emph {et~al.} (\bibinfo {collaboration} {CMS}),\ }\href
  {\doibase 10.1016/j.physletb.2012.08.021} {\bibfield  {journal} {\bibinfo
  {journal} {Phys. Lett.}\ }\textbf {\bibinfo {volume} {B716}},\ \bibinfo
  {pages} {30} (\bibinfo {year} {2012})},\ \Eprint
  {http://arxiv.org/abs/1207.7235} {arXiv:1207.7235 [hep-ex]} \BibitemShut
  {NoStop}%
\bibitem [{\citenamefont {Maiani}\ \emph {et~al.}(1978)\citenamefont {Maiani},
  \citenamefont {Parisi},\ and\ \citenamefont {Petronzio}}]{Maiani:1977cg}%
  \BibitemOpen
  \bibfield  {author} {\bibinfo {author} {\bibfnamefont {L.}~\bibnamefont
  {Maiani}}, \bibinfo {author} {\bibfnamefont {G.}~\bibnamefont {Parisi}}, \
  and\ \bibinfo {author} {\bibfnamefont {R.}~\bibnamefont {Petronzio}},\ }\href
  {\doibase 10.1016/0550-3213(78)90018-4} {\bibfield  {journal} {\bibinfo
  {journal} {Nucl. Phys.}\ }\textbf {\bibinfo {volume} {B136}},\ \bibinfo
  {pages} {115} (\bibinfo {year} {1978})}\BibitemShut {NoStop}%
\bibitem [{\citenamefont {Krasnikov}(1978)}]{Krasnikov:1978pu}%
  \BibitemOpen
  \bibfield  {author} {\bibinfo {author} {\bibfnamefont {N.~V.}\ \bibnamefont
  {Krasnikov}},\ }\href@noop {} {\bibfield  {journal} {\bibinfo  {journal}
  {Yad. Fiz.}\ }\textbf {\bibinfo {volume} {28}},\ \bibinfo {pages} {549}
  (\bibinfo {year} {1978})}\BibitemShut {NoStop}%
\bibitem [{\citenamefont {Lindner}(1986)}]{Lindner:1985uk}%
  \BibitemOpen
  \bibfield  {author} {\bibinfo {author} {\bibfnamefont {M.}~\bibnamefont
  {Lindner}},\ }\href {\doibase 10.1007/BF01479540} {\bibfield  {journal}
  {\bibinfo  {journal} {Z. Phys.}\ }\textbf {\bibinfo {volume} {C31}},\
  \bibinfo {pages} {295} (\bibinfo {year} {1986})}\BibitemShut {NoStop}%
\bibitem [{\citenamefont {Wetterich}(1987)}]{Wetterich:1987az}%
  \BibitemOpen
  \bibfield  {author} {\bibinfo {author} {\bibfnamefont {C.}~\bibnamefont
  {Wetterich}},\ }in\ \href
  {http://alice.cern.ch/format/showfull?sysnb=0094536} {\emph {\bibinfo
  {booktitle} {{Search for Scalar Particles: Experimental and Theoretical
  Aspects Trieste, Italy, July 23-24, 1987}}}}\ (\bibinfo {year}
  {1987})\BibitemShut {NoStop}%
\bibitem [{\citenamefont {Altarelli}\ and\ \citenamefont
  {Isidori}(1994)}]{Altarelli:1994rb}%
  \BibitemOpen
  \bibfield  {author} {\bibinfo {author} {\bibfnamefont {G.}~\bibnamefont
  {Altarelli}}\ and\ \bibinfo {author} {\bibfnamefont {G.}~\bibnamefont
  {Isidori}},\ }\href {\doibase 10.1016/0370-2693(94)91458-3} {\bibfield
  {journal} {\bibinfo  {journal} {Phys. Lett.}\ }\textbf {\bibinfo {volume}
  {B337}},\ \bibinfo {pages} {141} (\bibinfo {year} {1994})}\BibitemShut
  {NoStop}%
\bibitem [{\citenamefont {Schrempp}\ and\ \citenamefont
  {Wimmer}(1996)}]{Schrempp:1996fb}%
  \BibitemOpen
  \bibfield  {author} {\bibinfo {author} {\bibfnamefont {B.}~\bibnamefont
  {Schrempp}}\ and\ \bibinfo {author} {\bibfnamefont {M.}~\bibnamefont
  {Wimmer}},\ }\href {\doibase 10.1016/0146-6410(96)00059-2} {\bibfield
  {journal} {\bibinfo  {journal} {Prog. Part. Nucl. Phys.}\ }\textbf {\bibinfo
  {volume} {37}},\ \bibinfo {pages} {1} (\bibinfo {year} {1996})},\ \Eprint
  {http://arxiv.org/abs/hep-ph/9606386} {arXiv:hep-ph/9606386 [hep-ph]}
  \BibitemShut {NoStop}%
\bibitem [{\citenamefont {Hambye}\ and\ \citenamefont
  {Riesselmann}(1997)}]{Hambye:1996wb}%
  \BibitemOpen
  \bibfield  {author} {\bibinfo {author} {\bibfnamefont {T.}~\bibnamefont
  {Hambye}}\ and\ \bibinfo {author} {\bibfnamefont {K.}~\bibnamefont
  {Riesselmann}},\ }\href {\doibase 10.1103/PhysRevD.55.7255} {\bibfield
  {journal} {\bibinfo  {journal} {Phys. Rev.}\ }\textbf {\bibinfo {volume}
  {D55}},\ \bibinfo {pages} {7255} (\bibinfo {year} {1997})},\ \Eprint
  {http://arxiv.org/abs/hep-ph/9610272} {arXiv:hep-ph/9610272 [hep-ph]}
  \BibitemShut {NoStop}%
\bibitem [{\citenamefont {Wetterich}(1981)}]{Wetterich:1981ir}%
  \BibitemOpen
  \bibfield  {author} {\bibinfo {author} {\bibfnamefont {C.}~\bibnamefont
  {Wetterich}},\ }\href {\doibase 10.1016/0370-2693(81)90124-6} {\bibfield
  {journal} {\bibinfo  {journal} {Phys. Lett.}\ }\textbf {\bibinfo {volume}
  {B104}},\ \bibinfo {pages} {269} (\bibinfo {year} {1981})}\BibitemShut
  {NoStop}%
\bibitem [{\citenamefont {Bezrukov}\ and\ \citenamefont
  {Shaposhnikov}(2015)}]{Bezrukov:2014ina}%
  \BibitemOpen
  \bibfield  {author} {\bibinfo {author} {\bibfnamefont {F.}~\bibnamefont
  {Bezrukov}}\ and\ \bibinfo {author} {\bibfnamefont {M.}~\bibnamefont
  {Shaposhnikov}},\ }\href {\doibase 10.1134/S1063776115030152} {\bibfield
  {journal} {\bibinfo  {journal} {J. Exp. Theor. Phys.}\ }\textbf {\bibinfo
  {volume} {120}},\ \bibinfo {pages} {335} (\bibinfo {year} {2015})},\ \Eprint
  {http://arxiv.org/abs/1411.1923} {arXiv:1411.1923 [hep-ph]} \BibitemShut
  {NoStop}%
\bibitem [{\citenamefont {Krive}\ and\ \citenamefont
  {Linde}(1976)}]{Krive:1976sg}%
  \BibitemOpen
  \bibfield  {author} {\bibinfo {author} {\bibfnamefont {I.~V.}\ \bibnamefont
  {Krive}}\ and\ \bibinfo {author} {\bibfnamefont {A.~D.}\ \bibnamefont
  {Linde}},\ }\href {\doibase 10.1016/0550-3213(76)90573-3} {\bibfield
  {journal} {\bibinfo  {journal} {Nucl. Phys.}\ }\textbf {\bibinfo {volume}
  {B117}},\ \bibinfo {pages} {265} (\bibinfo {year} {1976})}\BibitemShut
  {NoStop}%
\bibitem [{\citenamefont {Hung}(1979)}]{Hung:1979dn}%
  \BibitemOpen
  \bibfield  {author} {\bibinfo {author} {\bibfnamefont {P.~Q.}\ \bibnamefont
  {Hung}},\ }\href {\doibase 10.1103/PhysRevLett.42.873} {\bibfield  {journal}
  {\bibinfo  {journal} {Phys. Rev. Lett.}\ }\textbf {\bibinfo {volume} {42}},\
  \bibinfo {pages} {873} (\bibinfo {year} {1979})}\BibitemShut {NoStop}%
\bibitem [{\citenamefont {Linde}(1980)}]{Linde:1979ny}%
  \BibitemOpen
  \bibfield  {author} {\bibinfo {author} {\bibfnamefont {A.~D.}\ \bibnamefont
  {Linde}},\ }\href {\doibase 10.1016/0370-2693(80)90318-4} {\bibfield
  {journal} {\bibinfo  {journal} {Phys. Lett.}\ }\textbf {\bibinfo {volume}
  {B92}},\ \bibinfo {pages} {119} (\bibinfo {year} {1980})}\BibitemShut
  {NoStop}%
\bibitem [{\citenamefont {Cabibbo}\ \emph {et~al.}(1979)\citenamefont
  {Cabibbo}, \citenamefont {Maiani}, \citenamefont {Parisi},\ and\
  \citenamefont {Petronzio}}]{Cabibbo:1979ay}%
  \BibitemOpen
  \bibfield  {author} {\bibinfo {author} {\bibfnamefont {N.}~\bibnamefont
  {Cabibbo}}, \bibinfo {author} {\bibfnamefont {L.}~\bibnamefont {Maiani}},
  \bibinfo {author} {\bibfnamefont {G.}~\bibnamefont {Parisi}}, \ and\ \bibinfo
  {author} {\bibfnamefont {R.}~\bibnamefont {Petronzio}},\ }\href {\doibase
  10.1016/0550-3213(79)90167-6} {\bibfield  {journal} {\bibinfo  {journal}
  {Nucl. Phys.}\ }\textbf {\bibinfo {volume} {B158}},\ \bibinfo {pages} {295}
  (\bibinfo {year} {1979})}\BibitemShut {NoStop}%
\bibitem [{\citenamefont {Politzer}\ and\ \citenamefont
  {Wolfram}(1979)}]{Politzer:1978ic}%
  \BibitemOpen
  \bibfield  {author} {\bibinfo {author} {\bibfnamefont {H.~D.}\ \bibnamefont
  {Politzer}}\ and\ \bibinfo {author} {\bibfnamefont {S.}~\bibnamefont
  {Wolfram}},\ }\href {\doibase 10.1016/0370-2693(79)90746-9} {\bibfield
  {journal} {\bibinfo  {journal} {Phys. Lett.}\ }\textbf {\bibinfo {volume}
  {B82}},\ \bibinfo {pages} {242} (\bibinfo {year} {1979})},\ \bibinfo {note}
  {[Erratum: Phys. Lett.83B,421(1979)]}\BibitemShut {NoStop}%
\bibitem [{\citenamefont {Kuti}\ \emph {et~al.}(1988)\citenamefont {Kuti},
  \citenamefont {Lin},\ and\ \citenamefont {Shen}}]{Kuti:1987nr}%
  \BibitemOpen
  \bibfield  {author} {\bibinfo {author} {\bibfnamefont {J.}~\bibnamefont
  {Kuti}}, \bibinfo {author} {\bibfnamefont {L.}~\bibnamefont {Lin}}, \ and\
  \bibinfo {author} {\bibfnamefont {Y.}~\bibnamefont {Shen}},\ }\href {\doibase
  10.1103/PhysRevLett.61.678} {\bibfield  {journal} {\bibinfo  {journal} {Phys.
  Rev. Lett.}\ }\textbf {\bibinfo {volume} {61}},\ \bibinfo {pages} {678}
  (\bibinfo {year} {1988})}\BibitemShut {NoStop}%
\bibitem [{\citenamefont {Sher}(1989)}]{Sher:1988mj}%
  \BibitemOpen
  \bibfield  {author} {\bibinfo {author} {\bibfnamefont {M.}~\bibnamefont
  {Sher}},\ }\href {\doibase 10.1016/0370-1573(89)90061-6} {\bibfield
  {journal} {\bibinfo  {journal} {Phys. Rept.}\ }\textbf {\bibinfo {volume}
  {179}},\ \bibinfo {pages} {273} (\bibinfo {year} {1989})}\BibitemShut
  {NoStop}%
\bibitem [{\citenamefont {Hasenfratz}\ \emph {et~al.}(1987)\citenamefont
  {Hasenfratz}, \citenamefont {Jansen}, \citenamefont {Lang}, \citenamefont
  {Neuhaus},\ and\ \citenamefont {Yoneyama}}]{Hasenfratz:1987eh}%
  \BibitemOpen
  \bibfield  {author} {\bibinfo {author} {\bibfnamefont {A.}~\bibnamefont
  {Hasenfratz}}, \bibinfo {author} {\bibfnamefont {K.}~\bibnamefont {Jansen}},
  \bibinfo {author} {\bibfnamefont {C.~B.}\ \bibnamefont {Lang}}, \bibinfo
  {author} {\bibfnamefont {T.}~\bibnamefont {Neuhaus}}, \ and\ \bibinfo
  {author} {\bibfnamefont {H.}~\bibnamefont {Yoneyama}},\ }\href {\doibase
  10.1016/0370-2693(87)91622-4} {\bibfield  {journal} {\bibinfo  {journal}
  {Phys. Lett.}\ }\textbf {\bibinfo {volume} {B199}},\ \bibinfo {pages} {531}
  (\bibinfo {year} {1987})}\BibitemShut {NoStop}%
\bibitem [{\citenamefont {Luscher}\ and\ \citenamefont
  {Weisz}(1989)}]{Luscher:1988uq}%
  \BibitemOpen
  \bibfield  {author} {\bibinfo {author} {\bibfnamefont {M.}~\bibnamefont
  {Luscher}}\ and\ \bibinfo {author} {\bibfnamefont {P.}~\bibnamefont
  {Weisz}},\ }\href {\doibase 10.1016/0550-3213(89)90637-8} {\bibfield
  {journal} {\bibinfo  {journal} {Nucl. Phys.}\ }\textbf {\bibinfo {volume}
  {B318}},\ \bibinfo {pages} {705} (\bibinfo {year} {1989})}\BibitemShut
  {NoStop}%
\bibitem [{\citenamefont {Lindner}\ \emph {et~al.}(1989)\citenamefont
  {Lindner}, \citenamefont {Sher},\ and\ \citenamefont
  {Zaglauer}}]{Lindner:1988ww}%
  \BibitemOpen
  \bibfield  {author} {\bibinfo {author} {\bibfnamefont {M.}~\bibnamefont
  {Lindner}}, \bibinfo {author} {\bibfnamefont {M.}~\bibnamefont {Sher}}, \
  and\ \bibinfo {author} {\bibfnamefont {H.~W.}\ \bibnamefont {Zaglauer}},\
  }\href {\doibase 10.1016/0370-2693(89)90540-6} {\bibfield  {journal}
  {\bibinfo  {journal} {Phys. Lett.}\ }\textbf {\bibinfo {volume} {B228}},\
  \bibinfo {pages} {139} (\bibinfo {year} {1989})}\BibitemShut {NoStop}%
\bibitem [{\citenamefont {Ford}\ \emph {et~al.}(1993)\citenamefont {Ford},
  \citenamefont {Jones}, \citenamefont {Stephenson},\ and\ \citenamefont
  {Einhorn}}]{Ford:1992mv}%
  \BibitemOpen
  \bibfield  {author} {\bibinfo {author} {\bibfnamefont {C.}~\bibnamefont
  {Ford}}, \bibinfo {author} {\bibfnamefont {D.~R.~T.}\ \bibnamefont {Jones}},
  \bibinfo {author} {\bibfnamefont {P.~W.}\ \bibnamefont {Stephenson}}, \ and\
  \bibinfo {author} {\bibfnamefont {M.~B.}\ \bibnamefont {Einhorn}},\ }\href
  {\doibase 10.1016/0550-3213(93)90206-5} {\bibfield  {journal} {\bibinfo
  {journal} {Nucl. Phys.}\ }\textbf {\bibinfo {volume} {B395}},\ \bibinfo
  {pages} {17} (\bibinfo {year} {1993})},\ \Eprint
  {http://arxiv.org/abs/hep-lat/9210033} {arXiv:hep-lat/9210033 [hep-lat]}
  \BibitemShut {NoStop}%
\bibitem [{\citenamefont {Heller}\ \emph {et~al.}(1993)\citenamefont {Heller},
  \citenamefont {Klomfass}, \citenamefont {Neuberger},\ and\ \citenamefont
  {Vranas}}]{Heller:1993yv}%
  \BibitemOpen
  \bibfield  {author} {\bibinfo {author} {\bibfnamefont {U.~M.}\ \bibnamefont
  {Heller}}, \bibinfo {author} {\bibfnamefont {M.}~\bibnamefont {Klomfass}},
  \bibinfo {author} {\bibfnamefont {H.}~\bibnamefont {Neuberger}}, \ and\
  \bibinfo {author} {\bibfnamefont {P.~M.}\ \bibnamefont {Vranas}},\ }\href
  {\doibase 10.1016/0550-3213(93)90559-8} {\bibfield  {journal} {\bibinfo
  {journal} {Nucl. Phys.}\ }\textbf {\bibinfo {volume} {B405}},\ \bibinfo
  {pages} {555} (\bibinfo {year} {1993})},\ \Eprint
  {http://arxiv.org/abs/hep-ph/9303215} {arXiv:hep-ph/9303215 [hep-ph]}
  \BibitemShut {NoStop}%
\bibitem [{\citenamefont {Arnold}(1989)}]{Arnold:1989cb}%
  \BibitemOpen
  \bibfield  {author} {\bibinfo {author} {\bibfnamefont {P.~B.}\ \bibnamefont
  {Arnold}},\ }\href {\doibase 10.1103/PhysRevD.40.613} {\bibfield  {journal}
  {\bibinfo  {journal} {Phys. Rev.}\ }\textbf {\bibinfo {volume} {D40}},\
  \bibinfo {pages} {613} (\bibinfo {year} {1989})}\BibitemShut {NoStop}%
\bibitem [{\citenamefont {Sher}(1993)}]{Sher:1993mf}%
  \BibitemOpen
  \bibfield  {author} {\bibinfo {author} {\bibfnamefont {M.}~\bibnamefont
  {Sher}},\ }\href {\doibase 10.1016/0370-2693(93)91586-C} {\bibfield
  {journal} {\bibinfo  {journal} {Phys. Lett.}\ }\textbf {\bibinfo {volume}
  {B317}},\ \bibinfo {pages} {159} (\bibinfo {year} {1993})},\ \bibinfo {note}
  {[Addendum: Phys. Lett.B331,448(1994)]},\ \Eprint
  {http://arxiv.org/abs/hep-ph/9307342} {arXiv:hep-ph/9307342 [hep-ph]}
  \BibitemShut {NoStop}%
\bibitem [{\citenamefont {Casas}\ \emph {et~al.}(1995)\citenamefont {Casas},
  \citenamefont {Espinosa},\ and\ \citenamefont {Quiros}}]{Casas:1994qy}%
  \BibitemOpen
  \bibfield  {author} {\bibinfo {author} {\bibfnamefont {J.~A.}\ \bibnamefont
  {Casas}}, \bibinfo {author} {\bibfnamefont {J.~R.}\ \bibnamefont {Espinosa}},
  \ and\ \bibinfo {author} {\bibfnamefont {M.}~\bibnamefont {Quiros}},\ }\href
  {\doibase 10.1016/0370-2693(94)01404-Z} {\bibfield  {journal} {\bibinfo
  {journal} {Phys. Lett.}\ }\textbf {\bibinfo {volume} {B342}},\ \bibinfo
  {pages} {171} (\bibinfo {year} {1995})},\ \Eprint
  {http://arxiv.org/abs/hep-ph/9409458} {arXiv:hep-ph/9409458 [hep-ph]}
  \BibitemShut {NoStop}%
\bibitem [{\citenamefont {Espinosa}\ and\ \citenamefont
  {Quiros}(1995)}]{Espinosa:1995se}%
  \BibitemOpen
  \bibfield  {author} {\bibinfo {author} {\bibfnamefont {J.~R.}\ \bibnamefont
  {Espinosa}}\ and\ \bibinfo {author} {\bibfnamefont {M.}~\bibnamefont
  {Quiros}},\ }\href {\doibase 10.1016/0370-2693(95)00572-3} {\bibfield
  {journal} {\bibinfo  {journal} {Phys. Lett.}\ }\textbf {\bibinfo {volume}
  {B353}},\ \bibinfo {pages} {257} (\bibinfo {year} {1995})},\ \Eprint
  {http://arxiv.org/abs/hep-ph/9504241} {arXiv:hep-ph/9504241 [hep-ph]}
  \BibitemShut {NoStop}%
\bibitem [{\citenamefont {Bergerhoff}\ \emph {et~al.}(1999)\citenamefont
  {Bergerhoff}, \citenamefont {Lindner},\ and\ \citenamefont
  {Weiser}}]{Bergerhoff:1999jj}%
  \BibitemOpen
  \bibfield  {author} {\bibinfo {author} {\bibfnamefont {B.}~\bibnamefont
  {Bergerhoff}}, \bibinfo {author} {\bibfnamefont {M.}~\bibnamefont {Lindner}},
  \ and\ \bibinfo {author} {\bibfnamefont {M.}~\bibnamefont {Weiser}},\ }\href
  {\doibase 10.1016/S0370-2693(99)01273-3} {\bibfield  {journal} {\bibinfo
  {journal} {Phys. Lett.}\ }\textbf {\bibinfo {volume} {B469}},\ \bibinfo
  {pages} {61} (\bibinfo {year} {1999})},\ \Eprint
  {http://arxiv.org/abs/hep-ph/9909261} {arXiv:hep-ph/9909261 [hep-ph]}
  \BibitemShut {NoStop}%
\bibitem [{\citenamefont {Isidori}\ \emph {et~al.}(2001)\citenamefont
  {Isidori}, \citenamefont {Ridolfi},\ and\ \citenamefont
  {Strumia}}]{Isidori:2001bm}%
  \BibitemOpen
  \bibfield  {author} {\bibinfo {author} {\bibfnamefont {G.}~\bibnamefont
  {Isidori}}, \bibinfo {author} {\bibfnamefont {G.}~\bibnamefont {Ridolfi}}, \
  and\ \bibinfo {author} {\bibfnamefont {A.}~\bibnamefont {Strumia}},\ }\href
  {\doibase 10.1016/S0550-3213(01)00302-9} {\bibfield  {journal} {\bibinfo
  {journal} {Nucl. Phys.}\ }\textbf {\bibinfo {volume} {B609}},\ \bibinfo
  {pages} {387} (\bibinfo {year} {2001})},\ \Eprint
  {http://arxiv.org/abs/hep-ph/0104016} {arXiv:hep-ph/0104016 [hep-ph]}
  \BibitemShut {NoStop}%
\bibitem [{\citenamefont {Ellis}\ \emph {et~al.}(2009)\citenamefont {Ellis},
  \citenamefont {Espinosa}, \citenamefont {Giudice}, \citenamefont {Hoecker},\
  and\ \citenamefont {Riotto}}]{Ellis:2009tp}%
  \BibitemOpen
  \bibfield  {author} {\bibinfo {author} {\bibfnamefont {J.}~\bibnamefont
  {Ellis}}, \bibinfo {author} {\bibfnamefont {J.~R.}\ \bibnamefont {Espinosa}},
  \bibinfo {author} {\bibfnamefont {G.~F.}\ \bibnamefont {Giudice}}, \bibinfo
  {author} {\bibfnamefont {A.}~\bibnamefont {Hoecker}}, \ and\ \bibinfo
  {author} {\bibfnamefont {A.}~\bibnamefont {Riotto}},\ }\href {\doibase
  10.1016/j.physletb.2009.07.054} {\bibfield  {journal} {\bibinfo  {journal}
  {Phys. Lett.}\ }\textbf {\bibinfo {volume} {B679}},\ \bibinfo {pages} {369}
  (\bibinfo {year} {2009})},\ \Eprint {http://arxiv.org/abs/0906.0954}
  {arXiv:0906.0954 [hep-ph]} \BibitemShut {NoStop}%
\bibitem [{\citenamefont {Holthausen}\ \emph {et~al.}(2012)\citenamefont
  {Holthausen}, \citenamefont {Lim},\ and\ \citenamefont
  {Lindner}}]{Holthausen:2011aa}%
  \BibitemOpen
  \bibfield  {author} {\bibinfo {author} {\bibfnamefont {M.}~\bibnamefont
  {Holthausen}}, \bibinfo {author} {\bibfnamefont {K.~S.}\ \bibnamefont {Lim}},
  \ and\ \bibinfo {author} {\bibfnamefont {M.}~\bibnamefont {Lindner}},\ }\href
  {\doibase 10.1007/JHEP02(2012)037} {\bibfield  {journal} {\bibinfo  {journal}
  {JHEP}\ }\textbf {\bibinfo {volume} {02}},\ \bibinfo {pages} {037} (\bibinfo
  {year} {2012})},\ \Eprint {http://arxiv.org/abs/1112.2415} {arXiv:1112.2415
  [hep-ph]} \BibitemShut {NoStop}%
\bibitem [{\citenamefont {Elias-Miro}\ \emph {et~al.}(2012)\citenamefont
  {Elias-Miro}, \citenamefont {Espinosa}, \citenamefont {Giudice},
  \citenamefont {Isidori}, \citenamefont {Riotto},\ and\ \citenamefont
  {Strumia}}]{EliasMiro:2011aa}%
  \BibitemOpen
  \bibfield  {author} {\bibinfo {author} {\bibfnamefont {J.}~\bibnamefont
  {Elias-Miro}}, \bibinfo {author} {\bibfnamefont {J.~R.}\ \bibnamefont
  {Espinosa}}, \bibinfo {author} {\bibfnamefont {G.~F.}\ \bibnamefont
  {Giudice}}, \bibinfo {author} {\bibfnamefont {G.}~\bibnamefont {Isidori}},
  \bibinfo {author} {\bibfnamefont {A.}~\bibnamefont {Riotto}}, \ and\ \bibinfo
  {author} {\bibfnamefont {A.}~\bibnamefont {Strumia}},\ }\href {\doibase
  10.1016/j.physletb.2012.02.013} {\bibfield  {journal} {\bibinfo  {journal}
  {Phys. Lett.}\ }\textbf {\bibinfo {volume} {B709}},\ \bibinfo {pages} {222}
  (\bibinfo {year} {2012})},\ \Eprint {http://arxiv.org/abs/1112.3022}
  {arXiv:1112.3022 [hep-ph]} \BibitemShut {NoStop}%
\bibitem [{\citenamefont {Degrassi}\ \emph {et~al.}(2012)\citenamefont
  {Degrassi}, \citenamefont {Di~Vita}, \citenamefont {Elias-Miro},
  \citenamefont {Espinosa}, \citenamefont {Giudice}, \citenamefont {Isidori},\
  and\ \citenamefont {Strumia}}]{Degrassi:2012ry}%
  \BibitemOpen
  \bibfield  {author} {\bibinfo {author} {\bibfnamefont {G.}~\bibnamefont
  {Degrassi}}, \bibinfo {author} {\bibfnamefont {S.}~\bibnamefont {Di~Vita}},
  \bibinfo {author} {\bibfnamefont {J.}~\bibnamefont {Elias-Miro}}, \bibinfo
  {author} {\bibfnamefont {J.~R.}\ \bibnamefont {Espinosa}}, \bibinfo {author}
  {\bibfnamefont {G.~F.}\ \bibnamefont {Giudice}}, \bibinfo {author}
  {\bibfnamefont {G.}~\bibnamefont {Isidori}}, \ and\ \bibinfo {author}
  {\bibfnamefont {A.}~\bibnamefont {Strumia}},\ }\href {\doibase
  10.1007/JHEP08(2012)098} {\bibfield  {journal} {\bibinfo  {journal} {JHEP}\
  }\textbf {\bibinfo {volume} {08}},\ \bibinfo {pages} {098} (\bibinfo {year}
  {2012})},\ \Eprint {http://arxiv.org/abs/1205.6497} {arXiv:1205.6497
  [hep-ph]} \BibitemShut {NoStop}%
\bibitem [{\citenamefont {Alekhin}\ \emph {et~al.}(2012)\citenamefont
  {Alekhin}, \citenamefont {Djouadi},\ and\ \citenamefont
  {Moch}}]{Alekhin:2012py}%
  \BibitemOpen
  \bibfield  {author} {\bibinfo {author} {\bibfnamefont {S.}~\bibnamefont
  {Alekhin}}, \bibinfo {author} {\bibfnamefont {A.}~\bibnamefont {Djouadi}}, \
  and\ \bibinfo {author} {\bibfnamefont {S.}~\bibnamefont {Moch}},\ }\href
  {\doibase 10.1016/j.physletb.2012.08.024} {\bibfield  {journal} {\bibinfo
  {journal} {Phys. Lett.}\ }\textbf {\bibinfo {volume} {B716}},\ \bibinfo
  {pages} {214} (\bibinfo {year} {2012})},\ \Eprint
  {http://arxiv.org/abs/1207.0980} {arXiv:1207.0980 [hep-ph]} \BibitemShut
  {NoStop}%
\bibitem [{\citenamefont {Masina}(2013)}]{Masina:2012tz}%
  \BibitemOpen
  \bibfield  {author} {\bibinfo {author} {\bibfnamefont {I.}~\bibnamefont
  {Masina}},\ }\href {\doibase 10.1103/PhysRevD.87.053001} {\bibfield
  {journal} {\bibinfo  {journal} {Phys. Rev.}\ }\textbf {\bibinfo {volume}
  {D87}},\ \bibinfo {pages} {053001} (\bibinfo {year} {2013})},\ \Eprint
  {http://arxiv.org/abs/1209.0393} {arXiv:1209.0393 [hep-ph]} \BibitemShut
  {NoStop}%
\bibitem [{\citenamefont {Buttazzo}\ \emph {et~al.}(2013)\citenamefont
  {Buttazzo}, \citenamefont {Degrassi}, \citenamefont {Giardino}, \citenamefont
  {Giudice}, \citenamefont {Sala}, \citenamefont {Salvio},\ and\ \citenamefont
  {Strumia}}]{Buttazzo:2013uya}%
  \BibitemOpen
  \bibfield  {author} {\bibinfo {author} {\bibfnamefont {D.}~\bibnamefont
  {Buttazzo}}, \bibinfo {author} {\bibfnamefont {G.}~\bibnamefont {Degrassi}},
  \bibinfo {author} {\bibfnamefont {P.~P.}\ \bibnamefont {Giardino}}, \bibinfo
  {author} {\bibfnamefont {G.~F.}\ \bibnamefont {Giudice}}, \bibinfo {author}
  {\bibfnamefont {F.}~\bibnamefont {Sala}}, \bibinfo {author} {\bibfnamefont
  {A.}~\bibnamefont {Salvio}}, \ and\ \bibinfo {author} {\bibfnamefont
  {A.}~\bibnamefont {Strumia}},\ }\href {\doibase 10.1007/JHEP12(2013)089}
  {\bibfield  {journal} {\bibinfo  {journal} {JHEP}\ }\textbf {\bibinfo
  {volume} {12}},\ \bibinfo {pages} {089} (\bibinfo {year} {2013})},\ \Eprint
  {http://arxiv.org/abs/1307.3536} {arXiv:1307.3536 [hep-ph]} \BibitemShut
  {NoStop}%
\bibitem [{\citenamefont {Gabrielli}\ \emph {et~al.}(2014)\citenamefont
  {Gabrielli}, \citenamefont {Heikinheimo}, \citenamefont {Kannike},
  \citenamefont {Racioppi}, \citenamefont {Raidal},\ and\ \citenamefont
  {Spethmann}}]{Gabrielli:2013hma}%
  \BibitemOpen
  \bibfield  {author} {\bibinfo {author} {\bibfnamefont {E.}~\bibnamefont
  {Gabrielli}}, \bibinfo {author} {\bibfnamefont {M.}~\bibnamefont
  {Heikinheimo}}, \bibinfo {author} {\bibfnamefont {K.}~\bibnamefont
  {Kannike}}, \bibinfo {author} {\bibfnamefont {A.}~\bibnamefont {Racioppi}},
  \bibinfo {author} {\bibfnamefont {M.}~\bibnamefont {Raidal}}, \ and\ \bibinfo
  {author} {\bibfnamefont {C.}~\bibnamefont {Spethmann}},\ }\href {\doibase
  10.1103/PhysRevD.89.015017} {\bibfield  {journal} {\bibinfo  {journal} {Phys.
  Rev.}\ }\textbf {\bibinfo {volume} {D89}},\ \bibinfo {pages} {015017}
  (\bibinfo {year} {2014})},\ \Eprint {http://arxiv.org/abs/1309.6632}
  {arXiv:1309.6632 [hep-ph]} \BibitemShut {NoStop}%
\bibitem [{\citenamefont {Bednyakov}\ \emph {et~al.}(2015)\citenamefont
  {Bednyakov}, \citenamefont {Kniehl}, \citenamefont {Pikelner},\ and\
  \citenamefont {Veretin}}]{Bednyakov:2015sca}%
  \BibitemOpen
  \bibfield  {author} {\bibinfo {author} {\bibfnamefont {A.~V.}\ \bibnamefont
  {Bednyakov}}, \bibinfo {author} {\bibfnamefont {B.~A.}\ \bibnamefont
  {Kniehl}}, \bibinfo {author} {\bibfnamefont {A.~F.}\ \bibnamefont
  {Pikelner}}, \ and\ \bibinfo {author} {\bibfnamefont {O.~L.}\ \bibnamefont
  {Veretin}},\ }\href {\doibase 10.1103/PhysRevLett.115.201802} {\bibfield
  {journal} {\bibinfo  {journal} {Phys. Rev. Lett.}\ }\textbf {\bibinfo
  {volume} {115}},\ \bibinfo {pages} {201802} (\bibinfo {year} {2015})},\
  \Eprint {http://arxiv.org/abs/1507.08833} {arXiv:1507.08833 [hep-ph]}
  \BibitemShut {NoStop}%
\bibitem [{\citenamefont {Gies}\ \emph {et~al.}(2014)\citenamefont {Gies},
  \citenamefont {Gneiting},\ and\ \citenamefont {Sondenheimer}}]{Gies:2013fua}%
  \BibitemOpen
  \bibfield  {author} {\bibinfo {author} {\bibfnamefont {H.}~\bibnamefont
  {Gies}}, \bibinfo {author} {\bibfnamefont {C.}~\bibnamefont {Gneiting}}, \
  and\ \bibinfo {author} {\bibfnamefont {R.}~\bibnamefont {Sondenheimer}},\
  }\href {\doibase 10.1103/PhysRevD.89.045012} {\bibfield  {journal} {\bibinfo
  {journal} {Phys. Rev.}\ }\textbf {\bibinfo {volume} {D89}},\ \bibinfo {pages}
  {045012} (\bibinfo {year} {2014})},\ \Eprint {http://arxiv.org/abs/1308.5075}
  {arXiv:1308.5075 [hep-ph]} \BibitemShut {NoStop}%
\bibitem [{\citenamefont {Gies}\ and\ \citenamefont
  {Sondenheimer}(2015)}]{Gies:2014xha}%
  \BibitemOpen
  \bibfield  {author} {\bibinfo {author} {\bibfnamefont {H.}~\bibnamefont
  {Gies}}\ and\ \bibinfo {author} {\bibfnamefont {R.}~\bibnamefont
  {Sondenheimer}},\ }\href {\doibase 10.1140/epjc/s10052-015-3284-1} {\bibfield
   {journal} {\bibinfo  {journal} {Eur. Phys. J.}\ }\textbf {\bibinfo {volume}
  {C75}},\ \bibinfo {pages} {68} (\bibinfo {year} {2015})},\ \Eprint
  {http://arxiv.org/abs/1407.8124} {arXiv:1407.8124 [hep-ph]} \BibitemShut
  {NoStop}%
\bibitem [{\citenamefont {Eichhorn}\ \emph {et~al.}(2015)\citenamefont
  {Eichhorn}, \citenamefont {Gies}, \citenamefont {Jaeckel}, \citenamefont
  {Plehn}, \citenamefont {Scherer},\ and\ \citenamefont
  {Sondenheimer}}]{Eichhorn:2015kea}%
  \BibitemOpen
  \bibfield  {author} {\bibinfo {author} {\bibfnamefont {A.}~\bibnamefont
  {Eichhorn}}, \bibinfo {author} {\bibfnamefont {H.}~\bibnamefont {Gies}},
  \bibinfo {author} {\bibfnamefont {J.}~\bibnamefont {Jaeckel}}, \bibinfo
  {author} {\bibfnamefont {T.}~\bibnamefont {Plehn}}, \bibinfo {author}
  {\bibfnamefont {M.~M.}\ \bibnamefont {Scherer}}, \ and\ \bibinfo {author}
  {\bibfnamefont {R.}~\bibnamefont {Sondenheimer}},\ }\href {\doibase
  10.1007/JHEP04(2015)022} {\bibfield  {journal} {\bibinfo  {journal} {JHEP}\
  }\textbf {\bibinfo {volume} {04}},\ \bibinfo {pages} {022} (\bibinfo {year}
  {2015})},\ \Eprint {http://arxiv.org/abs/1501.02812} {arXiv:1501.02812
  [hep-ph]} \BibitemShut {NoStop}%
\bibitem [{\citenamefont {Hegde}\ \emph {et~al.}(2014)\citenamefont {Hegde},
  \citenamefont {Jansen}, \citenamefont {Lin},\ and\ \citenamefont
  {Nagy}}]{Hegde:2013mks}%
  \BibitemOpen
  \bibfield  {author} {\bibinfo {author} {\bibfnamefont {P.}~\bibnamefont
  {Hegde}}, \bibinfo {author} {\bibfnamefont {K.}~\bibnamefont {Jansen}},
  \bibinfo {author} {\bibfnamefont {C.~J.~D.}\ \bibnamefont {Lin}}, \ and\
  \bibinfo {author} {\bibfnamefont {A.}~\bibnamefont {Nagy}},\ }\bibfield
  {booktitle} {\emph {\bibinfo {booktitle} {{Proceedings, 31st International
  Symposium on Lattice Field Theory (Lattice 2013)}}},\ }\href@noop {}
  {\bibfield  {journal} {\bibinfo  {journal} {PoS}\ }\textbf {\bibinfo {volume}
  {LATTICE2013}},\ \bibinfo {pages} {058} (\bibinfo {year} {2014})},\ \Eprint
  {http://arxiv.org/abs/1310.6260} {arXiv:1310.6260 [hep-lat]} \BibitemShut
  {NoStop}%
\bibitem [{\citenamefont {Chu}\ \emph {et~al.}(2015)\citenamefont {Chu},
  \citenamefont {Jansen}, \citenamefont {Knippschild}, \citenamefont {Lin},\
  and\ \citenamefont {Nagy}}]{Chu:2015nha}%
  \BibitemOpen
  \bibfield  {author} {\bibinfo {author} {\bibfnamefont {D.~Y.~J.}\
  \bibnamefont {Chu}}, \bibinfo {author} {\bibfnamefont {K.}~\bibnamefont
  {Jansen}}, \bibinfo {author} {\bibfnamefont {B.}~\bibnamefont {Knippschild}},
  \bibinfo {author} {\bibfnamefont {C.~J.~D.}\ \bibnamefont {Lin}}, \ and\
  \bibinfo {author} {\bibfnamefont {A.}~\bibnamefont {Nagy}},\ }\href {\doibase
  10.1016/j.physletb.2015.03.050} {\bibfield  {journal} {\bibinfo  {journal}
  {Phys. Lett.}\ }\textbf {\bibinfo {volume} {B744}},\ \bibinfo {pages} {146}
  (\bibinfo {year} {2015})},\ \Eprint {http://arxiv.org/abs/1501.05440}
  {arXiv:1501.05440 [hep-lat]} \BibitemShut {NoStop}%
\bibitem [{\citenamefont {Chu}\ \emph {et~al.}(2014)\citenamefont {Chu},
  \citenamefont {Jansen}, \citenamefont {Knippschild}, \citenamefont {Lin},
  \citenamefont {Nagai},\ and\ \citenamefont {Nagy}}]{Chu:2015ula}%
  \BibitemOpen
  \bibfield  {author} {\bibinfo {author} {\bibfnamefont {D.~Y.~J.}\
  \bibnamefont {Chu}}, \bibinfo {author} {\bibfnamefont {K.}~\bibnamefont
  {Jansen}}, \bibinfo {author} {\bibfnamefont {B.}~\bibnamefont {Knippschild}},
  \bibinfo {author} {\bibfnamefont {C.~J.~D.}\ \bibnamefont {Lin}}, \bibinfo
  {author} {\bibfnamefont {K.-I.}\ \bibnamefont {Nagai}}, \ and\ \bibinfo
  {author} {\bibfnamefont {A.}~\bibnamefont {Nagy}},\ }\bibfield  {booktitle}
  {\emph {\bibinfo {booktitle} {{Proceedings, 32nd International Symposium on
  Lattice Field Theory (Lattice 2014)}}},\ }\href@noop {} {\bibfield  {journal}
  {\bibinfo  {journal} {PoS}\ }\textbf {\bibinfo {volume} {LATTICE2014}},\
  \bibinfo {pages} {278} (\bibinfo {year} {2014})},\ \Eprint
  {http://arxiv.org/abs/1501.00306} {arXiv:1501.00306 [hep-lat]} \BibitemShut
  {NoStop}%
\bibitem [{\citenamefont {Akerlund}\ and\ \citenamefont
  {de~Forcrand}(2015)}]{Akerlund:2015fya}%
  \BibitemOpen
  \bibfield  {author} {\bibinfo {author} {\bibfnamefont {O.}~\bibnamefont
  {Akerlund}}\ and\ \bibinfo {author} {\bibfnamefont {P.}~\bibnamefont
  {de~Forcrand}},\ }\href@noop {} {\  (\bibinfo {year} {2015})},\ \Eprint
  {http://arxiv.org/abs/1508.07959} {arXiv:1508.07959 [hep-lat]} \BibitemShut
  {NoStop}%
\bibitem [{\citenamefont {Eichhorn}\ and\ \citenamefont
  {Scherer}(2014)}]{Eichhorn:2014qka}%
  \BibitemOpen
  \bibfield  {author} {\bibinfo {author} {\bibfnamefont {A.}~\bibnamefont
  {Eichhorn}}\ and\ \bibinfo {author} {\bibfnamefont {M.~M.}\ \bibnamefont
  {Scherer}},\ }\href {\doibase 10.1103/PhysRevD.90.025023} {\bibfield
  {journal} {\bibinfo  {journal} {Phys. Rev.}\ }\textbf {\bibinfo {volume}
  {D90}},\ \bibinfo {pages} {025023} (\bibinfo {year} {2014})},\ \Eprint
  {http://arxiv.org/abs/1404.5962} {arXiv:1404.5962 [hep-ph]} \BibitemShut
  {NoStop}%
\bibitem [{\citenamefont {Jakovac}\ \emph
  {et~al.}(2015{\natexlab{a}})\citenamefont {Jakovac}, \citenamefont
  {Kaposvari},\ and\ \citenamefont {Patkos}}]{Jakovac:2015kka}%
  \BibitemOpen
  \bibfield  {author} {\bibinfo {author} {\bibfnamefont {A.}~\bibnamefont
  {Jakovac}}, \bibinfo {author} {\bibfnamefont {I.}~\bibnamefont {Kaposvari}},
  \ and\ \bibinfo {author} {\bibfnamefont {A.}~\bibnamefont {Patkos}},\
  }\href@noop {} {\  (\bibinfo {year} {2015}{\natexlab{a}})},\ \Eprint
  {http://arxiv.org/abs/1508.06774} {arXiv:1508.06774 [hep-th]} \BibitemShut
  {NoStop}%
\bibitem [{\citenamefont {Jakovac}\ \emph
  {et~al.}(2015{\natexlab{b}})\citenamefont {Jakovac}, \citenamefont
  {Kaposvari},\ and\ \citenamefont {Patkos}}]{Jakovac:2015iqa}%
  \BibitemOpen
  \bibfield  {author} {\bibinfo {author} {\bibfnamefont {A.}~\bibnamefont
  {Jakovac}}, \bibinfo {author} {\bibfnamefont {I.}~\bibnamefont {Kaposvari}},
  \ and\ \bibinfo {author} {\bibfnamefont {A.}~\bibnamefont {Patkos}},\
  }\href@noop {} {\  (\bibinfo {year} {2015}{\natexlab{b}})},\ \Eprint
  {http://arxiv.org/abs/1510.05782} {arXiv:1510.05782 [hep-th]} \BibitemShut
  {NoStop}%
\bibitem [{\citenamefont {Datta}\ \emph {et~al.}(1996)\citenamefont {Datta},
  \citenamefont {Young},\ and\ \citenamefont {Zhang}}]{Datta:1996ni}%
  \BibitemOpen
  \bibfield  {author} {\bibinfo {author} {\bibfnamefont {A.}~\bibnamefont
  {Datta}}, \bibinfo {author} {\bibfnamefont {B.~L.}\ \bibnamefont {Young}}, \
  and\ \bibinfo {author} {\bibfnamefont {X.}~\bibnamefont {Zhang}},\ }\href
  {\doibase 10.1016/0370-2693(96)00830-1} {\bibfield  {journal} {\bibinfo
  {journal} {Phys. Lett.}\ }\textbf {\bibinfo {volume} {B385}},\ \bibinfo
  {pages} {225} (\bibinfo {year} {1996})},\ \Eprint
  {http://arxiv.org/abs/hep-ph/9604312} {arXiv:hep-ph/9604312 [hep-ph]}
  \BibitemShut {NoStop}%
\bibitem [{\citenamefont {Barbieri}\ and\ \citenamefont
  {Strumia}(1999)}]{Barbieri:1999tm}%
  \BibitemOpen
  \bibfield  {author} {\bibinfo {author} {\bibfnamefont {R.}~\bibnamefont
  {Barbieri}}\ and\ \bibinfo {author} {\bibfnamefont {A.}~\bibnamefont
  {Strumia}},\ }\href {\doibase 10.1016/S0370-2693(99)00882-5} {\bibfield
  {journal} {\bibinfo  {journal} {Phys. Lett.}\ }\textbf {\bibinfo {volume}
  {B462}},\ \bibinfo {pages} {144} (\bibinfo {year} {1999})},\ \Eprint
  {http://arxiv.org/abs/hep-ph/9905281} {arXiv:hep-ph/9905281 [hep-ph]}
  \BibitemShut {NoStop}%
\bibitem [{\citenamefont {Grzadkowski}\ and\ \citenamefont
  {Wudka}(2002)}]{Grzadkowski:2001vb}%
  \BibitemOpen
  \bibfield  {author} {\bibinfo {author} {\bibfnamefont {B.}~\bibnamefont
  {Grzadkowski}}\ and\ \bibinfo {author} {\bibfnamefont {J.}~\bibnamefont
  {Wudka}},\ }\href {\doibase 10.1103/PhysRevLett.88.041802} {\bibfield
  {journal} {\bibinfo  {journal} {Phys. Rev. Lett.}\ }\textbf {\bibinfo
  {volume} {88}},\ \bibinfo {pages} {041802} (\bibinfo {year} {2002})},\
  \Eprint {http://arxiv.org/abs/hep-ph/0106233} {arXiv:hep-ph/0106233 [hep-ph]}
  \BibitemShut {NoStop}%
\bibitem [{\citenamefont {Burgess}\ \emph {et~al.}(2002)\citenamefont
  {Burgess}, \citenamefont {Di~Clemente},\ and\ \citenamefont
  {Espinosa}}]{Burgess:2001tj}%
  \BibitemOpen
  \bibfield  {author} {\bibinfo {author} {\bibfnamefont {C.~P.}\ \bibnamefont
  {Burgess}}, \bibinfo {author} {\bibfnamefont {V.}~\bibnamefont
  {Di~Clemente}}, \ and\ \bibinfo {author} {\bibfnamefont {J.~R.}\ \bibnamefont
  {Espinosa}},\ }\href {\doibase 10.1088/1126-6708/2002/01/041} {\bibfield
  {journal} {\bibinfo  {journal} {JHEP}\ }\textbf {\bibinfo {volume} {01}},\
  \bibinfo {pages} {041} (\bibinfo {year} {2002})},\ \Eprint
  {http://arxiv.org/abs/hep-ph/0201160} {arXiv:hep-ph/0201160 [hep-ph]}
  \BibitemShut {NoStop}%
\bibitem [{\citenamefont {Barger}\ \emph {et~al.}(2003)\citenamefont {Barger},
  \citenamefont {Han}, \citenamefont {Langacker}, \citenamefont {McElrath},\
  and\ \citenamefont {Zerwas}}]{Barger:2003rs}%
  \BibitemOpen
  \bibfield  {author} {\bibinfo {author} {\bibfnamefont {V.}~\bibnamefont
  {Barger}}, \bibinfo {author} {\bibfnamefont {T.}~\bibnamefont {Han}},
  \bibinfo {author} {\bibfnamefont {P.}~\bibnamefont {Langacker}}, \bibinfo
  {author} {\bibfnamefont {B.}~\bibnamefont {McElrath}}, \ and\ \bibinfo
  {author} {\bibfnamefont {P.}~\bibnamefont {Zerwas}},\ }\href {\doibase
  10.1103/PhysRevD.67.115001} {\bibfield  {journal} {\bibinfo  {journal} {Phys.
  Rev.}\ }\textbf {\bibinfo {volume} {D67}},\ \bibinfo {pages} {115001}
  (\bibinfo {year} {2003})},\ \Eprint {http://arxiv.org/abs/hep-ph/0301097}
  {arXiv:hep-ph/0301097 [hep-ph]} \BibitemShut {NoStop}%
\bibitem [{\citenamefont {Blum}\ \emph {et~al.}(2015)\citenamefont {Blum},
  \citenamefont {D'Agnolo},\ and\ \citenamefont {Fan}}]{Blum:2015rpa}%
  \BibitemOpen
  \bibfield  {author} {\bibinfo {author} {\bibfnamefont {K.}~\bibnamefont
  {Blum}}, \bibinfo {author} {\bibfnamefont {R.~T.}\ \bibnamefont {D'Agnolo}},
  \ and\ \bibinfo {author} {\bibfnamefont {J.}~\bibnamefont {Fan}},\ }\href
  {\doibase 10.1007/JHEP03(2015)166} {\bibfield  {journal} {\bibinfo  {journal}
  {JHEP}\ }\textbf {\bibinfo {volume} {03}},\ \bibinfo {pages} {166} (\bibinfo
  {year} {2015})},\ \Eprint {http://arxiv.org/abs/1502.01045} {arXiv:1502.01045
  [hep-ph]} \BibitemShut {NoStop}%
\bibitem [{\citenamefont {Borchardt}\ and\ \citenamefont
  {Knorr}(2015)}]{Borchardt:2015rxa}%
  \BibitemOpen
  \bibfield  {author} {\bibinfo {author} {\bibfnamefont {J.}~\bibnamefont
  {Borchardt}}\ and\ \bibinfo {author} {\bibfnamefont {B.}~\bibnamefont
  {Knorr}},\ }\href {\doibase 10.1103/PhysRevD.91.105011} {\bibfield  {journal}
  {\bibinfo  {journal} {Phys. Rev.}\ }\textbf {\bibinfo {volume} {D91}},\
  \bibinfo {pages} {105011} (\bibinfo {year} {2015})},\ \Eprint
  {http://arxiv.org/abs/1502.07511} {arXiv:1502.07511 [hep-th]} \BibitemShut
  {NoStop}%
\bibitem [{\citenamefont {Borchardt}\ and\ \citenamefont
  {Knorr}()}]{Borchardt:2016}%
  \BibitemOpen
  \bibfield  {author} {\bibinfo {author} {\bibfnamefont {J.}~\bibnamefont
  {Borchardt}}\ and\ \bibinfo {author} {\bibfnamefont {B.}~\bibnamefont
  {Knorr}},\ }\href@noop {} {\enquote {\bibinfo {title} {{Solving functional
  flow equations with pseudo-spectral methods }},}\ }\bibinfo {note} {In
  preparation}\BibitemShut {NoStop}%
\bibitem [{\citenamefont {Holland}\ and\ \citenamefont
  {Kuti}(2004)}]{Holland:2003jr}%
  \BibitemOpen
  \bibfield  {author} {\bibinfo {author} {\bibfnamefont {K.}~\bibnamefont
  {Holland}}\ and\ \bibinfo {author} {\bibfnamefont {J.}~\bibnamefont {Kuti}},\
  }\bibfield  {booktitle} {\emph {\bibinfo {booktitle} {{Lattice hadron
  physics. Proceedings, 2nd Topical Workshop, LHP 2003, Cairns, Australia, July
  22-30, 2003}}},\ }\href {\doibase 10.1016/S0920-5632(03)02706-3} {\bibfield
  {journal} {\bibinfo  {journal} {Nucl. Phys. Proc. Suppl.}\ }\textbf {\bibinfo
  {volume} {129}},\ \bibinfo {pages} {765} (\bibinfo {year} {2004})},\ \bibinfo
  {note} {[,765(2003)]},\ \Eprint {http://arxiv.org/abs/hep-lat/0308020}
  {arXiv:hep-lat/0308020 [hep-lat]} \BibitemShut {NoStop}%
\bibitem [{\citenamefont {Holland}(2005)}]{Holland:2004sd}%
  \BibitemOpen
  \bibfield  {author} {\bibinfo {author} {\bibfnamefont {K.}~\bibnamefont
  {Holland}},\ }\bibfield  {booktitle} {\emph {\bibinfo {booktitle} {{Lattice
  field theory. Proceedings, 22nd International Symposium, Lattice 2004,
  Batavia, USA, June 21-26, 2004}}},\ }\href {\doibase
  10.1016/j.nuclphysbps.2004.11.293} {\bibfield  {journal} {\bibinfo  {journal}
  {Nucl. Phys. Proc. Suppl.}\ }\textbf {\bibinfo {volume} {140}},\ \bibinfo
  {pages} {155} (\bibinfo {year} {2005})},\ \bibinfo {note} {[,155(2004)]},\
  \Eprint {http://arxiv.org/abs/hep-lat/0409112} {arXiv:hep-lat/0409112
  [hep-lat]} \BibitemShut {NoStop}%
\bibitem [{\citenamefont {Branchina}\ and\ \citenamefont
  {Faivre}(2005)}]{Branchina:2005tu}%
  \BibitemOpen
  \bibfield  {author} {\bibinfo {author} {\bibfnamefont {V.}~\bibnamefont
  {Branchina}}\ and\ \bibinfo {author} {\bibfnamefont {H.}~\bibnamefont
  {Faivre}},\ }\href {\doibase 10.1103/PhysRevD.72.065017} {\bibfield
  {journal} {\bibinfo  {journal} {Phys. Rev.}\ }\textbf {\bibinfo {volume}
  {D72}},\ \bibinfo {pages} {065017} (\bibinfo {year} {2005})},\ \Eprint
  {http://arxiv.org/abs/hep-th/0503188} {arXiv:hep-th/0503188 [hep-th]}
  \BibitemShut {NoStop}%
\bibitem [{\citenamefont {Krajewski}\ and\ \citenamefont
  {Lalak}(2015)}]{Krajewski:2014vea}%
  \BibitemOpen
  \bibfield  {author} {\bibinfo {author} {\bibfnamefont {T.}~\bibnamefont
  {Krajewski}}\ and\ \bibinfo {author} {\bibfnamefont {Z.}~\bibnamefont
  {Lalak}},\ }\href {\doibase 10.1103/PhysRevD.92.075009} {\bibfield  {journal}
  {\bibinfo  {journal} {Phys. Rev.}\ }\textbf {\bibinfo {volume} {D92}},\
  \bibinfo {pages} {075009} (\bibinfo {year} {2015})},\ \Eprint
  {http://arxiv.org/abs/1411.6435} {arXiv:1411.6435 [hep-ph]} \BibitemShut
  {NoStop}%
\bibitem [{\citenamefont {Coleman}\ and\ \citenamefont
  {Weinberg}(1973)}]{Coleman:1973jx}%
  \BibitemOpen
  \bibfield  {author} {\bibinfo {author} {\bibfnamefont {S.~R.}\ \bibnamefont
  {Coleman}}\ and\ \bibinfo {author} {\bibfnamefont {E.~J.}\ \bibnamefont
  {Weinberg}},\ }\href {\doibase 10.1103/PhysRevD.7.1888} {\bibfield  {journal}
  {\bibinfo  {journal} {Phys. Rev.}\ }\textbf {\bibinfo {volume} {D7}},\
  \bibinfo {pages} {1888} (\bibinfo {year} {1973})}\BibitemShut {NoStop}%
\bibitem [{\citenamefont {Andreassen}\ \emph {et~al.}(2014)\citenamefont
  {Andreassen}, \citenamefont {Frost},\ and\ \citenamefont
  {Schwartz}}]{Andreassen:2014gha}%
  \BibitemOpen
  \bibfield  {author} {\bibinfo {author} {\bibfnamefont {A.}~\bibnamefont
  {Andreassen}}, \bibinfo {author} {\bibfnamefont {W.}~\bibnamefont {Frost}}, \
  and\ \bibinfo {author} {\bibfnamefont {M.~D.}\ \bibnamefont {Schwartz}},\
  }\href {\doibase 10.1103/PhysRevLett.113.241801} {\bibfield  {journal}
  {\bibinfo  {journal} {Phys. Rev. Lett.}\ }\textbf {\bibinfo {volume} {113}},\
  \bibinfo {pages} {241801} (\bibinfo {year} {2014})},\ \Eprint
  {http://arxiv.org/abs/1408.0292} {arXiv:1408.0292 [hep-ph]} \BibitemShut
  {NoStop}%
\bibitem [{\citenamefont {Andreassen}\ \emph {et~al.}(2015)\citenamefont
  {Andreassen}, \citenamefont {Frost},\ and\ \citenamefont
  {Schwartz}}]{Andreassen:2014eha}%
  \BibitemOpen
  \bibfield  {author} {\bibinfo {author} {\bibfnamefont {A.}~\bibnamefont
  {Andreassen}}, \bibinfo {author} {\bibfnamefont {W.}~\bibnamefont {Frost}}, \
  and\ \bibinfo {author} {\bibfnamefont {M.~D.}\ \bibnamefont {Schwartz}},\
  }\href {\doibase 10.1103/PhysRevD.91.016009} {\bibfield  {journal} {\bibinfo
  {journal} {Phys. Rev.}\ }\textbf {\bibinfo {volume} {D91}},\ \bibinfo {pages}
  {016009} (\bibinfo {year} {2015})},\ \Eprint {http://arxiv.org/abs/1408.0287}
  {arXiv:1408.0287 [hep-ph]} \BibitemShut {NoStop}%
\bibitem [{\citenamefont {Litim}(2000)}]{Litim:2000ci}%
  \BibitemOpen
  \bibfield  {author} {\bibinfo {author} {\bibfnamefont {D.~F.}\ \bibnamefont
  {Litim}},\ }\href {\doibase 10.1016/S0370-2693(00)00748-6} {\bibfield
  {journal} {\bibinfo  {journal} {Phys. Lett.}\ }\textbf {\bibinfo {volume}
  {B486}},\ \bibinfo {pages} {92} (\bibinfo {year} {2000})},\ \Eprint
  {http://arxiv.org/abs/hep-th/0005245} {arXiv:hep-th/0005245 [hep-th]}
  \BibitemShut {NoStop}%
\bibitem [{\citenamefont {Litim}(2001)}]{Litim:2001up}%
  \BibitemOpen
  \bibfield  {author} {\bibinfo {author} {\bibfnamefont {D.~F.}\ \bibnamefont
  {Litim}},\ }\href {\doibase 10.1103/PhysRevD.64.105007} {\bibfield  {journal}
  {\bibinfo  {journal} {Phys. Rev.}\ }\textbf {\bibinfo {volume} {D64}},\
  \bibinfo {pages} {105007} (\bibinfo {year} {2001})},\ \Eprint
  {http://arxiv.org/abs/hep-th/0103195} {arXiv:hep-th/0103195 [hep-th]}
  \BibitemShut {NoStop}%
\bibitem [{\citenamefont {Fodor}\ \emph {et~al.}(2007)\citenamefont {Fodor},
  \citenamefont {Holland}, \citenamefont {Kuti}, \citenamefont {Nogradi},\ and\
  \citenamefont {Schroeder}}]{Fodor:2007fn}%
  \BibitemOpen
  \bibfield  {author} {\bibinfo {author} {\bibfnamefont {Z.}~\bibnamefont
  {Fodor}}, \bibinfo {author} {\bibfnamefont {K.}~\bibnamefont {Holland}},
  \bibinfo {author} {\bibfnamefont {J.}~\bibnamefont {Kuti}}, \bibinfo {author}
  {\bibfnamefont {D.}~\bibnamefont {Nogradi}}, \ and\ \bibinfo {author}
  {\bibfnamefont {C.}~\bibnamefont {Schroeder}},\ }\bibfield  {booktitle}
  {\emph {\bibinfo {booktitle} {{Proceedings, 25th International Symposium on
  Lattice field theory (Lattice 2007)}}},\ }\href@noop {} {\bibfield  {journal}
  {\bibinfo  {journal} {PoS}\ }\textbf {\bibinfo {volume} {LAT2007}},\ \bibinfo
  {pages} {056} (\bibinfo {year} {2007})},\ \Eprint
  {http://arxiv.org/abs/0710.3151} {arXiv:0710.3151 [hep-lat]} \BibitemShut
  {NoStop}%
\bibitem [{\citenamefont {Gerhold}\ and\ \citenamefont
  {Jansen}(2007{\natexlab{a}})}]{Gerhold:2007yb}%
  \BibitemOpen
  \bibfield  {author} {\bibinfo {author} {\bibfnamefont {P.}~\bibnamefont
  {Gerhold}}\ and\ \bibinfo {author} {\bibfnamefont {K.}~\bibnamefont
  {Jansen}},\ }\href {\doibase 10.1088/1126-6708/2007/09/041} {\bibfield
  {journal} {\bibinfo  {journal} {JHEP}\ }\textbf {\bibinfo {volume} {09}},\
  \bibinfo {pages} {041} (\bibinfo {year} {2007}{\natexlab{a}})},\ \Eprint
  {http://arxiv.org/abs/0705.2539} {arXiv:0705.2539 [hep-lat]} \BibitemShut
  {NoStop}%
\bibitem [{\citenamefont {Gerhold}\ and\ \citenamefont
  {Jansen}(2007{\natexlab{b}})}]{Gerhold:2007gx}%
  \BibitemOpen
  \bibfield  {author} {\bibinfo {author} {\bibfnamefont {P.}~\bibnamefont
  {Gerhold}}\ and\ \bibinfo {author} {\bibfnamefont {K.}~\bibnamefont
  {Jansen}},\ }\href {\doibase 10.1088/1126-6708/2007/10/001} {\bibfield
  {journal} {\bibinfo  {journal} {JHEP}\ }\textbf {\bibinfo {volume} {10}},\
  \bibinfo {pages} {001} (\bibinfo {year} {2007}{\natexlab{b}})},\ \Eprint
  {http://arxiv.org/abs/0707.3849} {arXiv:0707.3849 [hep-lat]} \BibitemShut
  {NoStop}%
\bibitem [{\citenamefont {Gerhold}\ and\ \citenamefont
  {Jansen}(2009)}]{Gerhold:2009ub}%
  \BibitemOpen
  \bibfield  {author} {\bibinfo {author} {\bibfnamefont {P.}~\bibnamefont
  {Gerhold}}\ and\ \bibinfo {author} {\bibfnamefont {K.}~\bibnamefont
  {Jansen}},\ }\href {\doibase 10.1088/1126-6708/2009/07/025} {\bibfield
  {journal} {\bibinfo  {journal} {JHEP}\ }\textbf {\bibinfo {volume} {07}},\
  \bibinfo {pages} {025} (\bibinfo {year} {2009})},\ \Eprint
  {http://arxiv.org/abs/0902.4135} {arXiv:0902.4135 [hep-lat]} \BibitemShut
  {NoStop}%
\bibitem [{\citenamefont {Gerhold}\ \emph {et~al.}(2011)\citenamefont
  {Gerhold}, \citenamefont {Jansen},\ and\ \citenamefont
  {Kallarackal}}]{Gerhold:2010wv}%
  \BibitemOpen
  \bibfield  {author} {\bibinfo {author} {\bibfnamefont {P.}~\bibnamefont
  {Gerhold}}, \bibinfo {author} {\bibfnamefont {K.}~\bibnamefont {Jansen}}, \
  and\ \bibinfo {author} {\bibfnamefont {J.}~\bibnamefont {Kallarackal}},\
  }\href {\doibase 10.1007/JHEP01(2011)143} {\bibfield  {journal} {\bibinfo
  {journal} {JHEP}\ }\textbf {\bibinfo {volume} {01}},\ \bibinfo {pages} {143}
  (\bibinfo {year} {2011})},\ \Eprint {http://arxiv.org/abs/1011.1648}
  {arXiv:1011.1648 [hep-lat]} \BibitemShut {NoStop}%
\bibitem [{\citenamefont {Bulava}\ \emph
  {et~al.}(2013{\natexlab{a}})\citenamefont {Bulava}, \citenamefont {Gerhold},
  \citenamefont {Jansen}, \citenamefont {Kallarackal}, \citenamefont
  {Knippschild}, \citenamefont {Lin}, \citenamefont {Nagai}, \citenamefont
  {Nagy},\ and\ \citenamefont {Ogawa}}]{Bulava:2012rb}%
  \BibitemOpen
  \bibfield  {author} {\bibinfo {author} {\bibfnamefont {J.}~\bibnamefont
  {Bulava}}, \bibinfo {author} {\bibfnamefont {P.}~\bibnamefont {Gerhold}},
  \bibinfo {author} {\bibfnamefont {K.}~\bibnamefont {Jansen}}, \bibinfo
  {author} {\bibfnamefont {J.}~\bibnamefont {Kallarackal}}, \bibinfo {author}
  {\bibfnamefont {B.}~\bibnamefont {Knippschild}}, \bibinfo {author}
  {\bibfnamefont {C.~J.~D.}\ \bibnamefont {Lin}}, \bibinfo {author}
  {\bibfnamefont {K.-I.}\ \bibnamefont {Nagai}}, \bibinfo {author}
  {\bibfnamefont {A.}~\bibnamefont {Nagy}}, \ and\ \bibinfo {author}
  {\bibfnamefont {K.}~\bibnamefont {Ogawa}},\ }\href {\doibase
  10.1155/2013/875612} {\bibfield  {journal} {\bibinfo  {journal} {Adv. High
  Energy Phys.}\ }\textbf {\bibinfo {volume} {2013}},\ \bibinfo {pages}
  {875612} (\bibinfo {year} {2013}{\natexlab{a}})},\ \Eprint
  {http://arxiv.org/abs/1210.1798} {arXiv:1210.1798 [hep-lat]} \BibitemShut
  {NoStop}%
\bibitem [{\citenamefont {Bulava}\ \emph
  {et~al.}(2013{\natexlab{b}})\citenamefont {Bulava}, \citenamefont {Jansen},\
  and\ \citenamefont {Nagy}}]{Bulava:2013ep}%
  \BibitemOpen
  \bibfield  {author} {\bibinfo {author} {\bibfnamefont {J.}~\bibnamefont
  {Bulava}}, \bibinfo {author} {\bibfnamefont {K.}~\bibnamefont {Jansen}}, \
  and\ \bibinfo {author} {\bibfnamefont {A.}~\bibnamefont {Nagy}},\ }\href
  {\doibase 10.1016/j.physletb.2013.04.041} {\bibfield  {journal} {\bibinfo
  {journal} {Phys. Lett.}\ }\textbf {\bibinfo {volume} {B723}},\ \bibinfo
  {pages} {95} (\bibinfo {year} {2013}{\natexlab{b}})},\ \Eprint
  {http://arxiv.org/abs/1301.3416} {arXiv:1301.3416 [hep-lat]} \BibitemShut
  {NoStop}%
\bibitem [{\citenamefont {Djouadi}\ and\ \citenamefont
  {Lenz}(2012)}]{Djouadi:2012ae}%
  \BibitemOpen
  \bibfield  {author} {\bibinfo {author} {\bibfnamefont {A.}~\bibnamefont
  {Djouadi}}\ and\ \bibinfo {author} {\bibfnamefont {A.}~\bibnamefont {Lenz}},\
  }\href {\doibase 10.1016/j.physletb.2012.07.060} {\bibfield  {journal}
  {\bibinfo  {journal} {Phys. Lett.}\ }\textbf {\bibinfo {volume} {B715}},\
  \bibinfo {pages} {310} (\bibinfo {year} {2012})},\ \Eprint
  {http://arxiv.org/abs/1204.1252} {arXiv:1204.1252 [hep-ph]} \BibitemShut
  {NoStop}%
\bibitem [{\citenamefont {Hebecker}\ \emph {et~al.}(2013)\citenamefont
  {Hebecker}, \citenamefont {Knochel},\ and\ \citenamefont
  {Weigand}}]{Hebecker:2013lha}%
  \BibitemOpen
  \bibfield  {author} {\bibinfo {author} {\bibfnamefont {A.}~\bibnamefont
  {Hebecker}}, \bibinfo {author} {\bibfnamefont {A.~K.}\ \bibnamefont
  {Knochel}}, \ and\ \bibinfo {author} {\bibfnamefont {T.}~\bibnamefont
  {Weigand}},\ }\href {\doibase 10.1016/j.nuclphysb.2013.05.004} {\bibfield
  {journal} {\bibinfo  {journal} {Nucl. Phys.}\ }\textbf {\bibinfo {volume}
  {B874}},\ \bibinfo {pages} {1} (\bibinfo {year} {2013})},\ \Eprint
  {http://arxiv.org/abs/1304.2767} {arXiv:1304.2767 [hep-th]} \BibitemShut
  {NoStop}%
\bibitem [{\citenamefont {Wetterich}(1993)}]{Wetterich:1992yh}%
  \BibitemOpen
  \bibfield  {author} {\bibinfo {author} {\bibfnamefont {C.}~\bibnamefont
  {Wetterich}},\ }\href {\doibase 10.1016/0370-2693(93)90726-X} {\bibfield
  {journal} {\bibinfo  {journal} {Phys. Lett.}\ }\textbf {\bibinfo {volume}
  {B301}},\ \bibinfo {pages} {90} (\bibinfo {year} {1993})}\BibitemShut
  {NoStop}%
\bibitem [{\citenamefont {Berges}\ \emph {et~al.}(2002)\citenamefont {Berges},
  \citenamefont {Tetradis},\ and\ \citenamefont {Wetterich}}]{Berges:2000ew}%
  \BibitemOpen
  \bibfield  {author} {\bibinfo {author} {\bibfnamefont {J.}~\bibnamefont
  {Berges}}, \bibinfo {author} {\bibfnamefont {N.}~\bibnamefont {Tetradis}}, \
  and\ \bibinfo {author} {\bibfnamefont {C.}~\bibnamefont {Wetterich}},\ }\href
  {\doibase 10.1016/S0370-1573(01)00098-9} {\bibfield  {journal} {\bibinfo
  {journal} {Phys. Rept.}\ }\textbf {\bibinfo {volume} {363}},\ \bibinfo
  {pages} {223} (\bibinfo {year} {2002})},\ \Eprint
  {http://arxiv.org/abs/hep-ph/0005122} {arXiv:hep-ph/0005122 [hep-ph]}
  \BibitemShut {NoStop}%
\bibitem [{\citenamefont {Pawlowski}(2007)}]{Pawlowski:2005xe}%
  \BibitemOpen
  \bibfield  {author} {\bibinfo {author} {\bibfnamefont {J.~M.}\ \bibnamefont
  {Pawlowski}},\ }\href {\doibase 10.1016/j.aop.2007.01.007} {\bibfield
  {journal} {\bibinfo  {journal} {Annals Phys.}\ }\textbf {\bibinfo {volume}
  {322}},\ \bibinfo {pages} {2831} (\bibinfo {year} {2007})},\ \Eprint
  {http://arxiv.org/abs/hep-th/0512261} {arXiv:hep-th/0512261 [hep-th]}
  \BibitemShut {NoStop}%
\bibitem [{\citenamefont {Gies}(2012)}]{Gies:2006wv}%
  \BibitemOpen
  \bibfield  {author} {\bibinfo {author} {\bibfnamefont {H.}~\bibnamefont
  {Gies}},\ }\bibfield  {booktitle} {\emph {\bibinfo {booktitle} {{ECT* School
  on Renormalization Group and Effective Field Theory Approaches to Many-Body
  Systems Trento, Italy, February 27-March 10, 2006}}},\ }\href {\doibase
  10.1007/978-3-642-27320-9_6} {\bibfield  {journal} {\bibinfo  {journal}
  {Lect. Notes Phys.}\ }\textbf {\bibinfo {volume} {852}},\ \bibinfo {pages}
  {287} (\bibinfo {year} {2012})},\ \Eprint
  {http://arxiv.org/abs/hep-ph/0611146} {arXiv:hep-ph/0611146 [hep-ph]}
  \BibitemShut {NoStop}%
\bibitem [{\citenamefont {Delamotte}(2012)}]{Delamotte:2007pf}%
  \BibitemOpen
  \bibfield  {author} {\bibinfo {author} {\bibfnamefont {B.}~\bibnamefont
  {Delamotte}},\ }\href {\doibase 10.1007/978-3-642-27320-9_2} {\bibfield
  {journal} {\bibinfo  {journal} {Lect. Notes Phys.}\ }\textbf {\bibinfo
  {volume} {852}},\ \bibinfo {pages} {49} (\bibinfo {year} {2012})},\ \Eprint
  {http://arxiv.org/abs/cond-mat/0702365} {arXiv:cond-mat/0702365
  [cond-mat.stat-mech]} \BibitemShut {NoStop}%
\bibitem [{\citenamefont {Braun}(2012)}]{Braun:2011pp}%
  \BibitemOpen
  \bibfield  {author} {\bibinfo {author} {\bibfnamefont {J.}~\bibnamefont
  {Braun}},\ }\href {\doibase 10.1088/0954-3899/39/3/033001} {\bibfield
  {journal} {\bibinfo  {journal} {J. Phys.}\ }\textbf {\bibinfo {volume}
  {G39}},\ \bibinfo {pages} {033001} (\bibinfo {year} {2012})},\ \Eprint
  {http://arxiv.org/abs/1108.4449} {arXiv:1108.4449 [hep-ph]} \BibitemShut
  {NoStop}%
\bibitem [{\citenamefont {Gies}\ and\ \citenamefont
  {Scherer}(2010)}]{Gies:2009hq}%
  \BibitemOpen
  \bibfield  {author} {\bibinfo {author} {\bibfnamefont {H.}~\bibnamefont
  {Gies}}\ and\ \bibinfo {author} {\bibfnamefont {M.~M.}\ \bibnamefont
  {Scherer}},\ }\href {\doibase 10.1140/epjc/s10052-010-1256-z} {\bibfield
  {journal} {\bibinfo  {journal} {Eur. Phys. J.}\ }\textbf {\bibinfo {volume}
  {C66}},\ \bibinfo {pages} {387} (\bibinfo {year} {2010})},\ \Eprint
  {http://arxiv.org/abs/0901.2459} {arXiv:0901.2459 [hep-th]} \BibitemShut
  {NoStop}%
\bibitem [{\citenamefont {Vacca}\ and\ \citenamefont
  {Zambelli}(2015)}]{Vacca:2015nta}%
  \BibitemOpen
  \bibfield  {author} {\bibinfo {author} {\bibfnamefont {G.~P.}\ \bibnamefont
  {Vacca}}\ and\ \bibinfo {author} {\bibfnamefont {L.}~\bibnamefont
  {Zambelli}},\ }\href {\doibase 10.1103/PhysRevD.91.125003} {\bibfield
  {journal} {\bibinfo  {journal} {Phys. Rev.}\ }\textbf {\bibinfo {volume}
  {D91}},\ \bibinfo {pages} {125003} (\bibinfo {year} {2015})},\ \Eprint
  {http://arxiv.org/abs/1503.09136} {arXiv:1503.09136 [hep-th]} \BibitemShut
  {NoStop}%
\bibitem [{\citenamefont {Bohr}\ \emph {et~al.}(2001)\citenamefont {Bohr},
  \citenamefont {Schaefer},\ and\ \citenamefont {Wambach}}]{Bohr:2000gp}%
  \BibitemOpen
  \bibfield  {author} {\bibinfo {author} {\bibfnamefont {O.}~\bibnamefont
  {Bohr}}, \bibinfo {author} {\bibfnamefont {B.~J.}\ \bibnamefont {Schaefer}},
  \ and\ \bibinfo {author} {\bibfnamefont {J.}~\bibnamefont {Wambach}},\ }\href
  {\doibase 10.1142/S0217751X0100502X} {\bibfield  {journal} {\bibinfo
  {journal} {Int. J. Mod. Phys.}\ }\textbf {\bibinfo {volume} {A16}},\ \bibinfo
  {pages} {3823} (\bibinfo {year} {2001})},\ \Eprint
  {http://arxiv.org/abs/hep-ph/0007098} {arXiv:hep-ph/0007098 [hep-ph]}
  \BibitemShut {NoStop}%
\bibitem [{\citenamefont {Herbst}\ \emph {et~al.}(2013)\citenamefont {Herbst},
  \citenamefont {Pawlowski},\ and\ \citenamefont {Schaefer}}]{Herbst:2013ail}%
  \BibitemOpen
  \bibfield  {author} {\bibinfo {author} {\bibfnamefont {T.~K.}\ \bibnamefont
  {Herbst}}, \bibinfo {author} {\bibfnamefont {J.~M.}\ \bibnamefont
  {Pawlowski}}, \ and\ \bibinfo {author} {\bibfnamefont {B.-J.}\ \bibnamefont
  {Schaefer}},\ }\href {\doibase 10.1103/PhysRevD.88.014007} {\bibfield
  {journal} {\bibinfo  {journal} {Phys. Rev.}\ }\textbf {\bibinfo {volume}
  {D88}},\ \bibinfo {pages} {014007} (\bibinfo {year} {2013})},\ \Eprint
  {http://arxiv.org/abs/1302.1426} {arXiv:1302.1426 [hep-ph]} \BibitemShut
  {NoStop}%
\bibitem [{\citenamefont {Adams}\ \emph {et~al.}(1995)\citenamefont {Adams},
  \citenamefont {Berges}, \citenamefont {Bornholdt}, \citenamefont {Freire},
  \citenamefont {Tetradis},\ and\ \citenamefont {Wetterich}}]{Adams:1995cv}%
  \BibitemOpen
  \bibfield  {author} {\bibinfo {author} {\bibfnamefont {J.~A.}\ \bibnamefont
  {Adams}}, \bibinfo {author} {\bibfnamefont {J.}~\bibnamefont {Berges}},
  \bibinfo {author} {\bibfnamefont {S.}~\bibnamefont {Bornholdt}}, \bibinfo
  {author} {\bibfnamefont {F.}~\bibnamefont {Freire}}, \bibinfo {author}
  {\bibfnamefont {N.}~\bibnamefont {Tetradis}}, \ and\ \bibinfo {author}
  {\bibfnamefont {C.}~\bibnamefont {Wetterich}},\ }\href {\doibase
  10.1142/S0217732395002520} {\bibfield  {journal} {\bibinfo  {journal} {Mod.
  Phys. Lett.}\ }\textbf {\bibinfo {volume} {A10}},\ \bibinfo {pages} {2367}
  (\bibinfo {year} {1995})},\ \Eprint {http://arxiv.org/abs/hep-th/9507093}
  {arXiv:hep-th/9507093 [hep-th]} \BibitemShut {NoStop}%
\bibitem [{\citenamefont {Hofling}\ \emph {et~al.}(2002)\citenamefont
  {Hofling}, \citenamefont {Nowak},\ and\ \citenamefont
  {Wetterich}}]{Hofling:2002hj}%
  \BibitemOpen
  \bibfield  {author} {\bibinfo {author} {\bibfnamefont {F.}~\bibnamefont
  {Hofling}}, \bibinfo {author} {\bibfnamefont {C.}~\bibnamefont {Nowak}}, \
  and\ \bibinfo {author} {\bibfnamefont {C.}~\bibnamefont {Wetterich}},\ }\href
  {\doibase 10.1103/PhysRevB.66.205111} {\bibfield  {journal} {\bibinfo
  {journal} {Phys. Rev.}\ }\textbf {\bibinfo {volume} {B66}},\ \bibinfo {pages}
  {205111} (\bibinfo {year} {2002})},\ \Eprint
  {http://arxiv.org/abs/cond-mat/0203588} {arXiv:cond-mat/0203588 [cond-mat]}
  \BibitemShut {NoStop}%
\bibitem [{\citenamefont {Bonanno}\ and\ \citenamefont
  {Zappala}(2001)}]{Bonanno:2000yp}%
  \BibitemOpen
  \bibfield  {author} {\bibinfo {author} {\bibfnamefont {A.}~\bibnamefont
  {Bonanno}}\ and\ \bibinfo {author} {\bibfnamefont {D.}~\bibnamefont
  {Zappala}},\ }\href {\doibase 10.1016/S0370-2693(01)00273-8} {\bibfield
  {journal} {\bibinfo  {journal} {Phys. Lett.}\ }\textbf {\bibinfo {volume}
  {B504}},\ \bibinfo {pages} {181} (\bibinfo {year} {2001})},\ \Eprint
  {http://arxiv.org/abs/hep-th/0010095} {arXiv:hep-th/0010095 [hep-th]}
  \BibitemShut {NoStop}%
\bibitem [{\citenamefont {Boettcher}\ \emph {et~al.}(2015)\citenamefont
  {Boettcher}, \citenamefont {Braun}, \citenamefont {Herbst}, \citenamefont
  {Pawlowski}, \citenamefont {Roscher},\ and\ \citenamefont
  {Wetterich}}]{Boettcher:2014tfa}%
  \BibitemOpen
  \bibfield  {author} {\bibinfo {author} {\bibfnamefont {I.}~\bibnamefont
  {Boettcher}}, \bibinfo {author} {\bibfnamefont {J.}~\bibnamefont {Braun}},
  \bibinfo {author} {\bibfnamefont {T.~K.}\ \bibnamefont {Herbst}}, \bibinfo
  {author} {\bibfnamefont {J.~M.}\ \bibnamefont {Pawlowski}}, \bibinfo {author}
  {\bibfnamefont {D.}~\bibnamefont {Roscher}}, \ and\ \bibinfo {author}
  {\bibfnamefont {C.}~\bibnamefont {Wetterich}},\ }\href {\doibase
  10.1103/PhysRevA.91.013610} {\bibfield  {journal} {\bibinfo  {journal} {Phys.
  Rev.}\ }\textbf {\bibinfo {volume} {A91}},\ \bibinfo {pages} {013610}
  (\bibinfo {year} {2015})},\ \Eprint {http://arxiv.org/abs/1409.5070}
  {arXiv:1409.5070 [cond-mat.quant-gas]} \BibitemShut {NoStop}%
\bibitem [{\citenamefont {O'Raifeartaigh}\ \emph {et~al.}(1986)\citenamefont
  {O'Raifeartaigh}, \citenamefont {Wipf},\ and\ \citenamefont
  {Yoneyama}}]{ORaifeartaigh:1986hi}%
  \BibitemOpen
  \bibfield  {author} {\bibinfo {author} {\bibfnamefont {L.}~\bibnamefont
  {O'Raifeartaigh}}, \bibinfo {author} {\bibfnamefont {A.}~\bibnamefont
  {Wipf}}, \ and\ \bibinfo {author} {\bibfnamefont {H.}~\bibnamefont
  {Yoneyama}},\ }\href {\doibase 10.1016/S0550-3213(86)80031-1} {\bibfield
  {journal} {\bibinfo  {journal} {Nucl. Phys.}\ }\textbf {\bibinfo {volume}
  {B271}},\ \bibinfo {pages} {653} (\bibinfo {year} {1986})}\BibitemShut
  {NoStop}%
\bibitem [{\citenamefont {Litim}\ \emph {et~al.}(2006)\citenamefont {Litim},
  \citenamefont {Pawlowski},\ and\ \citenamefont {Vergara}}]{Litim:2006nn}%
  \BibitemOpen
  \bibfield  {author} {\bibinfo {author} {\bibfnamefont {D.~F.}\ \bibnamefont
  {Litim}}, \bibinfo {author} {\bibfnamefont {J.~M.}\ \bibnamefont
  {Pawlowski}}, \ and\ \bibinfo {author} {\bibfnamefont {L.}~\bibnamefont
  {Vergara}},\ }\href@noop {} {\  (\bibinfo {year} {2006})},\ \Eprint
  {http://arxiv.org/abs/hep-th/0602140} {arXiv:hep-th/0602140 [hep-th]}
  \BibitemShut {NoStop}%
\bibitem [{\citenamefont {Boyd}(2000)}]{Boyd:ChebyFourier}%
  \BibitemOpen
  \bibfield  {author} {\bibinfo {author} {\bibfnamefont {J.~P.}\ \bibnamefont
  {Boyd}},\ }\href@noop {} {\emph {\bibinfo {title} {{Chebyshev and Fourier
  Spectral Methods}}}},\ \bibinfo {edition} {2nd}\ ed.\ (\bibinfo  {publisher}
  {Dover Publications},\ \bibinfo {year} {2000})\BibitemShut {NoStop}%
\bibitem [{\citenamefont {Robson}\ and\ \citenamefont
  {Prytz}(1993)}]{Robson:1993}%
  \BibitemOpen
  \bibfield  {author} {\bibinfo {author} {\bibfnamefont {R.}~\bibnamefont
  {Robson}}\ and\ \bibinfo {author} {\bibfnamefont {A.}~\bibnamefont {Prytz}},\
  }\href@noop {} {\bibfield  {journal} {\bibinfo  {journal} {Australian Journal
  of Physics}\ }\textbf {\bibinfo {volume} {46}} (\bibinfo {year}
  {1993})}\BibitemShut {NoStop}%
\bibitem [{\citenamefont {Fischer}\ and\ \citenamefont
  {Gies}(2004)}]{Fischer:2004uk}%
  \BibitemOpen
  \bibfield  {author} {\bibinfo {author} {\bibfnamefont {C.~S.}\ \bibnamefont
  {Fischer}}\ and\ \bibinfo {author} {\bibfnamefont {H.}~\bibnamefont {Gies}},\
  }\href {\doibase 10.1088/1126-6708/2004/10/048} {\bibfield  {journal}
  {\bibinfo  {journal} {JHEP}\ }\textbf {\bibinfo {volume} {10}},\ \bibinfo
  {pages} {048} (\bibinfo {year} {2004})},\ \Eprint
  {http://arxiv.org/abs/hep-ph/0408089} {arXiv:hep-ph/0408089 [hep-ph]}
  \BibitemShut {NoStop}%
\bibitem [{\citenamefont {Gneiting}()}]{Gneiting:2005}%
  \BibitemOpen
  \bibfield  {author} {\bibinfo {author} {\bibfnamefont {C.}~\bibnamefont
  {Gneiting}},\ }\href@noop {} {\enquote {\bibinfo {title} {{Diploma
  thesis}},}\ }\bibinfo {note} {Heidelberg 2005}\BibitemShut {NoStop}%
\bibitem [{\citenamefont {Bonanno}\ and\ \citenamefont
  {Lacagnina}(2004)}]{Bonanno:2004pq}%
  \BibitemOpen
  \bibfield  {author} {\bibinfo {author} {\bibfnamefont {A.}~\bibnamefont
  {Bonanno}}\ and\ \bibinfo {author} {\bibfnamefont {G.}~\bibnamefont
  {Lacagnina}},\ }\href {\doibase 10.1016/j.nuclphysb.2004.06.003} {\bibfield
  {journal} {\bibinfo  {journal} {Nucl. Phys.}\ }\textbf {\bibinfo {volume}
  {B693}},\ \bibinfo {pages} {36} (\bibinfo {year} {2004})},\ \Eprint
  {http://arxiv.org/abs/hep-th/0403176} {arXiv:hep-th/0403176 [hep-th]}
  \BibitemShut {NoStop}%
\bibitem [{\citenamefont {Peláez}\ and\ \citenamefont
  {Wschebor}(2015)}]{Pelaez:2015nsa}%
  \BibitemOpen
  \bibfield  {author} {\bibinfo {author} {\bibfnamefont {M.}~\bibnamefont
  {Peláez}}\ and\ \bibinfo {author} {\bibfnamefont {N.}~\bibnamefont
  {Wschebor}},\ }\href@noop {} {\  (\bibinfo {year} {2015})},\ \Eprint
  {http://arxiv.org/abs/1510.05709} {arXiv:1510.05709 [cond-mat.stat-mech]}
  \BibitemShut {NoStop}%
\bibitem [{\citenamefont {Coleman}(1977)}]{Coleman:1977py}%
  \BibitemOpen
  \bibfield  {author} {\bibinfo {author} {\bibfnamefont {S.~R.}\ \bibnamefont
  {Coleman}},\ }\href {\doibase 10.1103/PhysRevD.15.2929,
  10.1103/PhysRevD.16.1248} {\bibfield  {journal} {\bibinfo  {journal} {Phys.
  Rev.}\ }\textbf {\bibinfo {volume} {D15}},\ \bibinfo {pages} {2929} (\bibinfo
  {year} {1977})},\ \bibinfo {note} {[Erratum: Phys.
  Rev.D16,1248(1977)]}\BibitemShut {NoStop}%
\bibitem [{\citenamefont {Callan}\ and\ \citenamefont
  {Coleman}(1977)}]{Callan:1977pt}%
  \BibitemOpen
  \bibfield  {author} {\bibinfo {author} {\bibfnamefont {C.~G.}\ \bibnamefont
  {Callan}, \bibfnamefont {Jr.}}\ and\ \bibinfo {author} {\bibfnamefont
  {S.~R.}\ \bibnamefont {Coleman}},\ }\href {\doibase 10.1103/PhysRevD.16.1762}
  {\bibfield  {journal} {\bibinfo  {journal} {Phys. Rev.}\ }\textbf {\bibinfo
  {volume} {D16}},\ \bibinfo {pages} {1762} (\bibinfo {year}
  {1977})}\BibitemShut {NoStop}%
\bibitem [{\citenamefont {Strumia}\ \emph {et~al.}(1999)\citenamefont
  {Strumia}, \citenamefont {Tetradis},\ and\ \citenamefont
  {Wetterich}}]{Strumia:1998qq}%
  \BibitemOpen
  \bibfield  {author} {\bibinfo {author} {\bibfnamefont {A.}~\bibnamefont
  {Strumia}}, \bibinfo {author} {\bibfnamefont {N.}~\bibnamefont {Tetradis}}, \
  and\ \bibinfo {author} {\bibfnamefont {C.}~\bibnamefont {Wetterich}},\ }\href
  {\doibase 10.1016/S0370-2693(99)01158-2} {\bibfield  {journal} {\bibinfo
  {journal} {Phys. Lett.}\ }\textbf {\bibinfo {volume} {B467}},\ \bibinfo
  {pages} {279} (\bibinfo {year} {1999})},\ \Eprint
  {http://arxiv.org/abs/hep-ph/9808263} {arXiv:hep-ph/9808263 [hep-ph]}
  \BibitemShut {NoStop}%
\bibitem [{\citenamefont {Garbrecht}\ and\ \citenamefont
  {Millington}(2015{\natexlab{a}})}]{Garbrecht:2015oea}%
  \BibitemOpen
  \bibfield  {author} {\bibinfo {author} {\bibfnamefont {B.}~\bibnamefont
  {Garbrecht}}\ and\ \bibinfo {author} {\bibfnamefont {P.}~\bibnamefont
  {Millington}},\ }\href {\doibase 10.1103/PhysRevD.91.105021} {\bibfield
  {journal} {\bibinfo  {journal} {Phys. Rev.}\ }\textbf {\bibinfo {volume}
  {D91}},\ \bibinfo {pages} {105021} (\bibinfo {year} {2015}{\natexlab{a}})},\
  \Eprint {http://arxiv.org/abs/1501.07466} {arXiv:1501.07466 [hep-th]}
  \BibitemShut {NoStop}%
\bibitem [{\citenamefont {Garbrecht}\ and\ \citenamefont
  {Millington}(2015{\natexlab{b}})}]{Garbrecht:2015yza}%
  \BibitemOpen
  \bibfield  {author} {\bibinfo {author} {\bibfnamefont {B.}~\bibnamefont
  {Garbrecht}}\ and\ \bibinfo {author} {\bibfnamefont {P.}~\bibnamefont
  {Millington}},\ }\href {\doibase 10.1103/PhysRevD.92.125022} {\bibfield
  {journal} {\bibinfo  {journal} {Phys. Rev.}\ }\textbf {\bibinfo {volume}
  {D92}},\ \bibinfo {pages} {125022} (\bibinfo {year} {2015}{\natexlab{b}})},\
  \Eprint {http://arxiv.org/abs/1509.08480} {arXiv:1509.08480 [hep-ph]}
  \BibitemShut {NoStop}%
\bibitem [{\citenamefont {Branchina}\ and\ \citenamefont
  {Messina}(2013)}]{Branchina:2013jra}%
  \BibitemOpen
  \bibfield  {author} {\bibinfo {author} {\bibfnamefont {V.}~\bibnamefont
  {Branchina}}\ and\ \bibinfo {author} {\bibfnamefont {E.}~\bibnamefont
  {Messina}},\ }\href {\doibase 10.1103/PhysRevLett.111.241801} {\bibfield
  {journal} {\bibinfo  {journal} {Phys. Rev. Lett.}\ }\textbf {\bibinfo
  {volume} {111}},\ \bibinfo {pages} {241801} (\bibinfo {year} {2013})},\
  \Eprint {http://arxiv.org/abs/1307.5193} {arXiv:1307.5193 [hep-ph]}
  \BibitemShut {NoStop}%
\bibitem [{\citenamefont {Branchina}\ \emph {et~al.}(2014)\citenamefont
  {Branchina}, \citenamefont {Messina},\ and\ \citenamefont
  {Platania}}]{Branchina:2014usa}%
  \BibitemOpen
  \bibfield  {author} {\bibinfo {author} {\bibfnamefont {V.}~\bibnamefont
  {Branchina}}, \bibinfo {author} {\bibfnamefont {E.}~\bibnamefont {Messina}},
  \ and\ \bibinfo {author} {\bibfnamefont {A.}~\bibnamefont {Platania}},\
  }\href {\doibase 10.1007/JHEP09(2014)182} {\bibfield  {journal} {\bibinfo
  {journal} {JHEP}\ }\textbf {\bibinfo {volume} {09}},\ \bibinfo {pages} {182}
  (\bibinfo {year} {2014})},\ \Eprint {http://arxiv.org/abs/1407.4112}
  {arXiv:1407.4112 [hep-ph]} \BibitemShut {NoStop}%
\bibitem [{\citenamefont {Branchina}\ \emph {et~al.}(2015)\citenamefont
  {Branchina}, \citenamefont {Messina},\ and\ \citenamefont
  {Sher}}]{Branchina:2014rva}%
  \BibitemOpen
  \bibfield  {author} {\bibinfo {author} {\bibfnamefont {V.}~\bibnamefont
  {Branchina}}, \bibinfo {author} {\bibfnamefont {E.}~\bibnamefont {Messina}},
  \ and\ \bibinfo {author} {\bibfnamefont {M.}~\bibnamefont {Sher}},\ }\href
  {\doibase 10.1103/PhysRevD.91.013003} {\bibfield  {journal} {\bibinfo
  {journal} {Phys. Rev.}\ }\textbf {\bibinfo {volume} {D91}},\ \bibinfo {pages}
  {013003} (\bibinfo {year} {2015})},\ \Eprint {http://arxiv.org/abs/1408.5302}
  {arXiv:1408.5302 [hep-ph]} \BibitemShut {NoStop}%
\bibitem [{\citenamefont {Lalak}\ \emph {et~al.}(2014)\citenamefont {Lalak},
  \citenamefont {Lewicki},\ and\ \citenamefont {Olszewski}}]{Lalak:2014qua}%
  \BibitemOpen
  \bibfield  {author} {\bibinfo {author} {\bibfnamefont {Z.}~\bibnamefont
  {Lalak}}, \bibinfo {author} {\bibfnamefont {M.}~\bibnamefont {Lewicki}}, \
  and\ \bibinfo {author} {\bibfnamefont {P.}~\bibnamefont {Olszewski}},\ }\href
  {\doibase 10.1007/JHEP05(2014)119} {\bibfield  {journal} {\bibinfo  {journal}
  {JHEP}\ }\textbf {\bibinfo {volume} {05}},\ \bibinfo {pages} {119} (\bibinfo
  {year} {2014})},\ \Eprint {http://arxiv.org/abs/1402.3826} {arXiv:1402.3826
  [hep-ph]} \BibitemShut {NoStop}%
\bibitem [{\citenamefont {Loebbert}\ and\ \citenamefont
  {Plefka}(2015)}]{Loebbert:2015eea}%
  \BibitemOpen
  \bibfield  {author} {\bibinfo {author} {\bibfnamefont {F.}~\bibnamefont
  {Loebbert}}\ and\ \bibinfo {author} {\bibfnamefont {J.}~\bibnamefont
  {Plefka}},\ }\href@noop {} {\  (\bibinfo {year} {2015})},\ \Eprint
  {http://arxiv.org/abs/1502.03093} {arXiv:1502.03093 [hep-ph]} \BibitemShut
  {NoStop}%
\bibitem [{\citenamefont {Bhattacharjee}\ and\ \citenamefont
  {Majumdar}(2014)}]{Bhattacharjee:2012my}%
  \BibitemOpen
  \bibfield  {author} {\bibinfo {author} {\bibfnamefont {S.}~\bibnamefont
  {Bhattacharjee}}\ and\ \bibinfo {author} {\bibfnamefont {P.}~\bibnamefont
  {Majumdar}},\ }\href {\doibase 10.1016/j.nuclphysb.2014.05.031} {\bibfield
  {journal} {\bibinfo  {journal} {Nucl. Phys.}\ }\textbf {\bibinfo {volume}
  {B885}},\ \bibinfo {pages} {481} (\bibinfo {year} {2014})},\ \Eprint
  {http://arxiv.org/abs/1210.0497} {arXiv:1210.0497 [hep-th]} \BibitemShut
  {NoStop}%
\bibitem [{\citenamefont {Haba}\ \emph {et~al.}(2015)\citenamefont {Haba},
  \citenamefont {Kaneta}, \citenamefont {Takahashi},\ and\ \citenamefont
  {Yamaguchi}}]{Haba:2014qca}%
  \BibitemOpen
  \bibfield  {author} {\bibinfo {author} {\bibfnamefont {N.}~\bibnamefont
  {Haba}}, \bibinfo {author} {\bibfnamefont {K.}~\bibnamefont {Kaneta}},
  \bibinfo {author} {\bibfnamefont {R.}~\bibnamefont {Takahashi}}, \ and\
  \bibinfo {author} {\bibfnamefont {Y.}~\bibnamefont {Yamaguchi}},\ }\href
  {\doibase 10.1103/PhysRevD.91.016004} {\bibfield  {journal} {\bibinfo
  {journal} {Phys. Rev.}\ }\textbf {\bibinfo {volume} {D91}},\ \bibinfo {pages}
  {016004} (\bibinfo {year} {2015})},\ \Eprint {http://arxiv.org/abs/1408.5548}
  {arXiv:1408.5548 [hep-ph]} \BibitemShut {NoStop}%
\bibitem [{\citenamefont {Abe}\ \emph {et~al.}(2016)\citenamefont {Abe},
  \citenamefont {Horikoshi},\ and\ \citenamefont {Inami}}]{Abe:2016irv}%
  \BibitemOpen
  \bibfield  {author} {\bibinfo {author} {\bibfnamefont {Y.}~\bibnamefont
  {Abe}}, \bibinfo {author} {\bibfnamefont {M.}~\bibnamefont {Horikoshi}}, \
  and\ \bibinfo {author} {\bibfnamefont {T.}~\bibnamefont {Inami}},\
  }\href@noop {} {\  (\bibinfo {year} {2016})},\ \Eprint
  {http://arxiv.org/abs/1602.03792} {arXiv:1602.03792 [hep-ph]} \BibitemShut
  {NoStop}%
\bibitem [{\citenamefont {Akerlund}\ \emph {et~al.}(2015)\citenamefont
  {Akerlund}, \citenamefont {de~Forcrand},\ and\ \citenamefont
  {Steinbauer}}]{Akerlund:2015gfy}%
  \BibitemOpen
  \bibfield  {author} {\bibinfo {author} {\bibfnamefont {O.}~\bibnamefont
  {Akerlund}}, \bibinfo {author} {\bibfnamefont {P.}~\bibnamefont
  {de~Forcrand}}, \ and\ \bibinfo {author} {\bibfnamefont {J.}~\bibnamefont
  {Steinbauer}},\ }in\ \href
  {http://inspirehep.net/record/1404165/files/arXiv:1511.03867.pdf} {\emph
  {\bibinfo {booktitle} {{Proceedings, 33rd International Symposium on Lattice
  Field Theory (Lattice 2015)}}}}\ (\bibinfo {year} {2015})\ \Eprint
  {http://arxiv.org/abs/1511.03867} {arXiv:1511.03867 [hep-lat]} \BibitemShut
  {NoStop}%
\end{thebibliography}%

\end{document}